\documentclass[structabstract]{aa}  
\usepackage{graphicx,natbib}
\usepackage{lscape}
\usepackage{longtable}
\usepackage{txfonts}

\newcommand{\mi}{\relax \ifmmode {\mu{\mbox m}}\else $\mu$m\fi}
\newcommand{\hii}{\relax \ifmmode {\mbox H\,{\scshape ii}}\else H\,{\scshape ii}\fi}
\newcommand{\sii}{\relax \ifmmode {\mbox S\,{\scshape ii}}\else S\,{\scshape ii}\fi}
\newcommand{\siii}{\relax \ifmmode {\mbox S\,{\textsc {iii}}}\else S\,{\scshape iii}\fi}
\newcommand{\siv}{\relax \ifmmode {\mbox S\,{\textsc {iv}}}\else S\,{\scshape iv}\fi}
\newcommand{\nii}{\relax \ifmmode {\mbox N\,{\scshape ii}}\else N\,{\scshape ii}\fi}
\newcommand{\neii}{\relax \ifmmode {\mbox Ne\,{\textsc {ii}}}\else Ne\,{\scshape ii}\fi}
\newcommand{\neiii}{\relax \ifmmode {\mbox Ne\,{\textsc {iii}}}\else Ne\,{\scshape iii}\fi}
\newcommand{\oiii}{\relax \ifmmode {\mbox O\,{\scshape iii}}\else O\,{\scshape iii}\fi}
\newcommand{\oii}{\relax \ifmmode {\mbox O\,{\scshape ii}}\else O\,{\scshape ii}\fi}
\newcommand{\oi}{\relax \ifmmode {\mbox O\,{\scshape i}}\else O\,{\scshape i}\fi}
\newcommand{\ha}{\relax \ifmmode {\mbox H}\alpha\else H$\alpha$\fi}
\newcommand{\hep}{\relax \ifmmode {\mbox H}\epsilon\else H$\epsilon$\fi}
\newcommand{\hdel}{\relax \ifmmode {\mbox H}\delta\else H$\delta$\fi}
\newcommand{\hgam}{\relax \ifmmode {\mbox H}\gamma\else H$\gamma$\fi}

\newcommand{\pa}{\relax \ifmmode {\mbox Pa}\alpha\else Pa$\alpha$\fi}
\newcommand{\hb}{\relax \ifmmode {\mbox H}\beta\else H$\beta$\fi}
\newcommand{\rdostres}{\relax \ifmmode {\,\mbox{R}}_{\rm 23}\else \,\mbox{R}$_{\rm 23}$\fi}
\newcommand{\ergs}{\relax \ifmmode {\,\mbox{erg\,s}}^{-1}\else \,\mbox{erg\,s}$^{-1}$\fi}
\newcommand{\me}{\relax \ifmmode {\,}^{-1}\else \,$^{-1}$\fi}

\newcommand{\kms}{km\,s$^{-1}$}

\newcommand{\msun}{\relax \ifmmode {\,\mbox{M}}_{\odot}\else \,\mbox{M}$_{\odot}$\fi}
\newcommand{\zsun}{Z$_{\odot}$}
\newcommand{\cmtres}{\relax \ifmmode {\,\mbox{cm}}^{-3}\else \,\mbox{cm}$^{-3}$\fi}
\newcommand{\cmdos}{\relax \ifmmode {\,\mbox{cm}}^{-2}\else \,\mbox{cm}$^{-2}$\fi}
\newcommand{\cmseis}{\relax \ifmmode {\,\mbox{cm}}^{-6}\else \,\mbox{cm}$^{-6}$\fi}
\newcommand{\hi}{\relax \ifmmode {\mbox H\,{\scshape i}}\else H\,{\scshape i}\fi}

\newcommand{\Spi}{{\it Spitzer}}
\newcommand{\Her}{{\it Herschel}}
\newcommand{\DE}{{\tt DustEM}}
\newcommand{\ytot}{$Y_{\rm TOTAL}$}
\newcommand{\ypah}{$Y_{\rm PAH}$}
\newcommand{\yvsg}{$Y_{\rm VSG}$}
\newcommand{\ybg}{$Y_{\rm BG}$}

\usepackage[colorlinks]{hyperref}
\begin{document}

   \title{Dust Properties in H\,{\sc ii} Regions in M\,33}
   \author{
     M.\,Rela\~{n}o\inst{1,2}  \and
     R. Kennicutt\inst{3} \and
     U.\,Lisenfeld\inst{1,2} \and
     S.\,Verley\inst{1,2} \and
     I.\,Hermelo\inst{4} \and
     M.\,Boquien\inst{3,5} \and  
     M.\,Albrecht\inst{6}\and 
     C.\,Kramer\inst{4}\and 
     J.\,Braine\inst{7} \and 
     E. P\'erez-Montero\inst{8} \and 
     I. \,De Looze\inst{9,10} \and 
     M. Xilouris\inst{11} \and  
     A.\,Kov{\'a}cs\inst{12} \and 
     J.\,Staguhn\inst{13}}  
   \institute{
     Dept. F\'{i}sica Te\'orica y del Cosmos, Universidad de Granada, Spain -- \email{mrelano@ugr.es}
     \and
      Instituto Universitario Carlos I de F\'isica Te\'orica y Computacional, Universidad de Granada, 18071, Granada, Spain
     \and
     Institute of Astronomy, University of Cambridge, Madingley Road, Cambridge, CB3 0HA, UK
     \and
     Instituto Radioastronom\'{i}a Milim\'{e}trica, Av. Divina Pastora 7, N\'ucleo Central, E-18012 Granada, Spain
     \and
     Unidad de Astronom\'{i}a, Fac. Cs. B\'asicas, Universidad de Antofagasta, Avda. U. de Antofagasta 02800, Antofagasta, Chile
     \and
     Argelander-Institut f\"ur Astronomie, University of Bonn, Auf dem H\"ugel 71, D-53121 Bonn, Germany
     \and 
     Univ. Bordeaux, Laboratoire d'Astrophysique de Bordeaux, CNRS, LAB, UMR 5804, F-33270, Floirac, France
     \and
     Instituto de Astrof\'i­sica de Andaluc\'i­a - CSIC. Apdo. 3004, 18008, Granada, Spain
     \and
     Department of Physics and Astronomy, University College London, Gower Street, London WC1E 6BT, UK
     \and 
     Sterrenkundig Observatorium, Universiteit Gent, Krijgslaan 281 S9, B-9000 Gent, Belgium
     \and 
     Institute of Astronomy and Astrophysics, National Observatory of Athens, P. Penteli, 15236 Athens, Greece
     \and
     Department of Astronomy, University of Minnesota, 116 Church St SE, Minneapolis, MN 55414, USA 
    \and
    NASA Goddard Space Flight Center, Code 665, Greenbelt, MD 20771, USA
    }

   \date{Received ; accepted }

 
  \abstract
   {The infrared emission (IR) of the interstellar dust has been claimed to be a tracer of the star formation rate. However, the conversion of the IR emission into star formation rate can be strongly dependent on the physical properties of the dust, which are affected by the environmental conditions where the dust is embedded.}
   {We study here the dust properties of a set of \hii\ regions in the Local Group Galaxy M\,33 presenting different spatial configurations between the stars, gas and dust to understand the dust evolution under different environments.}
   {We model the SED of each region using the \DE\ tool and obtain the mass relative to hydrogen for Very Small Grains (\yvsg), Polycyclic Aromatic Hydrocarbons (\ypah) and Big Grains (\ybg). We furthermore perform a pixel-by-pixel SED modelling and derive maps of relative mass of each grain type for the whole surface of the two most luminous \hii\ regions in M\,33, NGC~604 and NGC~595.}
   {The relative mass of the VSGs (\yvsg/\ytot) changes with the morphology of the region:  \yvsg/\ytot\ is a factor of $\sim$1.7 higher for \hii\ regions classified as {{\it filled}} and {{\it mixed}} than for regions presenting a shell structure. The enhancement of VSGs within NGC~604 and NGC~595 is correlated to expansive gas structures with velocities $\geq$50\,\kms.  The gas-to-dust ratio derived for the \hii\ regions in our sample exhibits two regimes related to the \hi$\rm -H_{2}$ transition of the ISM. Regions corresponding to the \hi\ diffuse regime present a gas-to-dust ratio compatible with the expected value if we assume that the gas-to-dust ratio scales linearly with metallicity, while regions corresponding to a $\rm H_{2}$ molecular phase present a flatter  dust-gas surface density distribution. }
   {The fraction of VSGs can be affected by the conditions of the interstellar environment: strong shocks of $\sim$\,50-90\,\kms\ existing in the interior of the most luminous \hii\ regions can lead to fragmentation of BGs into smaller ones, while the more evolved shell and clear shell objects provide a more quiescent environment where reformation of dust BG grains might occur. The gas-to-dust variations found in this analysis might imply that grain coagulation and/or gas-phase metals incorporation to the dust mass is occurring in the interior of the \hii\ regions in M\,33.}

   \maketitle
%

\section{Introduction}\label{sec:intr}

The presence of interstellar dust in star-forming regions is supported by the observed mid-infrared (MIR) and far-infrared (FIR) emission in star-forming regions. In the Milky Way several studies have shown that dust is ubiquitous in star-forming regions and at the same time exhibits a wide range of properties \citep[see e.g.][]{2009PASP..121..213C}. Within a single galaxy the properties of dust emission change very strongly, from cold dust in the diffuse ISM to warm dust processed by the intense radiation field within the star-forming regions \citep{2012A&A...543A..74X}.  Since the pioneer study of \citet{Calzetti:2005p518} where a correlation between the 24\,\mi\ emission and the emission of the ionised gas was observed for star-forming regions in NGC~5194, several studies have probed the correlation between the IR dust emission and the emission of ionised gas in many galaxies with different properties \citep[see e.g.][]{2007ApJ...666..870C}. Dust appears to survive in the interior of the star-forming regions and its infrared emission is spatially correlated with the location of the newly formed stars.

Interstellar dust can also be affected by the environment in which it is located and thus exhibits different physical properties depending on the characteristics of the interstellar medium (ISM) that surrounds it. Dust does not remain with the same physical properties during its lifetime, as dust particles interact with the medium where they are embedded \citep[see][]{2004ASPC..309..347J}. Dust is a key for understanding star formation, its production/destruction mechanisms and how they are linked to star formation are not yet completely clear. This hampers our ability to trace accurately star formation or the gas reservoir, as this requires assumptions on the dust properties and size distribution.

There are several mechanisms affecting the evolution of the dust, some of them leading to a change in the physical properties and others to the destruction of a particular dust species \citep[see][for a review]{2004ASPC..309..347J}. 
(i) Photon-Grain interactions: high-energy UV-visible photons are absorbed and scattered leading to grain heating and subsequent thermal emission, (ii) Atom/Ion-Grain interactions: at high gas temperature, impinging atoms or ions can erode the grain and lead to sputtering of atoms from the grain. This mechanism depends on the relative gas-grain velocities. When the gas-grain velocity is due to the motion of the grain relative to the gas, as in a SN shock wave, the process is called non-thermal sputtering; and if the relative gas-grain velocity is due to the thermal velocity of the gas ions (e.g. in a hot gas at more than $\rm 10^{5}$\,K as it occurs in the hot post-shock gas in a SN bubble), the process is called thermal-sputtering. In general, sputtering affects the grain surfaces but cannot disrupt the grain cores unless complete grain destructions occurs. (iii) 
Low-energy grain-grain collisions lead to grain coagulation, increasing the mean grain size, while at higher energies grain fragmentation/disruption can occur. Shattering in grain-grain collisions preserves the total grain mass but alters the grain size distribution. 

\citet{1996ApJ...469..740J} \citep[see also][for dust processing with the new dust model from \citet{2013A&A...558A..62J}]{2014A&A...570A..32B} studied the effects of grain shattering in shocks through grain-grain collisions which produces fragmentation of all, or part of, a grain into smaller but distinct sub-grains. The result is a transfer of mass from large ($\sim200\,\AA$) grains into smaller ($\sim60\,\AA$) ones, which leads to a change of the grain size distribution, being the mass percentage of small grains formed by shattering up to 40\% higher for the case of grains with radii greater than 50\,\AA. 

Despite the theoretical studies regarding the dust evolution within different environments, little has been done from the observational point of view. Recently, \citet{2011ApJ...735....6P} and \citet{2014ApJ...784..147S} \citep[see also][]{2008A&A...491..797C} have presented evidence of dust evolution for some \hii\ regions in the Large Magellanic Cloud (LMC). Also, \citet{Bernard:2008p587} found a 70\,\mi\ excess in the SED of the LMC with respect to the Milky Way and proposed the production of a large fraction of very small grains (VSGs) via erosion of larger grains as the most plausible explanation for this excess.  

Although these studies show evidence of a change in the dust size distribution of individual \hii\ regions, a systematic study of a large \hii\ region sample covering a wide range of physical properties is needed in order to infer firm conclusions about the evolution of dust in different environments. \citet[][]{2013A&A...552A.140R}\defcitealias{2013A&A...552A.140R}{Paper I} \citepalias[hereafter][]{2013A&A...552A.140R} presented observational SEDs of a set of \hii\ regions in the nearby \citep[840\,kpc,][]{1991ApJ...372..455F} spiral galaxy M\,33 covering a wavelength range from the UV (GALEX) to the FIR (\Her). Using the \ha\ emission distribution of each object, the regions of the sample were classified as {\it filled}, {\it mixed}, {\it shell} and {\it clear shell} objects \citepalias[fig.~2 in][shows examples of each type]{2013A&A...552A.140R}. Each morphological classification might well represent an evolutionary stage of the \hii\ region from ages of a few Myr for {\it filled} and {\it mixed} regions, where gas (and dust) is co-spatial with the stellar cluster, to 5-10\,Myr for {\it shells} and {\it clear shells}, where the massive stellar winds and supernovae (SN) have created holes and shell structures around the stellar cluster \citep{2011ApJ...729...78W}. A different star-gas-dust spatial configuration \citep[see fig.\,3 in][]{Verley:2010p687} is also expected for each morphological type. \citetalias{2013A&A...552A.140R} studied trends in the observed SEDs with the morphology of the region and found that the FIR SED peak of regions with {\it shell} morphology seems to be located towards longer wavelengths, implying that the dust is cooler for this type of objects. Assuming a characteristic environment for each \hii\ region type (the star-gas-dust spatial configuration is different in {\it filled} compact regions and shell-like objects), we would expect an evolution of the physical properties of the dust in each morphological type.

Dust evolution processes can be studied through the variation of the dust size distribution and, in particular, with the ratio of the abundance of small to large grains \citep{2008A&A...491..797C}. Therefore, we present a study of the relative abundance of small to large grains for the \hii\ regions in M\,33 and analyse the results in terms of the region morphology. We estimate the abundances of the different grain types by modelling the observed SEDs with the dust emission tool \DE\ \citep{2011A&A...525A.103C}. \DE\ allows us to obtain the mass abundance relative to hydrogen of each dust grain population included in the model (several dust models can be considered by the software tool), as well as the intensity of the interstellar radiation field (ISRF) heating the dust. 

The paper is organised as follows. In Sect.\,\ref{sec:data} we present the observations and describe in detail the new data set that complements the SEDs already published in \citetalias{2013A&A...552A.140R}. 
In Sect.\,\ref{sec:met} we describe how the observed SEDs were derived and how the dust modelling is performed. The results are described in Sect.\,\ref{sec:results}. A detailed analysis of the two most luminous \hii\ regions in M\,33, NGC~604 and NGC~595, is presented in Sect.\,\ref{sec:604_595}. Finally, in Sect.\,\ref{sec:sum} we summarise the main conclusions of this paper.

\section{The data}\label{sec:data}
The observed SEDs of the \hii\ regions are built up using data from the FUV to FIR wavelength range. We use the same UV, \ha, \Spi\  data presented in \citetalias{2013A&A...552A.140R}, therefore we refer the reader to this paper for further description of these particular data. Here, we describe the new data set included in this study. 

\subsection{\Her\ data}\label{subsec:pacs}

In \citetalias{2013A&A...552A.140R} we used 100\,\mi\ and 160\,\mi\ PACS and 250\,\mi\ SPIRE data. For this paper, we use the latest version of 100\,\mi\ PACS data reprocessed with Scanamorphos v16 \citep{2013PASP..125.1126R} as described in \citet{Boquien:2011p764} with a resolution of 7\farcs7, and the new observed 70\,\mi\ and 160\,\mi\ PACS data, obtained in a follow-up open time cycle 2 programme of \Her\ \citep{2015A&A...578A...8B} with a resolution of 5\farcs5 and 11\farcs2. We take into account the difference of the astrometry into account  in the 100\,\mi\ image already reported in \citet{2015A&A...578A...8B} and change the coordinates of the image accordingly. 
The field of view of the new 70\,\mi\ and 160\,\mi\ PACS data is slightly smaller than the field of view of the data set in \citetalias{2013A&A...552A.140R}, and 4 regions of the original sample in \citetalias{2013A&A...552A.140R} are missing in the new PACS images. Therefore, the 70\,\mi\ and 160\,\mi\  fluxes of these regions were discarded from the analysis presented here. 

The SPIRE and the 100\,\mi\ PACS data were taken in the framework of the HerM33es open time key project \citep{Kramer:2010p688}. The 250\,\mi\ SPIRE image with a resolution of 21\farcs2 was obtained using the new version 10.3.0 of  the \Her\ Data Processing System \citep[HIPE,][]{2010ASPC..434..139O,2011ASPC..442..347O}.  For a description of the data reduction the reader is referred to \citet{2012A&A...543A..74X}. The calibration uncertainties are 5\% for the PACS images and 15\% for the  250\,\mi\ SPIRE image. Due to the low spatial resolution we did not use 350\,\mi\ and 500\,\mi\ SPIRE data.

\subsection{WISE data}\label{subsec:pacs}
The {\it Wide-field Infrared Survey Explorer} ({\it WISE}) \citep{2010AJ....140.1868W} observed M\,33 in four photometric bands: 3.4\,\mi\ ({\it W}1), 4.6\,\mi\ ({\it W}2), 12\,\mi\ ({\it W}3), and 22\,\mi\ ({\it W}4) with spatial resolutions of 6\farcs1, 6\farcs4, 6\farcs5, and 12\farcs0, respectively. 
We use the Image Mosaic Service {\it Montage}\footnote{\url {http://hachi.ipac.caltech.edu:8080/montage}} to obtain the final mosaic image of M\,33 in each band. After subtracting the global background, the {\it WISE} maps were converted from digital numbers to Jy using the photometric zero points given in table\,1 of the {\it Explanatory Supplement to the WISE All-Sky Data Release Products}\footnote{\url {http://wise2.ipac.caltech.edu/docs/release/allsky/expsup/sec2_3f.html}}. We follow \citet{2013AJ....145....6J} and apply the corrections recommended by them for extended source photometry. First, we apply an aperture correction that accounts for the Point Spread Function (PSF) profile fitting used in the {\it WISE} absolute photometric calibration. The corrections are 0.034, 0.041, -0.030 and 0.029 mag for the {\it W}1, {\it W}2, {\it W}3 and {\it W}4 bands, respectively. The second correction is a colour correction that we do not apply here, as it is already taken into account by the SED fitting process within the \DE\ code (see Sect.\,\ref{subsec:SEDmod}). The third correction, related to a discrepancy in the calibration between the {\it WISE} photometric standard blue stars and red galaxies, only applies to the {\it W}4 image and accounts for a factor of 0.92. The calibration accuracy of the {\it WISE} maps are 2.4\%, 2.8\%, 4.5\% and 5.7\%, for {\it W}1, {\it W}2, {\it W}3 and {\it W}4 images \citep{2011ApJ...735..112J}.

\subsection{LABOCA 870\,\mi\ data}\label{subsec:laboca} 
LABOCA (Large APEX BOlometer CAmera) is a multi-channel bolometer array for continuum observations \citep{2009A&A...497..945S} installed at APEX\footnote{This publication is based on data acquired with the Atacama Pathfinder 
Experiment (APEX) under programme IDs 085.F-0045, 086.F-9301 and 
089.C-0935. APEX is a collaboration between the Max-Planck-Institut f\"ur 
Radioastronomie, the European Southern Observatory, and the Onsala Space 
Observatory.
} (Atacama Pathfinder EXperiment) telescope covering a bandwidth of $\sim$\,150\,\mi\ around the central wavelength of 870\,\mi. The FWHM of the PSF is $\sim$19\farcs2 \citep{2009ApJ...707.1201W}, close to that of the 250\,\mi\ SPIRE band.  We refer the reader to \citet{2016arXiv160302125H} for a description of the data reduction of the LABOCA observations of M\,33. 10 \hii\ regions of the original sample in \citetalias{2013A&A...552A.140R} are located close to the edge of the LABOCA field of view and thus we were not able to perform an accurate photometry for these objects. Therefore, their fluxes in this band were not considered in the analysis of the SEDs. The uncertainty in the calibration is $\sim$12\%.

Observed LABOCA fluxes include emission from the CO rotational transition CO(3-2) at 870\,\mi\ and from thermal free-free emission. \citet{2016arXiv160302125H} estimate the CO(3-2) contribution to the total LABOCA flux of M\,33 to be $\sim$7.6\%. We adopt this value for the fluxes of the individual \hii\ regions. The thermal emission is estimated from the analysis of \citet{Tabatabaei:2007p664} at 3.6\,cm. These authors obtained from multi-wavelength data, using two different methods, a thermal fraction of 76.4\% and 99.2\% for the observed 3.6\,cm flux. Adopting the mean value of both methods and a spectral index of 0.1 we obtain a thermal contribution of 3\% to the LABOCA observed flux. We take into account both CO(3-2) and thermal contributions and subtract them from the observed LABOCA fluxes for each object. 

\subsection{GISMO 2\,mm data of NGC 604}\label{subsec:gismo}

The Goddard-IRAM Superconducting 2 Millimeter Observer (GISMO) continuum camera
on the IRAM 30m telescope \citep{2006SPIE.6275E..1DS} works at a central frequency of 150\,GHz,
with  $\Delta\nu/\nu$=0.15 (or a band-width of 22\,GHz at 150 GHz).
It comprises an array of 8$\times$16 pixels separated by 14\arcsec\ on the sky, with a resolution of 17.5\arcsec.

\begin{figure}
   \centering
  \includegraphics[width=0.49\textwidth]{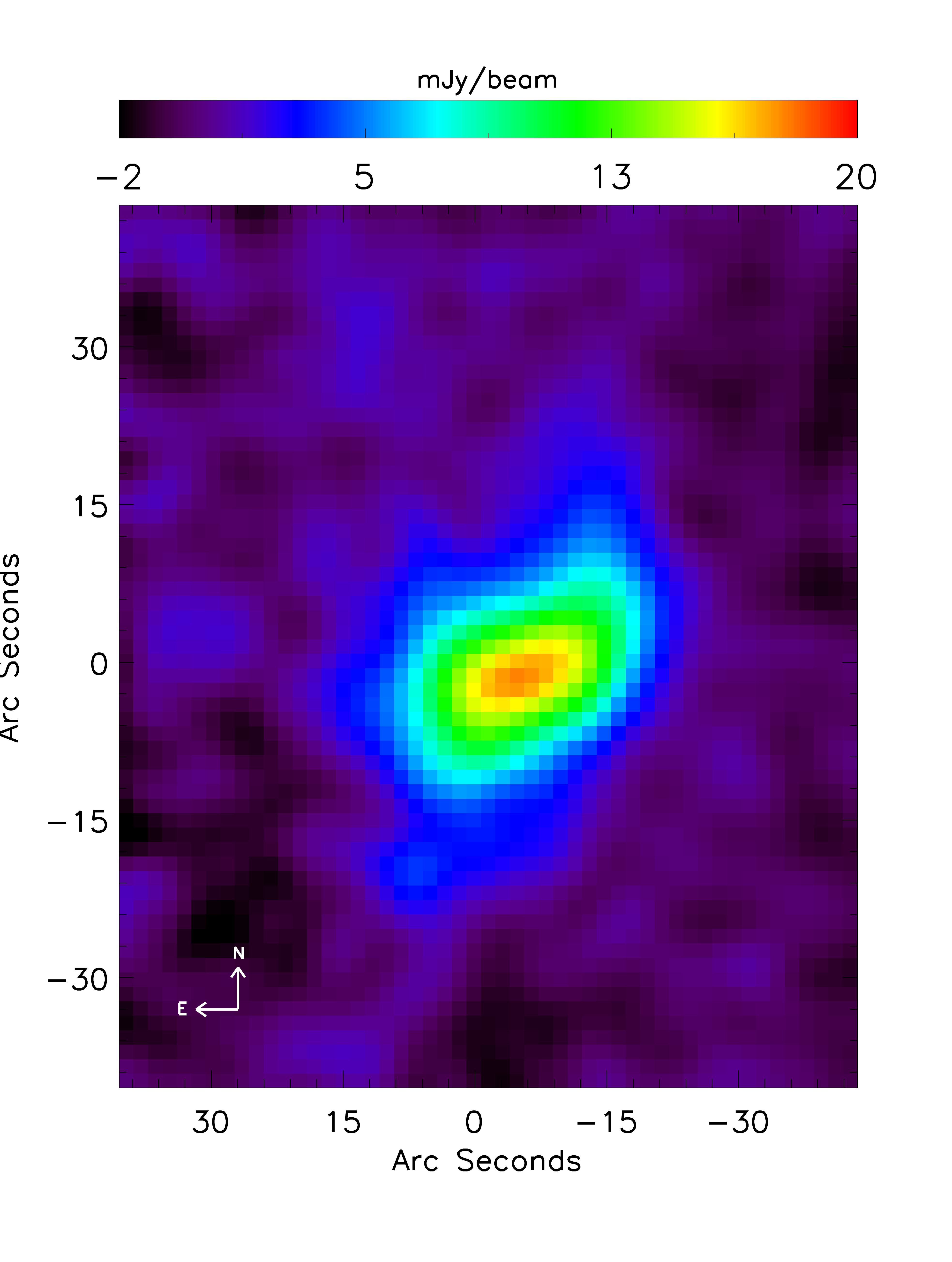}   
   \caption{2\,mm image of NGC~604 observed with the GISMO continuum camera. The final spatial resolution of the image is 21.0\arcsec\ to match the other data.}
              \label{fig:gismomap}%
    \end{figure}

Observations of NGC 604 were conducted in several shifts, in April 2012, and April, October,
and November 2013, during about 26 hours of total observing time by 181 on-the-fly scanning maps
in total power of 10\arcmin$\times$10\arcmin\ in size.
Typical scanning speeds were 44\arcsec/s.
To correct for atmospheric transmission, the IRAM taumeter was used,
which continuously measures the sky opacity at 225\,GHz.
The GISMO conversion factor, 29.3 counts/Jy, was derived from Mars, Uranus, and Neptune observations.
The relative calibration uncertainty is estimated to be $\sim$10\% (see the reports on the GISMO performance\footnote{\url {http://www.iram.es/IRAMES/mainWiki/\\Continuum/GISMO/Main\#Documentation_and_Publications}}).
To obtain the positions of the GISMO pixels on the sky and source gains,
Mars was mapped covering the source within each pixel.
The {\tt crush}\footnote{\url {http://www.submm.caltech.edu/~sharc/crush/}} data processing software (vers. 2.16-3) (Kovacs 2008)
was used with the options for faint emission.
NGC~604 is detected with a total flux density of 64.4\,mJy integrated over a 74.2\arcsec\ aperture radii.
The peak flux density is 18\,mJy/beam.
The rms was measured to be 0.55\,mJy/beam. The final map of NGC~604 is presented in Fig.~\ref{fig:gismomap}. 

The observed GISMO flux needs to be corrected for the thermal emission of the free electrons within the ionised gas. We use the results from  \citet{Tabatabaei:2007p664}  at 3.6\,cm to estimate this contribution as in Sect.\,\ref{subsec:laboca}. We obtain that 51\% of the observed flux in GISMO corresponds to thermal emission. We subtract this contribution from the observed GISMO flux, both for the global and the pixel-by-pixel analysis presented in Sect.\,\ref{sec:604_595}.

\subsection{CO and HI data}\label{subsec:cohi}
We use here the $^{12}$CO(J=2--1) (at 230.538\,GHz) velocity integrated intensity map presented in \citet{2014A&A...567A.118D}. The spatial resolution of the map is 12\arcsec\ with a pixel size of 3\arcsec. The HI (21\,cm) intensity map is presented in \citet{2010A&A...522A...3G}, it has a spatial resolution of 17\arcsec\ and a pixel size  of 4.5\arcsec.  Details of data reduction are given in \citet{2014A&A...567A.118D} and \citet{2010A&A...522A...3G}, for CO and HI intensity maps, respectively. 

\section{Methodology}\label{sec:met}
In this section we explain how the photometry was performed and describe the SED fitting procedure. 
\subsection{Photometry}\label{subsec:phot}
We aim to model the SED of the set of \hii\ regions studied in \citetalias{2013A&A...552A.140R} including the new PACS, {\it WISE}, and LABOCA data. Since we also use the new version of the SPIRE 250\,\mi\ data, we smooth and register all the images from FUV to LABOCA, CO and HI to the  resolution of the new version of the SPIRE 250\,\mi\ image. The final data set has a spatial resolution (FWHM) of 21.2\arcsec\ ($\sim$86\,pc) and a pixel size of 6\arcsec ($\sim$24\,pc). 

The total flux for each region is obtained using the IRAF task {\tt phot} and the central coordinates and aperture radii defined in table~1 of \citetalias{2013A&A...552A.140R}. The photometric aperture radii range from $\sim$50\,pc for {\it filled} regions to $\sim$250\,pc for {\it clear shell} objects \citepalias[see Fig.\,4 in][]{2013A&A...552A.140R}. The local background is defined and subtracted as it was done in \citetalias{2013A&A...552A.140R}. Regions with absolute fluxes lower than their errors are assigned an upper limit of 3 times the estimated uncertainty in the flux. The negative fluxes showing absolute values higher than the corresponding errors are discarded from the study. The final fluxes for all the bands used in this paper, as well as FUV, NUV and \ha\ fluxes, are given in Tables~\ref{tab:phot1} and ~\ref{tab:phot2}. 

We compared the fluxes obtained here from the updated data set with the fluxes already published in \citetalias{2013A&A...552A.140R}. For FUV, NUV, IRAC, and MIPS bands we find differences of $\leq$20\% mainly related to the subtraction of the local background (differences for the fluxes without subtracting the local background are $\sim$1\%). 
The lower the flux of the region, the larger the difference, which shows that for some regions the uncertainties are dominated by the estimate of the local background. 

For the PACS 100\,\mi\ and 160\,\mi\ images the differences in the fluxes are larger. The main reason is that there is a significant difference in the data reduction process of the last version (Scanamorphos v16) of these images, improving the drift removal \citep{2015A&A...578A...8B}. This can affect the fluxes for the low luminosity \hii\ regions in these bands (see Fig.~\ref{comp_pacs}).

\subsection{SED modelling}\label{subsec:SEDmod}
In order to study the properties of the dust in the sample of \hii\ regions we fit the observed SEDs with the dust emission tool \DE\ \citep{2011A&A...525A.103C}. The code allows a combination of various grain types as input and gives the emissivity per hydrogen atom ($\rm N_{H} = N_{HI}+2N_{H_{2}}$) for each grain type based on the incident radiation field strength and the grain physics in the optically thin limit. 
The free parameters are the mass relative to hydrogen of each dust population ($Y_{i}$), the intensity of a NIR continuum modelled using a black body with a temperature of 1000\,K and the ISRF parameter $\rm G_{0}$, which is the scale factor to the solar neighbourhood ISRF given in \citet{Mathis:1983p593}. The origin of the 1000\,K black body emission is  uncertain. It was observed by \citet{2003ApJ...588..199L} with ISO/ISOPHOT in a sample of 45 normal star-forming galaxies, it has also been observed in the spectra of reflection nebulae \citep{1983ApJ...271L..13S} and in the diffuse cirrus emission of the Milky Way \citep{2006A&A...453..969F}. We were not able to fit the SEDs of our regions without including this component, therefore we have included it and obtained the scale factor as an output parameter. Given an observed SED and the associated uncertainties, \DE\ performs the SED fitting using the MPFIT IDL minimisation routine \citep{2009ASPC..411..251M}, based on the Levenberg-Marquardt minimisation method. In the process \DE\ predicts the SED values that would be observed by the astronomical instruments taking into account the colour corrections for wide filters and the flux conventions used by the instruments.

\DE\ can handle an arbitrary number of grain types and size distributions. We adopt here the \citet{2011A&A...525A.103C} dust model ({\it Compiegne} dust model) which includes three dust components: Polycyclic Aromatic Hydrocarbons (PAHs) with radii $r\lesssim10\,\AA$, hydrogenated amorphous carbonaceus grains (amC) and amorphous silicates (aSil) with radii $r\gtrsim100\,\AA$. The population of amorphous carbon dust has been divided into small (SamC, $r\sim10-100\,\AA$) and large (LamC, $r\gtrsim100\,\AA$) grains. Besides, the model differentiates between ionised and neutral PAHs. We define $Y_{\rm VSG}=Y_{\rm SamC}$ and $Y_{\rm BG}=Y_{\rm LamC}+Y_{\rm aSil}$, in a similar way as it was done in \citet{2011A&A...531A..51F}. In the fitting process we keep the ratio between the aSil and LamC grains at the constant value given by the model, 5.38, in order to have an acceptable ratio between observational data points and input parameters. We compare the results of this dust model with the classical one proposed by \citet{1990A&A...237..215D}  (hereafter the {\it Desert} dust model), which consists of three dust grain types: PAHs, VSGs of carbonaceous material, and big grains (BGs) of astronomical silicates. The comparison is presented in Appendix\,\ref{app:compSED}. In general, the SED is also well fitted with the {\it Desert} dust model, except the 6-9\,\mi\ wavelength range where the model underpredicts the observed emission. This shows that {\it Desert} dust model might not be reliable to constrain the PAH abundance.

\subsection{Input radiation field}\label{sec:comp_isrf}

\begin{figure} 
   \centering
  \includegraphics[width=0.49\textwidth]{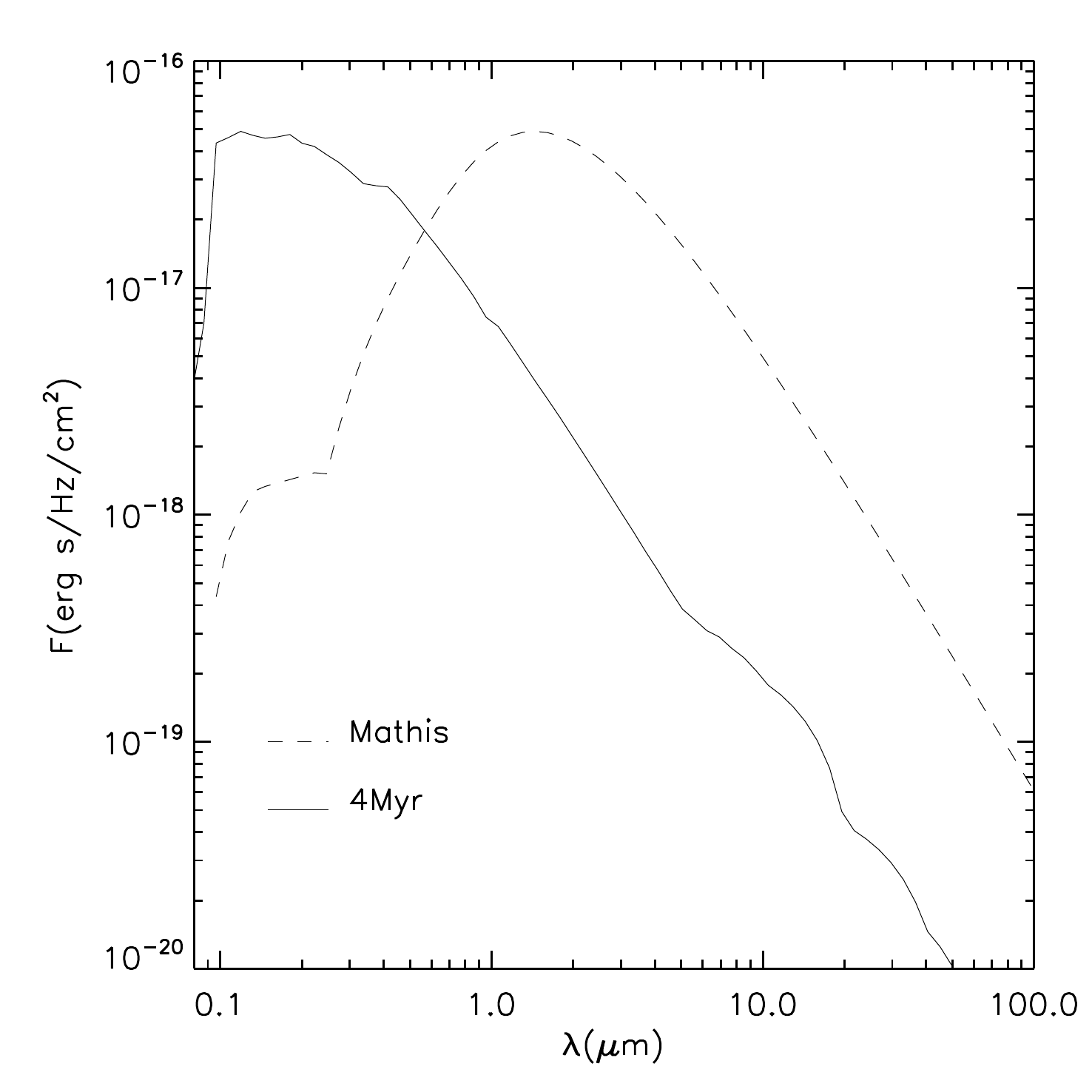}   
   \caption{Comparison of \citet{Mathis:1983p593} ISRF (dashed line) and an ISRF of a 4\,Myr star cluster (continuous line) (see text for details). Both ISRFs have been scaled to the same maximum intensity for a better comparison. The ISRF of a 4\,Myr star cluster has a significant contribution in the UV part of the spectrum.}
              \label{fig:isrf}%
    \end{figure}

A more realistic ISRF for a typical \hii\ region compared to the ISRF from  \citet{Mathis:1983p593} used as default in \DE\ is that of a young star cluster. We have generated the spectrum of a young star cluster of 4\,Myr using the online STARBURST99 application \citep{Leitherer:1999p491}. We assume a Kroupa initial mass function \citep{2001MNRAS.322..231K} and Geneva stellar tracks and obtain the spectrum for a fixed stellar mass of $10^{4}$\msun\ and 4\,Myr age. The stellar mass and the age are consistent with those expected from the \ha\ and FUV luminosities of the brightest \hii\ regions in M\,33 \citep[see Fig.~5 of][]{Relano:2009p558}. In Fig.~\ref{fig:isrf} we show the 4\,Myr (continuous line) and {\it Mathis} (dashed line) ISRFs. A 4\,Myr ISRF is harder, having a significant contribution in the UV part of the spectrum. 
  
\begin{figure*} 
   \centering
  \includegraphics[width=0.49\textwidth]{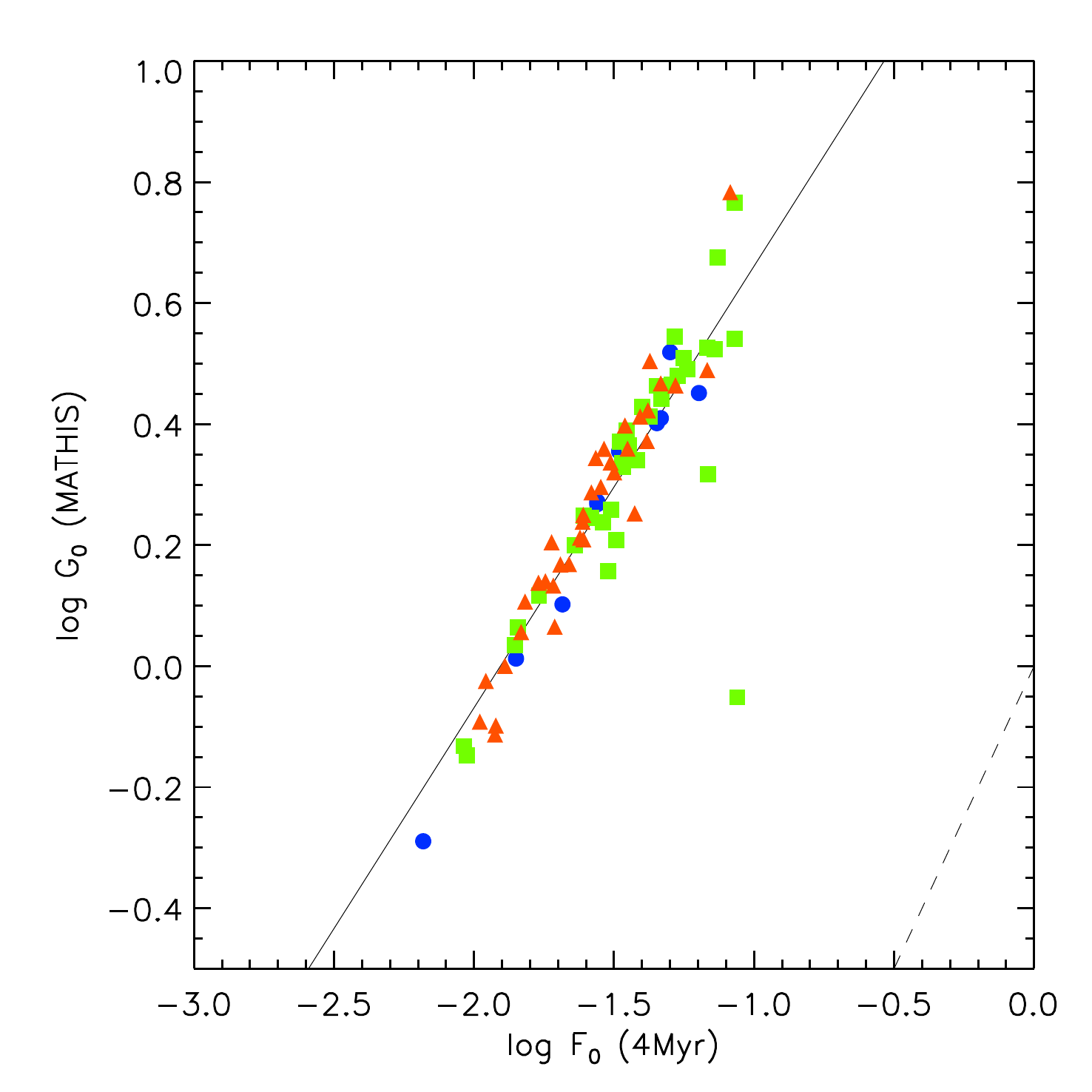}   
  \includegraphics[width=0.49\textwidth]{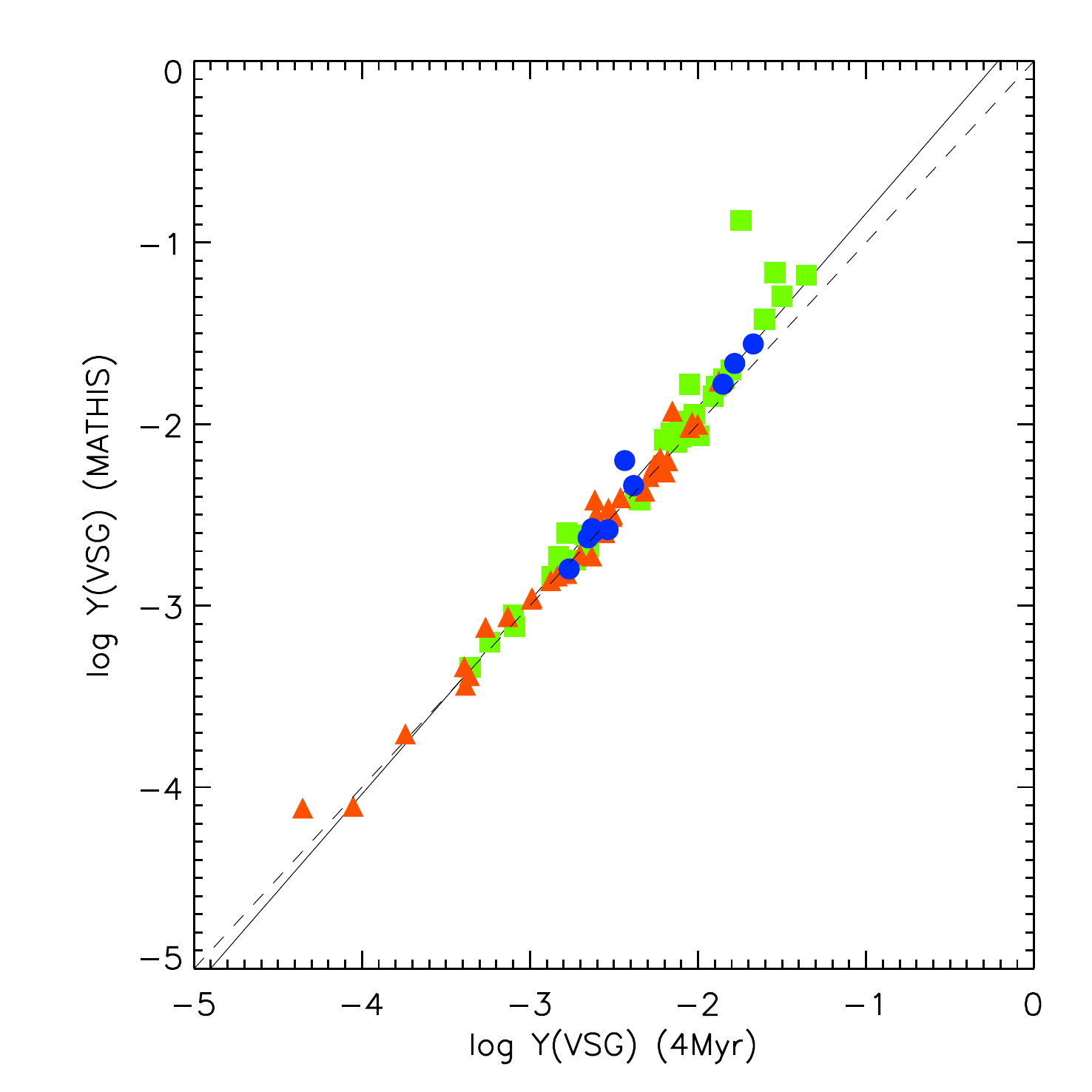} 
  \includegraphics[width=0.49\textwidth]{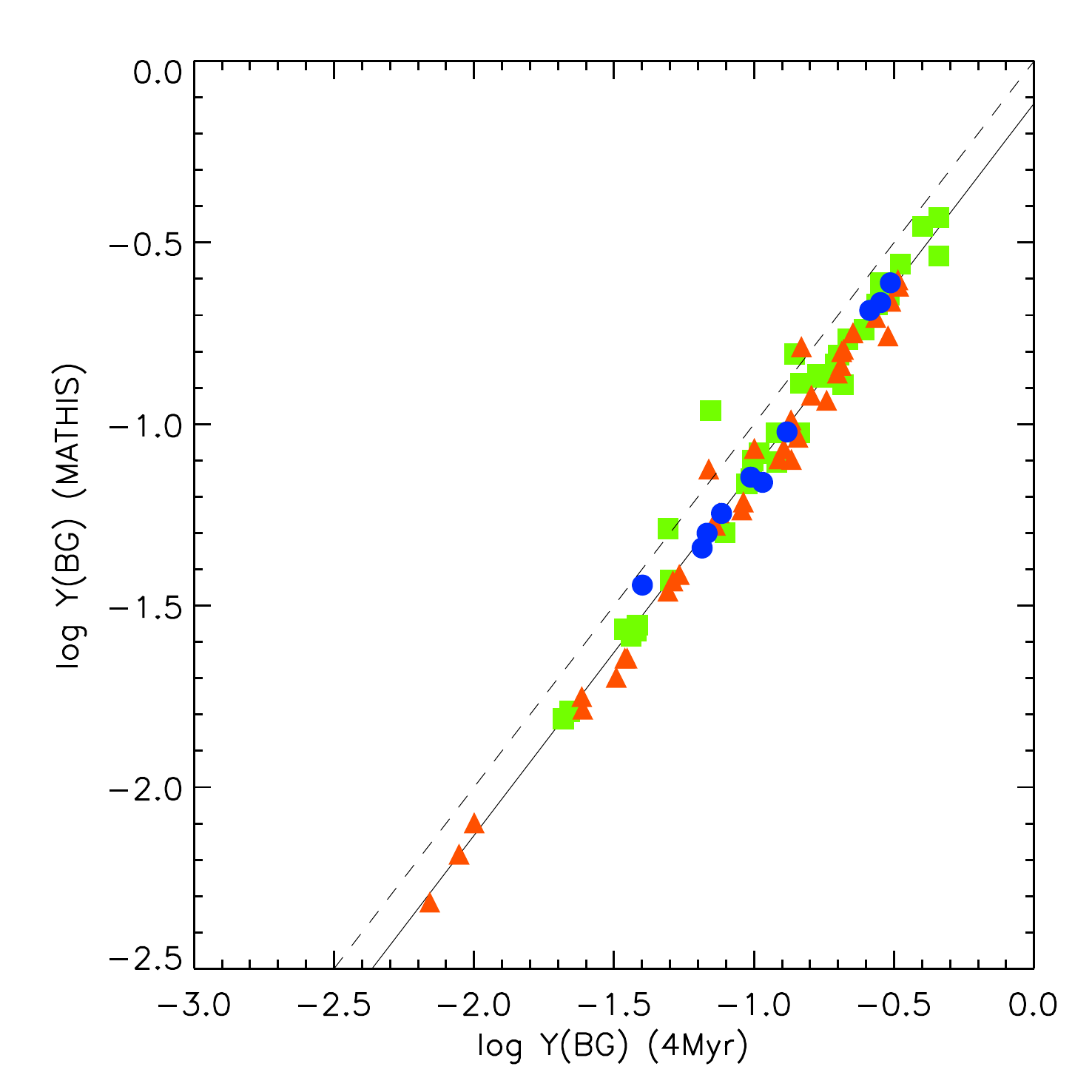} 
  \includegraphics[width=0.49\textwidth]{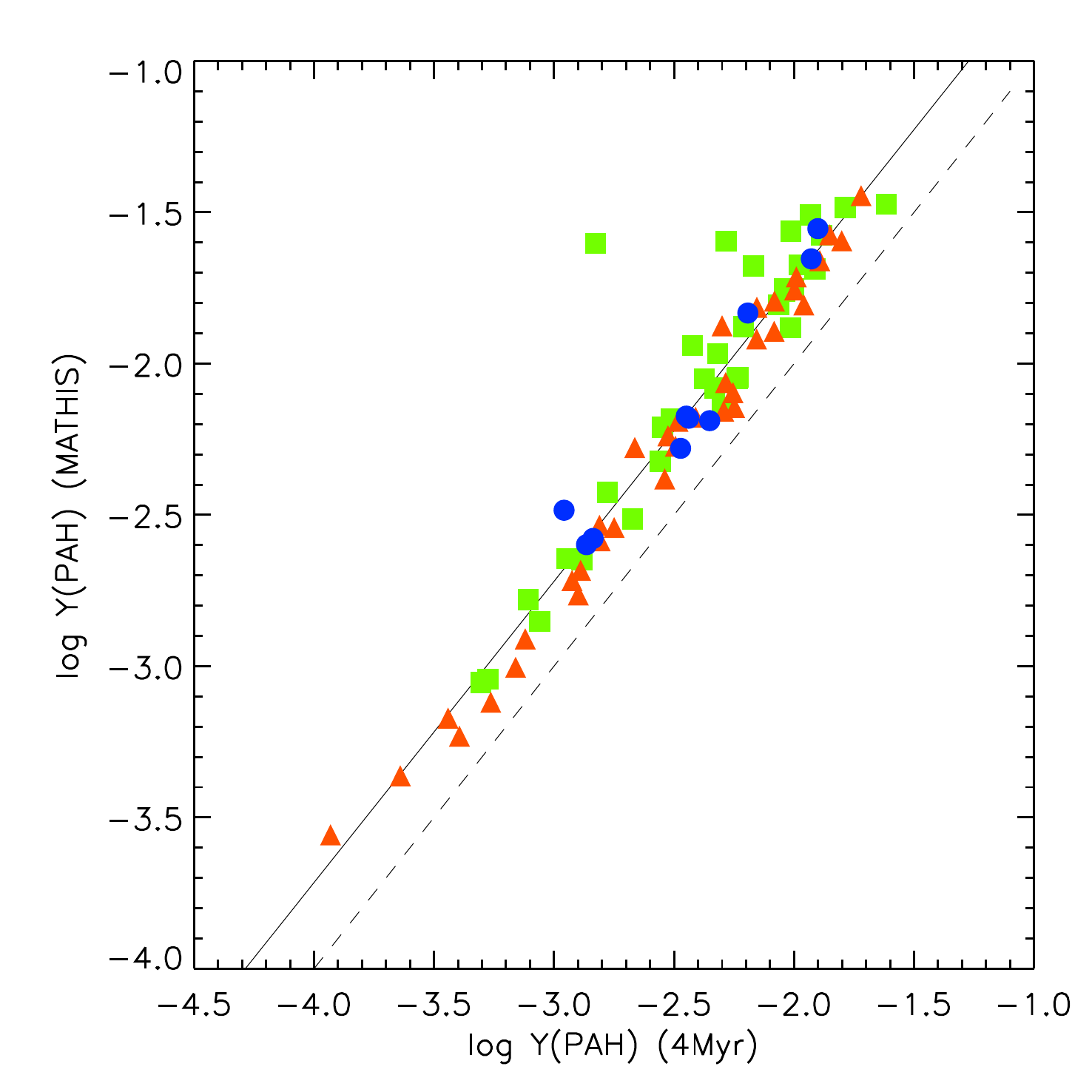}   
   \caption{Comparison of the \DE\ output parameters for the \hii\ regions in our sample using a 4\,Myr star cluster ISRF and a {\it Mathis} ISRF. {\it Top-left}: scale factor of the ISRF intensity, {\it top-right}: \yvsg, {\it bottom-left}: \ybg, and {\it bottom-right}: \ypah. The continuous line corresponds to the linear fit to the data and the dashed line is the one-to-one relation. The colour code corresponds to the morphology of the region classified in \citetalias{2013A&A...552A.140R}: blue circles are {\it filled} regions, green squares {\it mixed} ones, and red triangles {\it shells} or {\it clear shells}.}
              \label{fig:comp_abun}%
    \end{figure*}

The different shape of both ISRFs can influence the heating of the different types of dust grains. In Fig.~\ref{fig:comp_abun} we show the results of the \DE\ modelling for both ISRFs. \yvsg\ does not change significantly when a different ISRF is assumed and \ybg\ changes but the change is small and marginally located within the data dispersion. However, \ypah\ and the scale of the ISRF do change for {\it Mathis} and 4\,Myr star cluster ISRFs. \ypah\ for the case of {\it Mathis} ISRF is $\sim$3 times higher than \ypah\ when a 4\,Myr star cluster ISRF is assumed and  $\rm G_{0}$ is a $\sim$60 times higher than $\rm F_{0}$\footnote{We define $\rm F_{0}$ in the same way as $\rm G_{0}$ but for the 4\,Myr star cluster ISRF.}. The change in the scale parameter of the ISRF intensity is expected, as the 4\,Myr star cluster ISRF is harder than the {\it Mathis} one. The reason why \ypah\ is higher for the {\it Mathis} ISRF is related to the shape of the ISRF.  For the {\it Compiegne} dust model we expect, following the extinction curve given in \citet{2011A&A...525A.103C}, that the PAHs will absorb more radiation in the UV part of the spectrum than in the optical. Thus, for a 4\,Myr star cluster ISRF, which has a significant component in the UV wavelength range, the amount of PAH grains needed to explain the emission in the 3-10\,\mi\ range of the observed SED will be less than the one required for the {\it Mathis} ISRF. This shows that unless we are able to characterise the shape of the ISRF heating the dust within the regions we are not able to constrain completely the PAH abundance, there is a degeneracy between the PAH abundance and the scaling factor of the ISRF. However, we will be able to infer conclusions on the \ypah\ distribution within the region, as we will show in Sect.~\ref{sec:604_595}. In the following we will use a 4\,Myr star cluster ISRF to model the SEDs of our \hii\ region sample. 
 
\section{Statistical analysis of \hii\ regions}\label{sec:results}

\subsection{Results of the modelling}
\begin{figure*} 
   \centering
  \includegraphics[width=0.49\textwidth]{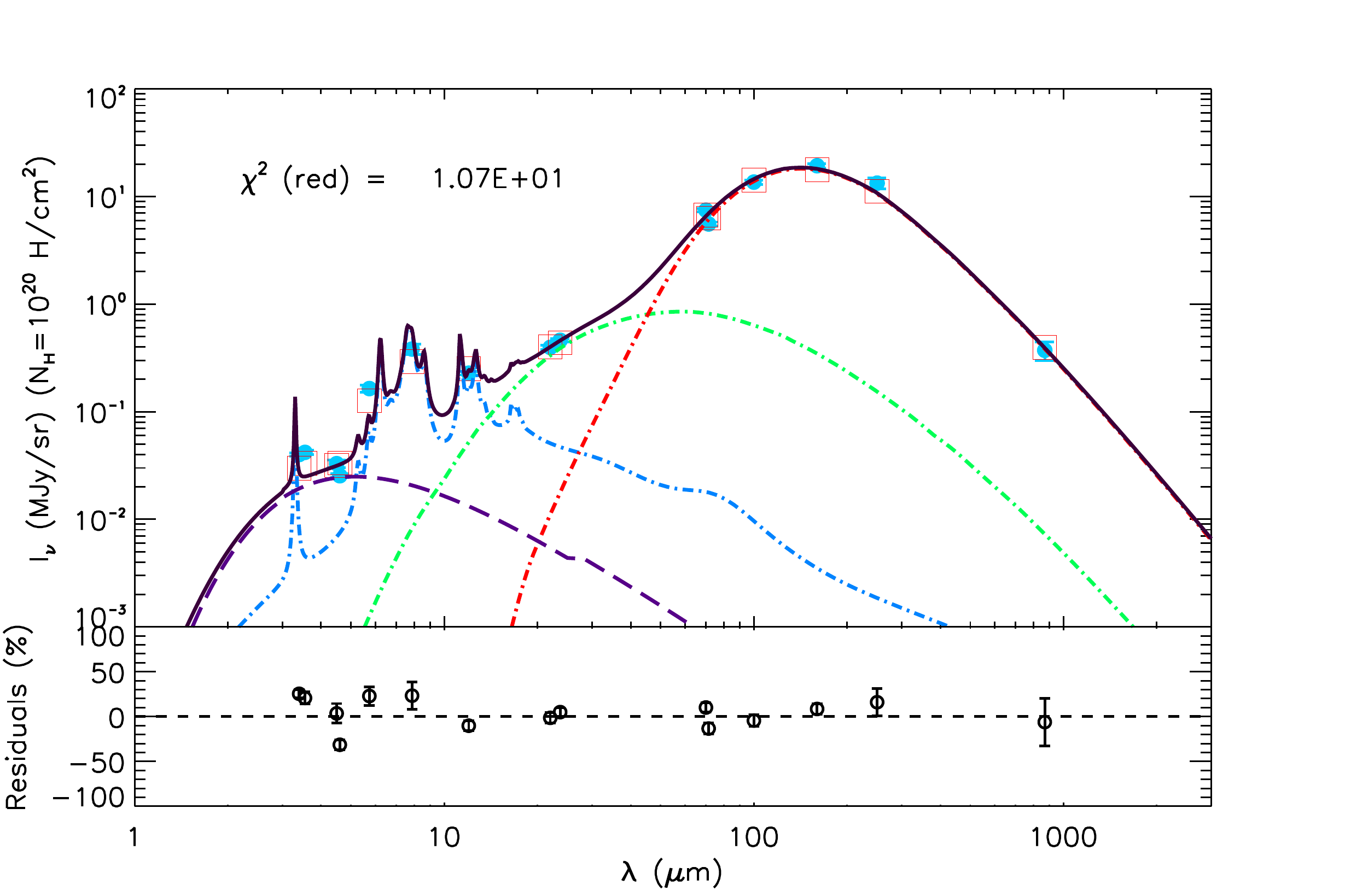} 
  \includegraphics[width=0.49\textwidth]{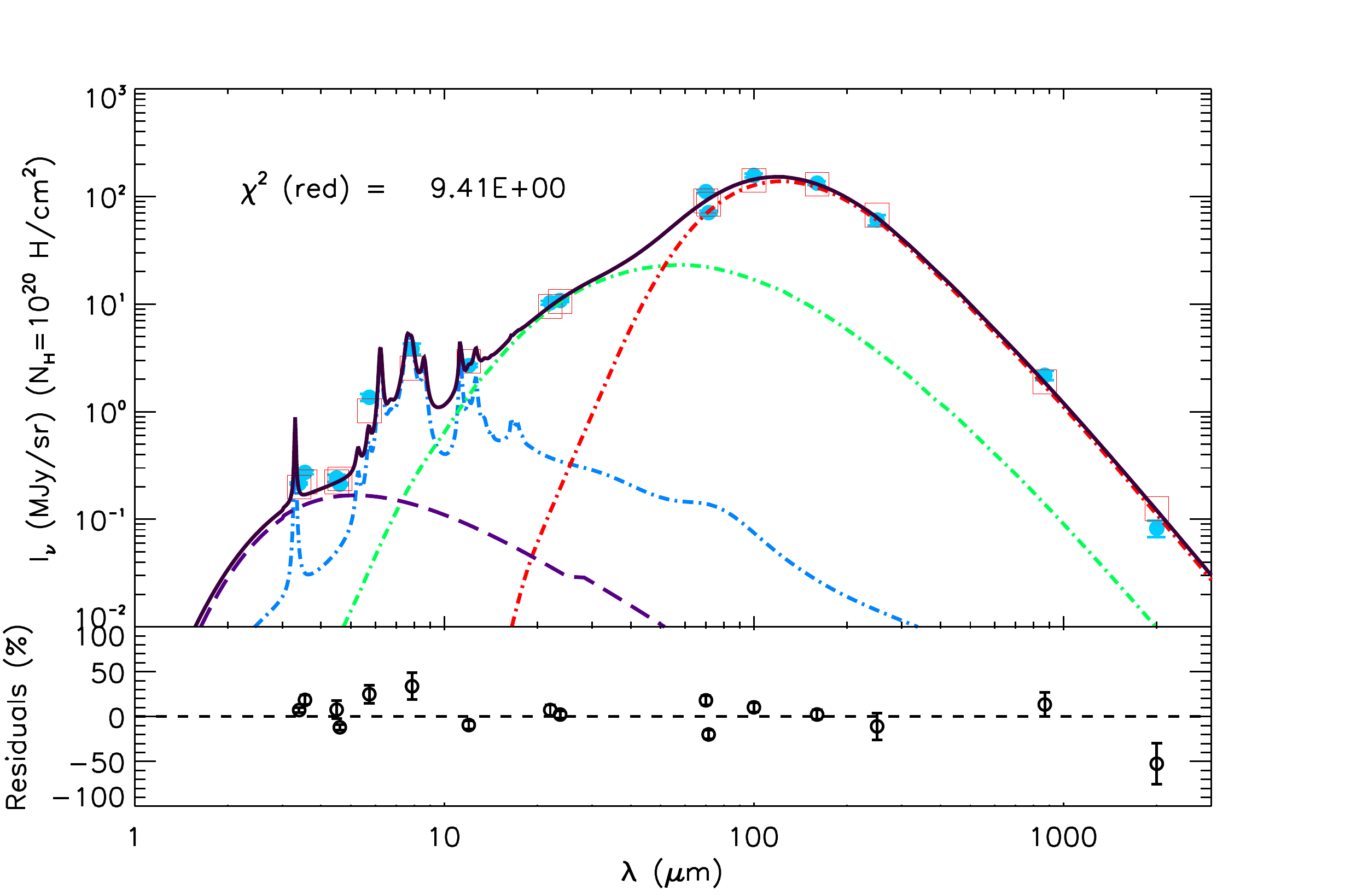}  
  \includegraphics[width=0.49\textwidth]{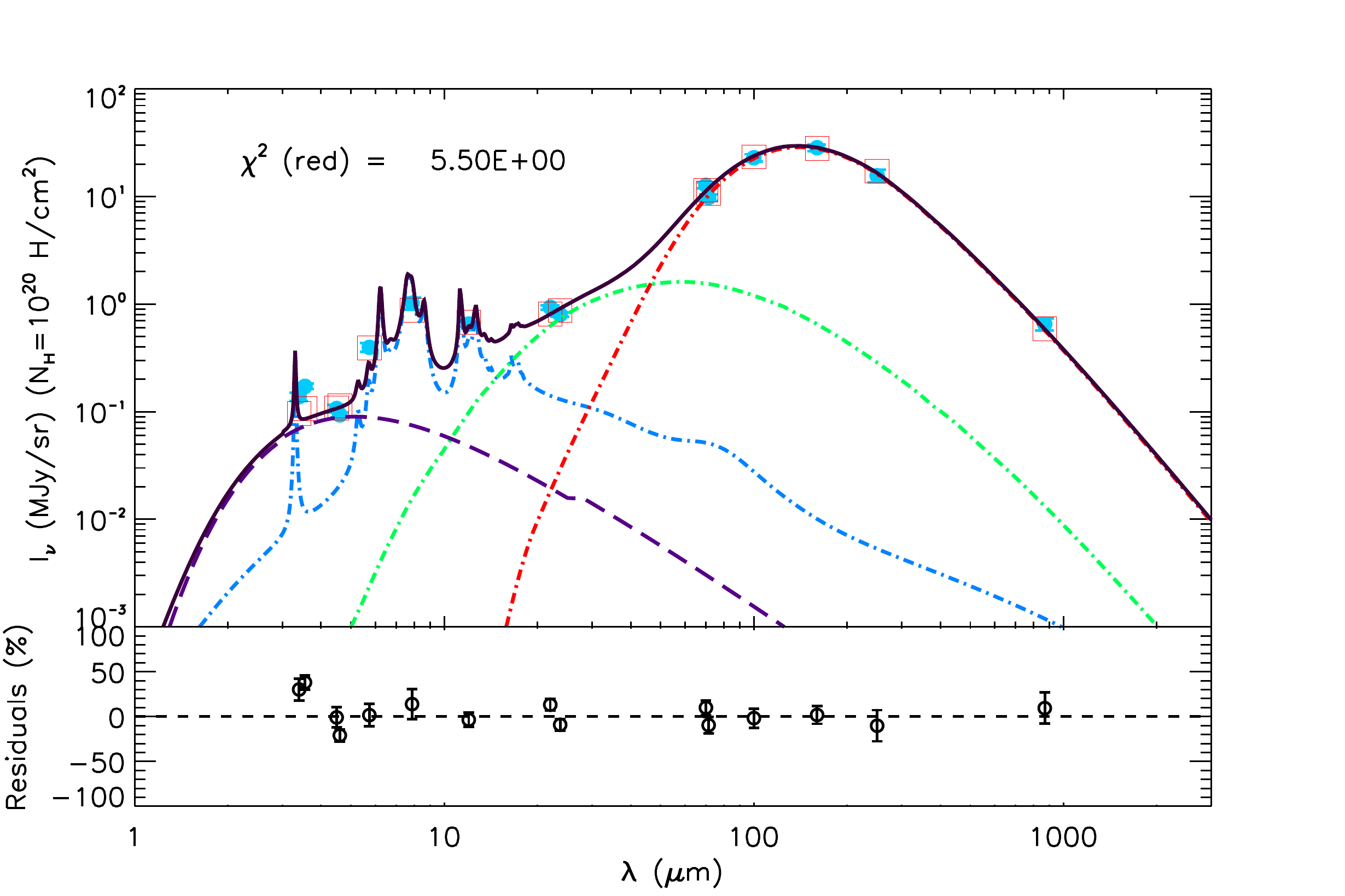} 
  \includegraphics[width=0.49\textwidth]{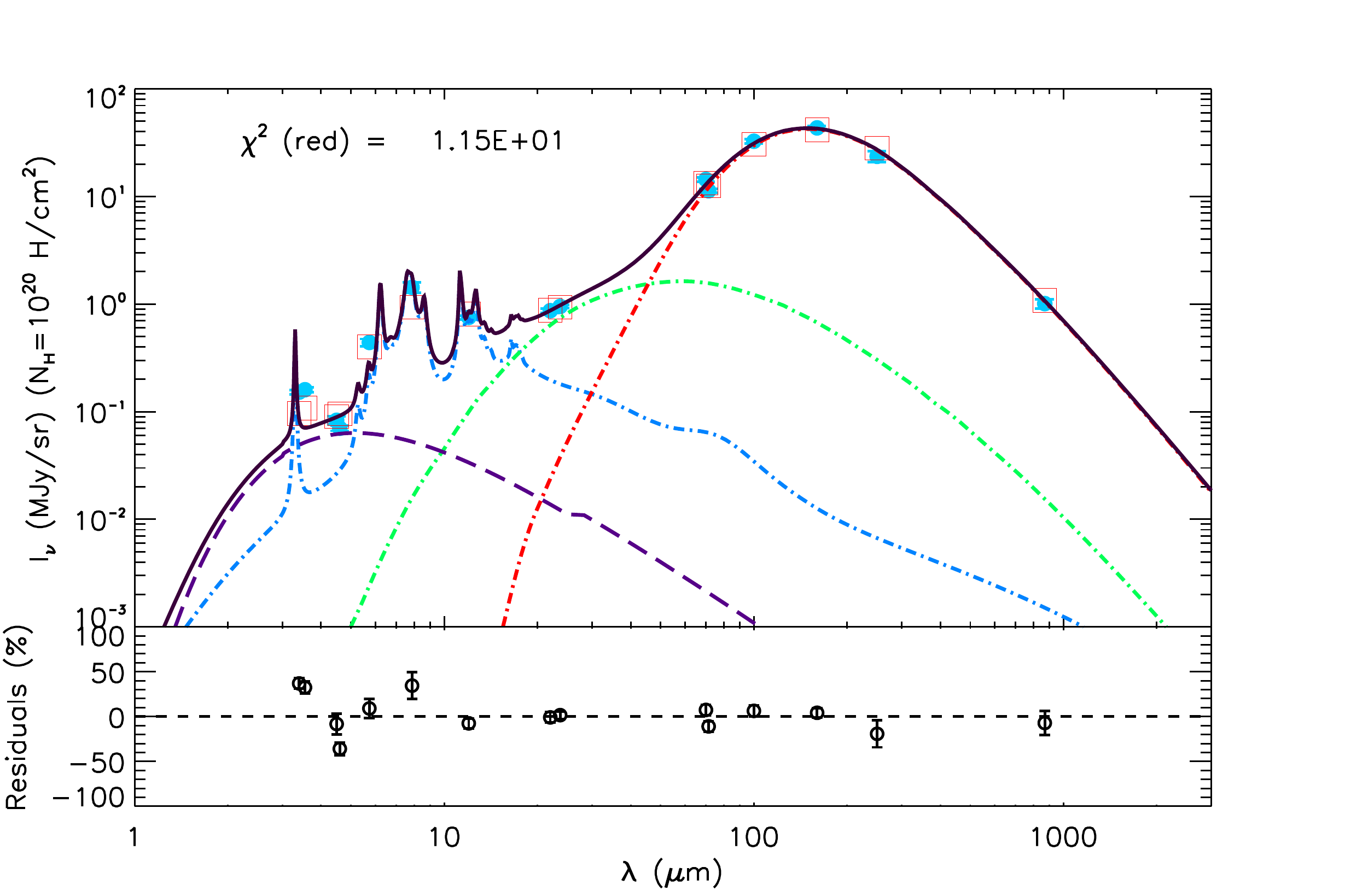}   
   \caption{Example of SED models for 4 regions of the sample catalogued as {\it filled} ({\it top-left}), {\it mixed} ({\it top-right}), {\it shell} ({\it bottom-left}), and {\it clear shell} ({\it bottom-right}). These regions correspond to reg.~5, reg.~98, reg.~45, and reg.~78 in Table B.1 in \citetalias{2013A&A...552A.140R}, respectively. Reg.~98 is the most luminous \hii\ region in M\,33, NGC~604. The best-fit SED ({\it continuous black line}) is obtained assuming a 4\,Myr star cluster ISRF and {\it Compiegne} dust model. {\it  Light blue points}: observed data with the errors, {\it red squares}: modelled broad-band fluxes, {\it dashed-dot blue line}: total (ionised and neutral) PAH emission, {\it dashed-dot green line}: VSG (SamC) grain emission, {\it dashed-dot red line}: BG (LamC and aSil grains) emission, and {\it dashed purple line}: NIR continuum.}
              \label{fig:sed}%
    \end{figure*}

In Fig.~\ref{fig:sed} we show some examples of SEDs modelled with the \DE\ code.  PAH emission dominates the SED at shorter wavelengths (3-12\,\mi), while the emission of the VSGs has its maximum between 24\,\mi\ and 100\,\mi. At longer wavelengths ($\lambda>100$\,\mi), the SED is dominated by the emission of the BGs. All the SEDs are fitted with residuals less than $\sim$50\% (see the bottom part of each figure). 

\begin{figure*} 
   \centering
\includegraphics[width=\textwidth]{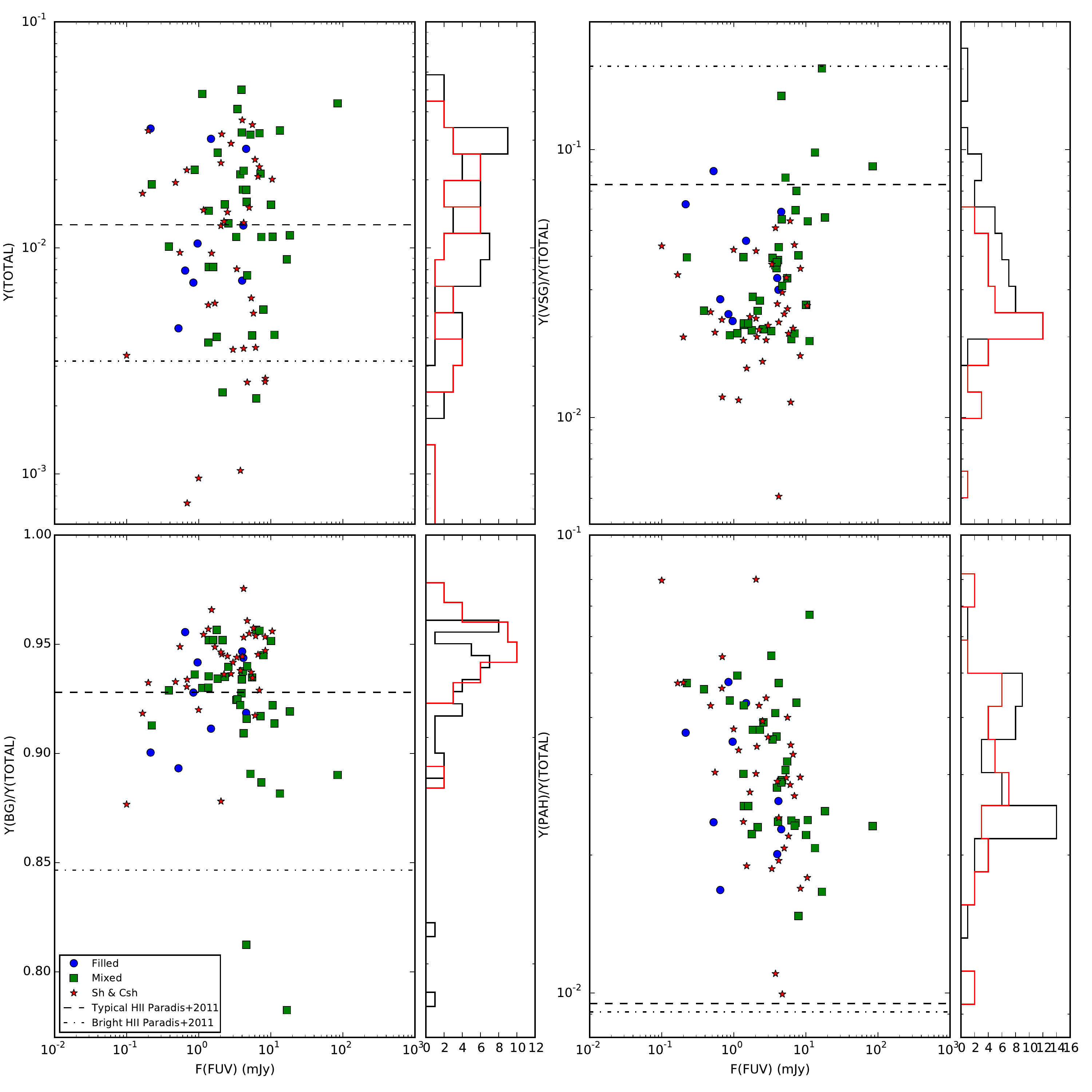}
   \caption{Dust mass abundances obtained fitting the SED of each region using a 4\,Myr ISRF and {\it Compiegne} dust model. Only fits with a $\chi^{2}_{\rm red}<20$ are taken into account. Blue circles are {\it filled} regions, green squares {\it mixed} ones, and red triangles {\it shells} or {\it clear shells}. {\it Dashed and dashed-dotted lines} correspond to the values obtained by \citet{2011ApJ...735....6P} who separate between {\it typical} \hii\ regions and regions being significantly brighter at \ha: for {\it typical} (dashed line) and {\it bright} (dot-dashed line) \hii\ regions in the LMC modelled assuming a 4\,Myr star cluster ISRF and the {\it Desert} dust model. Mean values of the relative errors in the data presented in the figures are: 11\%, 19\%, 16\% and 23\% for top-left, top-right, bottom-left and bottom-right figures, respectively. {\it Black histograms correspond to the {\it filled} and {\it mixed} regions and red ones to the  {\it shells} or {\it clear shells} regions.} }
              \label{fig:Y4myr}%
    \end{figure*}

In Fig.~\ref{fig:Y4myr} we analyse the relative dust masses for each grain type for the best fit with a 4\,Myr cluster ISRF. 
In the lower panels we plot the dust masses of the BGs and the PAHs relative to the total dust abundance for the best fit. BGs represent the highest fraction of dust mass in the regions ($\sim$90\%, and varying within a small range) independently of the morphological type. The relative mass of the PAHs is $\lesssim$5\% except for three objects. \citet{Draine:2007p588} found a value of 4.6\% for the relative mass of the PAHs in the Milky Way and \citet{2008ApJ...672..214G} showed that the PAH fraction varies significantly with the metallicity. The relative mass of the PAHs found here for most of our objects is consistent with the trend presented in Fig.~25 of  \citet{2008ApJ...672..214G} considering the metallicity of M\,33\footnote{Since M\,33 has a shallow metallicity gradient \citep{2011ApJ...730..129B}, we use a mean value between the extreme cases of the radial gradient as the characteristic metallicity for M\,33. Thus, for this study we assume $\rm Z_{M33}\sim0.5\,$\zsun.}. There are no variations with morphology but we observe a trend of lower \ypah/\ytot\ for regions with high FUV flux. We show here that the relative PAH mass abundance obtained using a 4\,Myr cluster ISRF and the {\it Desert} dust model (see Fig.~\ref{fig:sed:desert}) is lower than the one predicted by the {\it Compiegne} dust model.

\begin{table}
\caption{Statistics of \yvsg/\ytot,  \ypah/\ytot,  \ybg/\ytot\ for each morphological type of the \hii\ regions in our sample. The error bars are the standard error of the mean.} \label{tab:data}
\begin{center}
\begin{tabular}{l l l l }
\hline
Relative fractions &  \yvsg/\ytot\ &  \ypah/\ytot\ &  \ybg/\ytot\   \\
\hline
{\it Compiegne} dust model&&& \\
\hline
{\it Filled} & 0.04\,$\pm$\,0.01 & 0.03\,$\pm$\,0.01 & 0.93\,$\pm$\,0.30 \\
{\it Mixed} & 0.045\,$\pm$\,0.007 & 0.033\,$\pm$\,0.005 & 0.92\,$\pm$\,0.15 \\
{\it Sh \& Csh} & 0.026\,$\pm$\,0.004 & 0.033\,$\pm$\,0.005 & 0.94\,$\pm$\,0.16 \\ 
\hline
{\it Desert} dust model &&& \\
\hline
{\it Filled} & 0.09\,$\pm$\,0.03 & 0.009\,$\pm$\,0.003 & 0.91\,$\pm$\,0.30 \\
{\it Mixed} & 0.10\,$\pm$\,0.02 & 0.009\,$\pm$\,0.002 & 0.89\,$\pm$\,0.15 \\
{\it Sh \& Csh} & 0.048\,$\pm$\,0.008 & 0.012\,$\pm$\,0.002 & 0.94\,$\pm$\,0.16 \\ 
\hline
Paradis et al. 2011 &&& \\
\hline
Bright & 0.20\,$\pm$\,0.02 & 0.0091\,$\pm$\,0.001 & 0.85\,$\pm$\,0.01 \\
Typical & 0.074\,$\pm$\,0.007 & 0.0095\,$\pm$\,0.001 & 0.93\,$\pm$\,0.06 \\
\hline

\end{tabular}
\end{center}
\end{table}

The top-right panel of Fig.~\ref{fig:Y4myr} shows the relative dust mass abundance for the VSGs. The most interesting feature in this plot is that the relative dust mass abundance for the VSGs changes with the morphology of the region: red stars, corresponding to {\it shells} and {\it clear shells} are located in the lower part of the figure, while green squares and blue dots, corresponding to {\it filled} and {\it mixed} regions respectively, tend to be in the top part of the distribution. The  {\it shells} and {\it clear shells} do not present \yvsg/\ytot\ values higher than $\sim$0.05, while there is no {\it filled} nor {\it mixed} region with values of \yvsg/\ytot\ less than $\sim$0.02. This is clearly seen in the histogram at the right-hand side of the figure. In Table~\ref{tab:data} we give the mean values of the relative abundance of each grain type for each morphological classification. Regions classified as {\it shells} and {\it clear shells} have a factor of $\sim$1.7 smaller fraction of dust mass in the form of VSGs than the {\it filled} and {\it mixed} regions, and the trend is the same using the {\it Desert} dust model (see Fig.~\ref{fig:Y4myr:desert} and Table~\ref{tab:data}). 

The trend shown here agrees with the results presented in other studies. \citet{2011ApJ...735....6P} found, using the {\it Desert} dust model, an increase of the VSG relative abundance by a factor of 2-2.5 between {\it bright} and   {\it typical} regions in the LMC (see Table~\ref{tab:data} and Table~2 in \citet{2011ApJ...735....6P}). {\it Bright} \hii\ regions in \citet{2011ApJ...735....6P}  would correspond to our  {\it filled} and  {\it mixed} regions, while the {\it shell} objects present lower \ha\ surface brightness and can be included in the {\it typical} regions classification of \citet{2011ApJ...735....6P}. Regarding the PAHs, the values obtained here are slightly lower than those predicted by \citet{2011ApJ...735....6P}, but an agreement is found when we use the {\it Desert} dust model as in the \citet{2011ApJ...735....6P} study (see Fig.~\ref{fig:Y4myr:desert}). Caution needs to be taken when using the {\it Desert} dust model, as this model is unable to fit the SED within the 6-9\,\mi\ wavelength range, under-predicting systematically the observed PAH emission (see  Fig.~\ref{fig:sed:desert}). We found no trend of relative PAH abundance with the morphology, while \citet{2011ApJ...735....6P} found a factor of 0.96 between {\it bright} and {\it typical} regions in the LMC.

\citet{2014ApJ...784..147S} analysed the IR emission in small subregions of two classical \hii\ regions and two superbubbles in the LMC fitting the SEDs with \DE\ code and  using the {\it Desert} dust model. They found that the emission from VSGs increases at the inner locations of the two {\it classical} \hii\ regions with respect to the values obtained at the edges of the region, while for the two superbubbles the enhancement is not observed. We are observing the same trend, the {\it filled} and {\it mixed} regions, which would be {\it classical} in \citet{2014ApJ...784..147S} terminology, present higher fraction of VSGs than the {\it shells} and {\it clear shells}, corresponding to {\it superbubbles} in \citet{2014ApJ...784..147S}. The enhancement of the relative VSGs abundance in the center of the {\it classical} \hii\ regions corresponds to the location where the ISRF is higher and where the massive stars are situated, as is calculated in Sect.~3.2 in \citet{2014ApJ...784..147S}. As we will discuss in Sect.~\ref{sec:dustcoltemp}, we expect an enhancement of VSGs at the position of strong shocks where the massive stars are located.

\subsection{Morphology and dust properties}\label{sec:dustcoltemp}
We show here that \yvsg/\ytot\ is higher for {\it filled} and {\it mixed} regions by a factor of $\sim$1.7. This trend allows us to understand the change in dust properties in different environments and to shed light on the dust evolution in the ISM. \citet{1996ApJ...469..740J} showed that dust grain-grain collisions, normally referred to {\it shattering}, lead to fragmentation of the entire grain, or part of it, into smaller distinct subgrains. Fig.~13\,c of \citet{1996ApJ...469..740J} shows that large grains (radii $>$\,400\,\AA) are disrupted by a single 100\,\kms\ shock and that the grain mass is transferred to smaller grains. This {\it mass transfer} is seen in the resulting postshock grain size distribution (Fig.~17 in that paper). Moreover, the adoption of a different initial grain size spectrum does not greatly affect the final grain size distribution, as shown in \citet{1996ApJ...469..740J}. A re-evaluation of the dust processing effects in shocks has been performed by \citet{2014A&A...570A..32B}. These authors use the new dust model from \citet{2013A&A...558A..62J}, which assumes only two grain types (hydrogenated amorphous carbon and large amorphous silicate grains), and found that both carbonaceous and silicate grains undergo more destruction than in the previous studies, being the silicates more resilient to shocks than the carbonaceous grains. Since we keep the ratio of the silicates and carbonaceous fixed in our models we are not able to test this further. 

Our sample of \hii\ regions classified as {\it filled} and {\it mixed} have central clusters rich in massive stars producing strong stellar winds and creating high-velocity \citep[$\sim$\,50-90\,\kms, according to][]{Relano:2005p644} expansive structures in their interiors. Therefore, for these regions a higher fraction of \yvsg/\ytot\ compared to more quiescent environments is expected according to the \citet{1996ApJ...469..740J} dust evolution model. {\it Shell} and {\it clear shell} objects in our sample are in a more evolved and quiescent state. They have already swept up all the gas and dust in their close surroundings and they show prominent shell structures not only in \ha\ but also in the longer wavelengths 250\,\mi, 350\,\mi, and 500\,\mi\ of SPIRE \citep[see Fig.\,3 of][]{Verley:2010p687}. We expect these regions to have much lower shock velocities, as they are in a more evolutionary state, and therefore to present a lower relative fraction of VSGs. This agrees with the results presented in \citet{2014ApJ...784..147S}, who showed that for the most evolved \hii\ region in their sample, a superbubble named N70,  \yvsg/\ytot\ is lower than for the less evolved \hii\ regions in their sample. Another interpretation of the observed shell structures in the SPIRE bands could be the reformation of dust BGs via coagulation and accretion in dense clouds, as it has also been suggested by \citet{1996ApJ...469..740J}. \citet{2003A&A...398..551S} proposed grain-grain coagulation as the main mechanism to explain the deficit of the IRAS I(60\,\mi)/I(100\,\mi) flux ratio in the Taurus molecular complex compared to the diffuse ISM. The same region was studied by \citet{2013A&A...559A.133Y} using \DE\ who found that the observed SED at different locations within the complex could be explained by changes in the optical properties of the grains. \citet{2009A&A...506..745P} found an increase of the FIR emissivity in molecular clouds along the Galactic plane with significantly colder dust and interpreted it as coagulation of dust grains into fractal aggregates. Although reformation of BGs is not excluded as a mechanism to explain the shell structures observed in the SPIRE bands, detailed estimations of the dust temperature of the BGs and the density of the ISM in these regions have to be obtained in order to check the viability of this phenomenon \citep{2012A&A...548A..61K}. For example, \citet{2015A&A...579A..15K} found that coagulation and accretion produce significant changes in the dust properties that can explain the observed SED variations between diffuse and denser regions.

\subsection{Dust Temperature}\label{sec:MBB}

\begin{figure*} 
  \includegraphics[width=\textwidth]{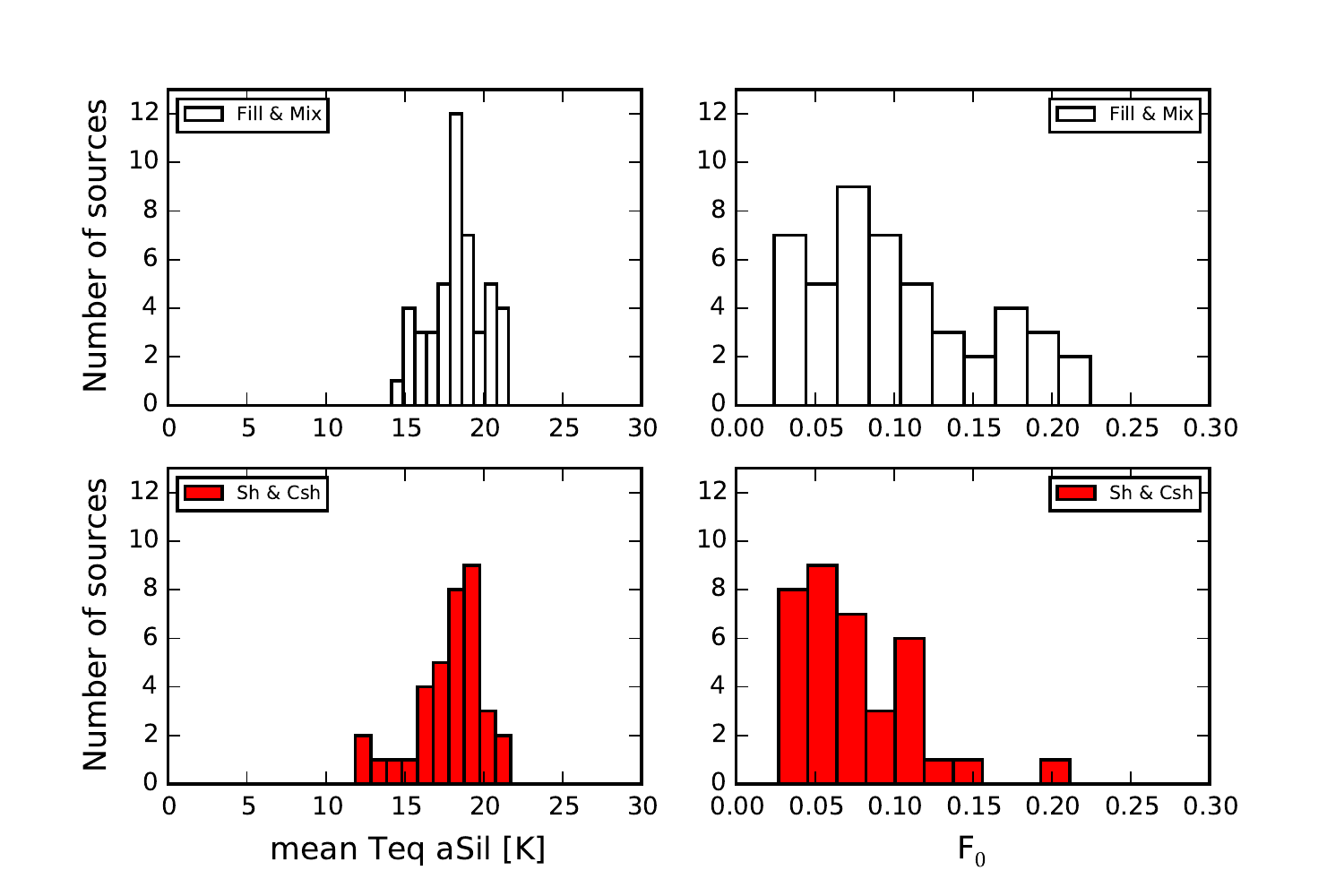} 
    \caption{Dust temperature (left) and $\rm F_{0}$ (right) distributions for {\it  filled} and {\it mixed} together, and {\it shell} morphologies. The dust temperature is defined as the mean of the equilibrium temperature of all the aSil grains with different sizes.}
              \label{fig:histTeqGo}
\end{figure*}

In order to study the temperature of the dust for each \hii\ region in our sample we make use of the equilibrium temperature provided by \DE.  The code provides, for each size, the equilibrium temperature of the LamC and aSil grains. For each grain type, we estimate the dust temperature as the mean value of the equilibrium temperature of all aSil grains with different sizes as in \citet{2013A&A...559A.133Y}.\footnote{Other authors use the equilibrium temperature of the most populated dust grain \citep{2011A&A...531A..51F}. We have tested that the conclusions of this section do not change using this other definition of dust temperature.} We make use of the aSil grains because they are more abundant than LamC ones and the relative fraction between the two of them is kept constant in our fitting (see Sect.\,\ref{subsec:SEDmod}). In the left hand panel of Fig.~\ref{fig:histTeqGo} we show the distribution of the dust temperature for each morphological classification. We can see that there is only a slight trend for the distribution of the {\it shells} and {\it clear shells} to extend to lower temperatures, while the disitribution for the {\it  filled} and {\it mixed} regions is narrower, having a peak at $\sim$18\,K. We find a mean dust temperature of $18.3\pm0.3$\,K and $17.8\pm0.4$\,K for {\it  filled} and {\it mixed} together, and {\it shell} type objects, respectively, which shows that there is no significant difference in the dust temperature for each classification.

We have also performed the same analysis using the classical definition of dust temperature from a modified blackbody (MBB) fitting. We use observational data in the 100-870\,\mi\ wavelength range, as the expected contribution of the stochastically heated grains in this wavelength range is negligible (for our set of \hii\ regions the best fits give a mean contribution of $\sim$10\% of the VSGs to the 100\,\mi\ band). We adopt the simple approach to keep $\beta$=2 fixed and restrict the analysis to those fits with $\chi^{2}_{\rm red}<50$\footnote{{$\chi^{2}_{\rm red}$ is defined as $\chi^{2}/n$, where $n$ is the number of degrees of freedom and $\chi^{2}=\sum(F_{obs}-F_{mod})^2/\sigma^2$}.} and having a model 70\,\mi\ flux below the observed one. In total we end up with a set of 71 \hii\ regions: 7 {\it filled}, 35 {\it mixed}, and 29 {\it shell} and {\it clear shell} objects. We find a mean dust temperature of $17.9\pm0.6$\,K, and $15.5\pm0.6$\,K for {\it  filled} and {\it mixed}, and {\it shell} type objects, respectively. 

We find that the equilibrium temperature for the BGs obtained with \DE\ does not show any trend with morphology, while the dust temperature derived from the MBB shows differences that are slightly larger than 3$\sigma$. Based on these results, we cannot conclude that there is a statistically significant difference in the dust temperature distributions, but we find evidence from the MBB analysis that \hii\ regions of {\it shell} type might have the lowest temperatures. This trend can be understood by the different star-to-dust geometries in the \hii\ regions, as within {\it  filled} and {\it mixed} regions the dust and gas is probably mixed and closer to the stars that heat the dust than in the {\it shells }and {\it clear shells} regions. Indeed, in the right panel of Fig.~\ref{fig:histTeqGo} we show the distribution of 
$\rm F_{0}$, a measure of the radiation field intensity (see Sect.\,\ref{sec:comp_isrf}), for both {\it  filled} and {\it mixed}, and {\it shell} type regions, showing that {\it shell} objects are lacking the high values of $\rm F_{0}$ that are present in some {\it  filled} and {\it mixed} types. We find mean values of $\rm F_{0}$ of $0.103\pm0.008$, and $0.076\pm0.006$ for {\it  filled} and {\it mixed}, and {\it shell} type objects, respectively. 

\begin{figure*} 
  \includegraphics[width=\textwidth]{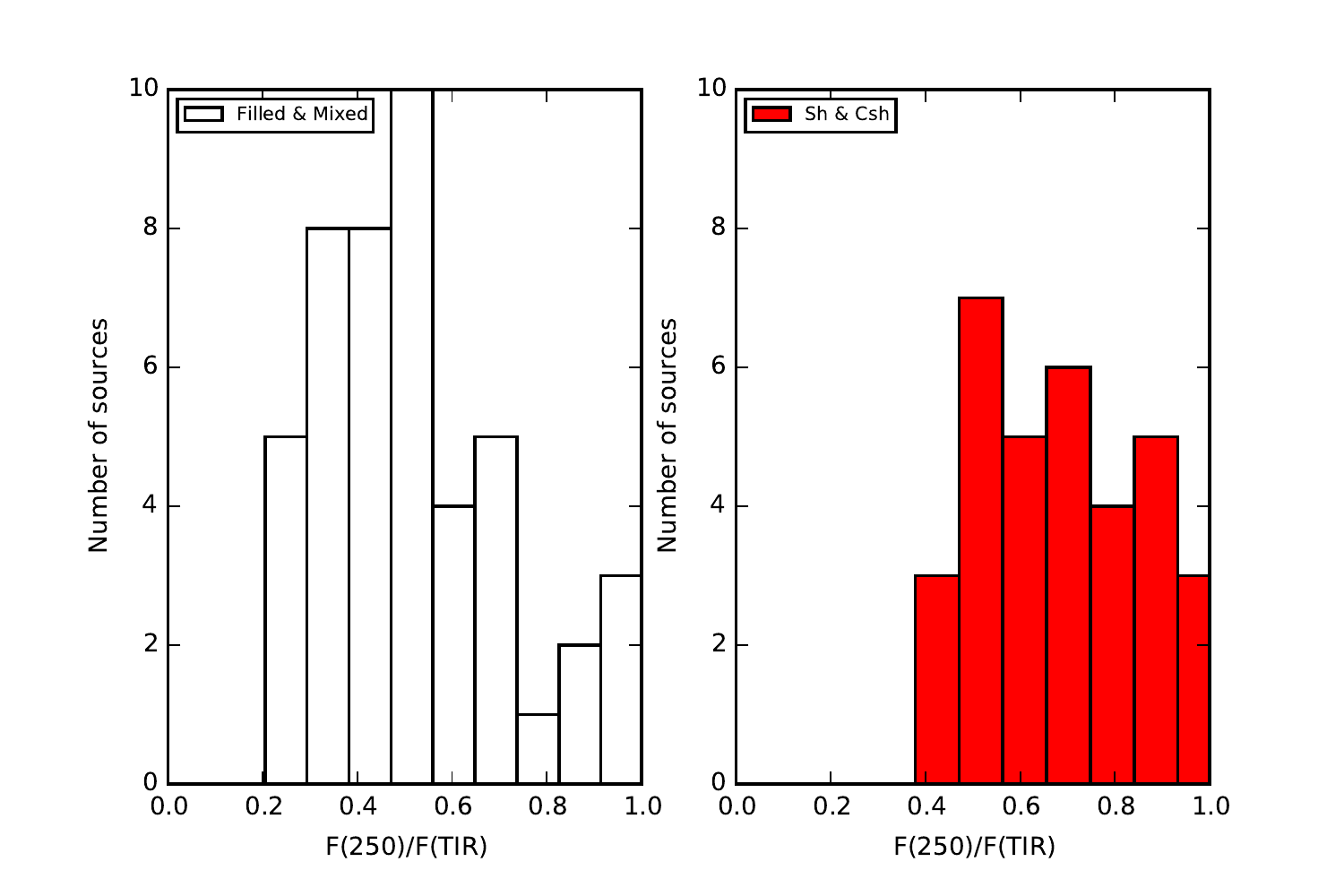} 
    \caption{250\,\mi\ emission relative to the TIR emission distribution for both {\it  filled} and {\it mixed} together, and {\it shell} morphologies. The TIR emission is obtained using the prescription given by \citet{Boquien:2011p764} including SPIRE and \Her\ data from 24 to 250\,\mi.}
              \label{fig:hist250_TIR}
\end{figure*}

We cannot rule out other mechanisms producing this slight difference in the dust temperature, as, e.g., a higher fraction of BGs in the  {\it shell} regions  that would imply cooler grains with respect to the {\it  filled} and {\it mixed} objects. Indeed, we have already shown that the relative fraction of VSGs is lower in  the {\it  filled} and {\it mixed} regions (see Fig.~\ref{fig:Y4myr}). It is also worth to note that the emission at 250\,\mi\ relative to the TIR emission, an indicator of the dust temperature and/or the relative amount of BGs, presents higher values for {\it shell} regions than for  {\it  filled} and {\it mixed} objects (see Fig.~\ref{fig:hist250_TIR}): the mean value of the 
250\,\mi/TIR ratio for {\it shell} regions is 0.73$\pm$0.04, while for {\it  filled} and {\it mixed} objects is 0.53$\pm$0.03.

The trends shown here are consistent with previous results given in the literature. \citetalias{2013A&A...552A.140R} presented evidence of a far-IR peak shifted towards longer wavelengths in the SED of \hii\ regions classified as {\it shell} and {\it clear shell} objects, which points to a colder dust in this type of objects. \citet{Verley:2010p687} show examples of {\it  filled} and  {\it clear shell} regions where the dust emission distribution observed in different bands of \Spi\ and \Her\ differs from band to band. {\it  Filled} regions tend to show emission in all IR bands, while {\it clear shell} objects do not show any emission at 24\,\mi\ and the emission in the SPIRE bands is delineating the shell structure of the region. The authors suggest an evolutionary scenario where the dust is affected by the expansion of the gas within the region. The  {\it  filled} regions would be younger. The stellar winds would have not been able to create a cavity yet, therefore the dust and ionised gas are mixed within the region. They present emission at 24\,\mi\ inside the region indicating that the dust is warm for these objects. The {\it clear shell} objects, however, would be more evolved. The expanding structure would have been created and the dust would have been swept together with the ionised gas. The lack of emission at 24\,\mi\ and the presence of SPIRE emission at the shells shows that the dust is cooler for these objects.

\subsection{Gas-to-dust mass ratio}\label{sec:GTD}
Using the available HI and CO data we are able to derive the gas mass within the aperture of each \hii\ region in the sample. The \DE\ fits provide us the dust mass for each region and therefore we can obtain the gas-to-dust mass ratio in our sample. The molecular mass is obtained using the $^{12}$CO(J=2--1) and HI intensity maps presented in \citet{2014A&A...567A.118D} and \citet{2010A&A...522A...3G}, respectively. The data were smoothed and registered to the SPIRE 250\,\mi\ spatial resolution and pixel size, and photometry was extracted following the procedure explained in Sect.~\ref{subsec:phot}. We use a CO-to-$\rm H_{2}$ conversion factor of $X(\rm CO)=\frac{N_{H_{2}}}{I_{CO(1-0)}}=4.0\times10^{20}\,cm^{-2}/(K\, km\,s^{-1}$), which is the one applied by \citet{2014A&A...567A.118D}  and consistent with the metallicity of M\,33. We then apply Eq.~3 from \citet{2014A&A...567A.118D}, which assumes a constant $\rm I_{CO(2-1)}/I_{CO(1-0)}$,  to derive the molecular hydrogen mass $\rm M_{H_{2}}$ for each \hii\ region. The HI mass is obtained using a conversion factor for HI of $\rm 1.8\times10^{18}\,cm^{-2}/(K\,km\,s^{-1})$, as in \citet{2010A&A...522A...3G}. In both cases we take into account the fraction of helium mass with a correction of 37\%. The total gas mass is the sum of the neutral HI and the molecular $\rm H_{2}$ mass ($\rm M_{tot}=M_{HI}+M_{H_{2}}$). 

\begin{figure} 
   \centering
  \includegraphics[width=0.49\textwidth]{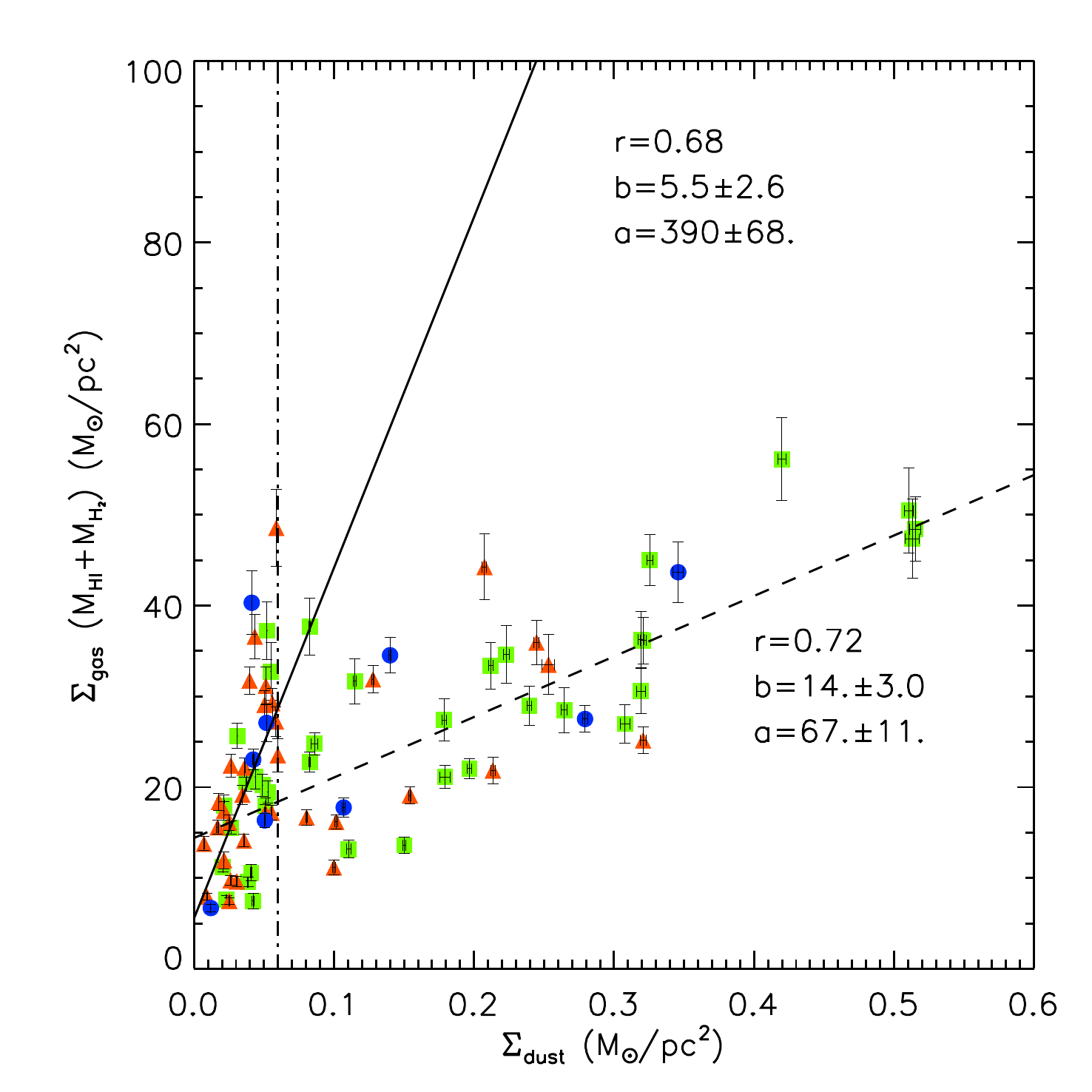} 
   \caption{Total gas surface density versus the dust surface density for the \hii\ regions in our sample with fits presenting $\chi^{2}_{\rm red}<20$. We have also excluded 5 regions that  do not show CO emission. The colour code represents the morphological classification of the sample, as in Fig.~\ref{fig:Y4myr}. The continuous line represents the fit including only regions with $\rm \Sigma_{dust}<0.06\,M_{\sun}/pc^{2}$, while the dashed line shows the fit for regions with $\rm \Sigma_{dust}\geq0.06\,M_{\sun}/pc^{2}$. $a$, $b$, and $r$ represent the slope, intercept and correlation coefficient of the linear fits.}
              \label{fig:SDdustgas}%
    \end{figure}

In Fig.~\ref{fig:SDdustgas} we show the total gas surface density $\rm \Sigma_{gas} (M_{HI}+M_{H_{2}})$ versus the dust surface density $\rm \Sigma_{dust}$ for the \hii\ regions in our sample. As in the other sections in the paper we use only those regions showing reliable fits ($\chi^{2}_{\rm red}<20$), and we also exclude 5 regions not presenting CO emission within the defined apertures. In general, the distribution in Fig.~\ref{fig:SDdustgas} tends to flatten at high $\rm \Sigma_{dust}$, as it has been recently observed in the N11 star-forming complex of the LMC \citep{2015arXiv151107457G}.  Analysing the distribution in more detail we see that two regimes separated at a value of $\rm \Sigma_{dust}=0.06\,M_{\sun}/pc^{2}$ seem to coexist, as was already shown for the LMC and SMC in \citet[][]{2014ApJ...797...86R}. The distributions of the $\rm \Sigma_{gas} (M_{HI}+M_{H_{2}})$ and $\rm \Sigma_{dust}$ are different for values higher and lower than $\rm \Sigma_{dust}=0.06\,M_{\sun}/pc^{2}$ with a p-value from the KS test of less than 1\% in both cases. $\rm \Sigma_{dust}\sim0.06\,M_{\sun}/pc^{2}$ is a reference value to separate the diffuse atomic and molecular ISM at the metallicity of M\,33. This value corresponds to a visual extinction of $\rm A_{V}\sim0.4$\,mag \citep[see][]{2014ApJ...797...86R} and \citet{2009ApJ...693..216K} shows that the \hi$\rm -H_{2}$ transition occurs for $\rm A_{V}=0.2-0.4$\,mag at $\rm Z\sim0.5\,$\zsun(see their Fig.\,1). Regions with $\rm \Sigma_{dust}<0.06\,M_{\sun}/pc^{2}$ are in the \hi\ diffuse regime, where a significant fraction of $\rm H_{2}$ molecular gas is dissociated, while regions with $\rm \Sigma_{dust}>0.06\,M_{\sun}/pc^{2}$ correspond to a molecular phase where the $\rm H_{2}$ is not dissociated. In our sample of regions, we obtain median values of $\rm M_{H_2}/M_{HI}=0.41\pm0.06$ and $\rm M_{H_2}/M_{HI}=0.88\pm0.08$, for the diffuse atomic and molecular regimes, confirming
that $\rm \Sigma_{dust}=0.06\,M_{\sun}/pc^{2}$ indeed separates molecular and atomic dominated regions.

We treat the two regimes separately and perform linear fits in each one (see Fig.~\ref{fig:SDdustgas}). The derived slopes account for the gas-to-dust ratio in both regimes. The slope for the \hi\ diffuse regime, $390\pm68$, is closer to the value of the gas-to-dust ratio expected from the metallicity of M\,33, \citep[gas-to-dust\,$\sim$\,325, following the empirical broken power law given by][]{2014A&A...563A..31R} than the gas-to-dust ratio derived for the molecular regime, $67\pm11$. Note, however, that other gas-to-dust ratios derived for M\,33 using modified black body fitting range from 250 to 180, e.g. \citet[][]{Kramer:2010p688} and \citet{2016arXiv160302125H}. All of this shows the importance of using physically-motivated dust models as the results might change significantly depending on the used model. The slope obtained from the fit in the molecular regime gives a value $\sim$5 times lower than the one derived in the  \hi\ diffuse regime, in agreement with the difference found between the  slopes of both regimes observed in the LMC by \citet[][]{2014ApJ...797...86R}. 

These authors present a detailed study of the possibilities causing the different slopes. One of the options is a different $X(\rm CO)$ factor for the diffuse atomic and molecular regimes. However, in order to bring the slopes of both regimes into agreement with the gas-to-dust ratio expected from the metallicity in M\,33 we would need a higher $X(\rm CO)$ factor in the molecular dense regime. Recent models from \citet{2011MNRAS.412.1686S} show the opposite, the $X(\rm CO)$ factor would decrease in denser regions. Other physical explanations proposed by \citet[][]{2014ApJ...797...86R} are the existence of grain coagulation and/or accretion of gas-phase metals onto dust grains. Grain coagulation would imply a constant dust mass fraction, as in this process it is the dust size distribution which is altered, keeping the total dust mass constant. Coagulation of VSGs into BGs can produce an enhancement in the dust opacity at long wavelengths \citep{2012A&A...548A..61K,2015A&A...579A..15K}, and thus the emissivity of the BGs would be enhanced in the FIR part of the SED. The result would be an overestimation of the dust mass. In the grain accretion process, gas-phase metals will be incorporated to the dust mass and thus will produce an increase of the dust-to-gas ratio. In other to shed light onto the physical mechanisms that can cause the difference in the slopes for both regimes, we plan to perform a detailed study of the dust-to-gas ratio variations in M\,33 using \DE\ models selecting an appropriate sample of spatial regions in this galaxy, characteristic of the \hi\ and molecular diffuse regimes (Rela\~no et al. in prep). 

\section{Spatially resolved analysis: NGC~604 and NGC~595}\label{sec:604_595}
The spatial resolution of \Her\ data allows us to analyse the interior of the largest and most luminous \hii\ regions in M\,33: NGC~604 and NGC~595. The \ha\ emission of NGC~604 reveals the presence of large cavities of ionised gas \citep[see Fig.~2 in][]{Relano:2009p558} produced by the stellar winds coming from the central star clusters. Stellar population studies in this region present evidence of Wolf Rayet (WR) stars \citep{Hunter:1996p605} and red supergiants (RSGs) \citep{Eldridge:2011p756}, showing two distinct formation episodes with ages 3.2$\pm$1.0\,Myr and 12.4$\pm$2.1\,Myr \citep{Eldridge:2011p756}. 

NGC~595 is the second most luminous \hii\ region in M\,33. The \ha\ emission presents an open arc structure with a bright knot in the centre where the most massive stars are located \citep{2010MNRAS.402.1635R}. The region also hosts numerous OB-type and RSG stars \citep{Malumuth:1996p494}, as well as WR stars \citep{Drissen:2008p481}.  \citet{Malumuth:1996p494} derived an age of 4.5$\pm$1.0\,Myr for the stellar population of NGC~595, which was confirmed later by \citet{Pellerin:2006p499}. \citet{2011MNRAS.412..675P} performed a detailed modelling of NGC~595 using CLOUDY \citep{1998PASP..110..761F} in a set of concentric elliptical annuli. These authors were able to fit the optical integral field spectroscopic observations from \citet{2010MNRAS.402.1635R} and mid-infrared observations from \Spi. The 8\,\mi/24\,\mi\ ratio was successfully fitted with a dust-to-gas ratio which increases radially from the center to the outer parts of the region. The spatial distribution of the dust emission at 24\,\mi\ correlates with the emission at \ha\ for both \hii\ regions, showing that there is hot dust mixed with the ionised gas in both objects \citep[][]{Relano:2009p558}.  

  \begin{figure*} 
   \centering
   \includegraphics[width=\textwidth]{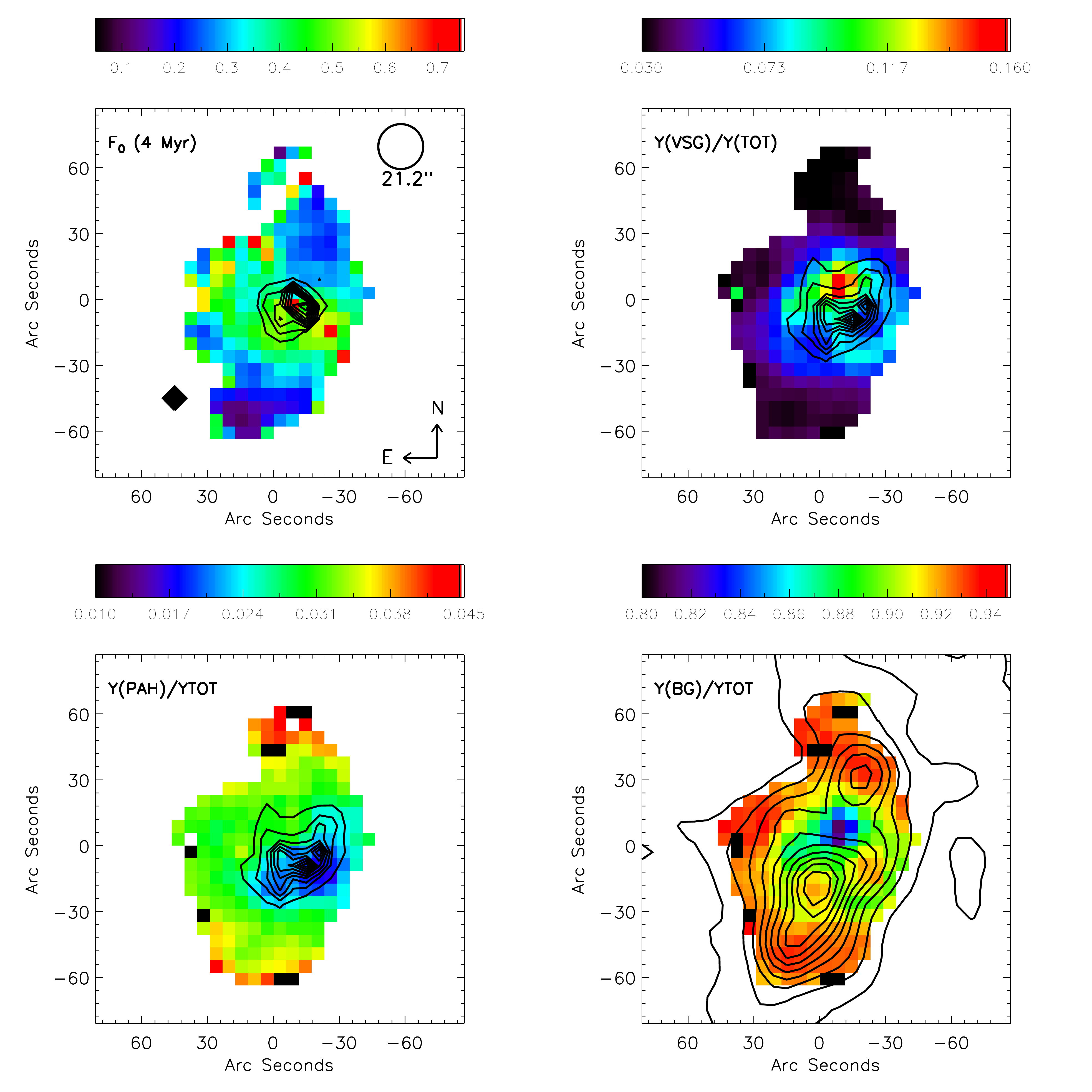} 
         \caption{$F_{0}$ (top-left), \yvsg/\ytot\ (top-right), \ypah/\ytot\ (bottom-left), and \ybg/\ytot\ (bottom-right) distribution maps of NGC~604 obtained using \DE\ with an ISRF representing a 4\,Myr star cluster given. The contours are FUV, \ha, \ha, and CO emission for the $F_{0}$, \yvsg/\ytot, \ypah/\ytot, and \ybg/\ytot\ distribution maps, respectively. The spatial resolution of the contours are 4.4\arcsec, 6.6\arcsec, 3.0\arcsec, and 20\arcsec, respectively.}
              \label{fig:yallrel_4Myr604}%
\end{figure*}

  \begin{figure*} 
   \centering
    \includegraphics[width=\textwidth]{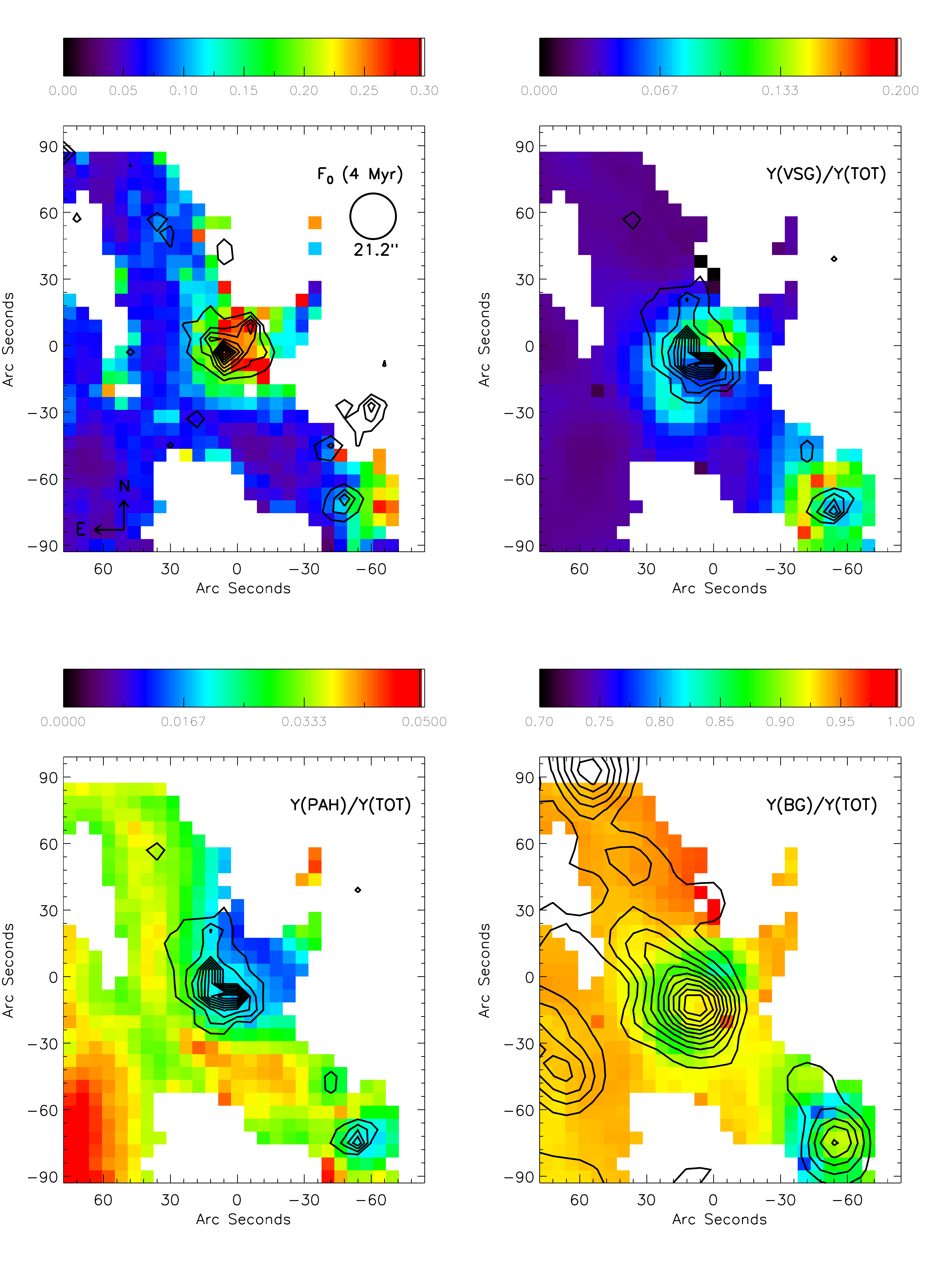}
    \vspace{-1cm}
      \caption{$F_{0}$ (top-left), \yvsg/\ytot\ (top-right), \ypah/\ytot\ (bottom-left), and \ybg/\ytot\ (bottom-right) distribution maps of NGC~595 obtained using \DE\ with an ISRF representing a 4\,Myr star cluster, as it was done in previous sections. The contours are FUV, \ha, \ha,  and CO emission for the $F_{0}$, \yvsg/\ytot, \ypah/\ytot, and \ybg/\ytot\ distribution maps, respectively. The spatial resolution of the contours are 4.4\arcsec, 6.6\arcsec, 3.0\arcsec, and 20\arcsec, respectively.}
              \label{fig:yallrel_4Myr595}%
\end{figure*}

For the whole surface of NGC~604 and NGC~595 we perform a  pixel-by-pixel SED modelling using \DE\ and derive maps for each free parameter in the fit. We impose a limit in the surface brightness at 250\,\mi\ to constrain the spatial emission of the \hii\ regions and to discard pixels with low surface brightness having high uncertainties in their fluxes. We assume a 4\,Myr and $10^{4}$\msun\ star cluster ISRF  which accounts for the age and stellar mass of the two regions. 

\subsection{Maps of the relative mass abundance of grains}
In Figs.~\ref{fig:yallrel_4Myr604} and ~\ref{fig:yallrel_4Myr595} we show the results of the SED fitting on a pixel-by-pixel basis. 
$F_{0}$ correlates with FUV emission, which seems plausible as $F_{0}$ is a measurement of the ISRF intensity coming from the stars within the region. The spatial distribution of the \yvsg/\ytot\ map in both regions resembles the form of arcs and shells bordering the \ha\ emission, and the relative fraction of PAHs, \ypah/\ytot, tends to be higher around the region. Finally, \ybg/\ytot\ distribution at the central areas of both regions tends to show high values at the location of high emission in CO. All these maps show that the relative mass abundance of each grain type changes within the region depending on the local properties of the environment.

\subsection{Signature of dust evolution within the regions}

\begin{figure*}
   \centering
      \includegraphics[width=0.49\textwidth]{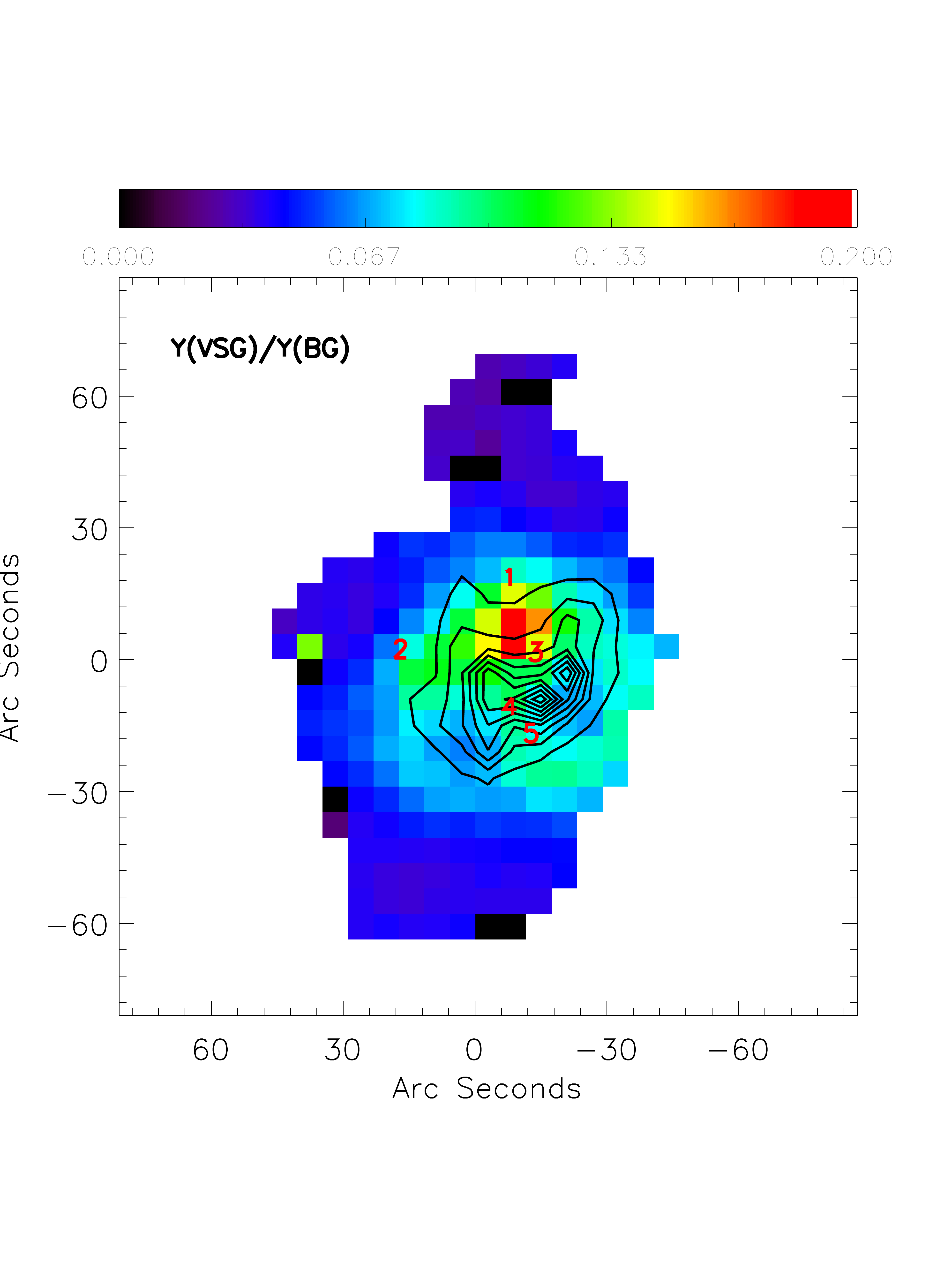}
     \includegraphics[width=0.49\textwidth]{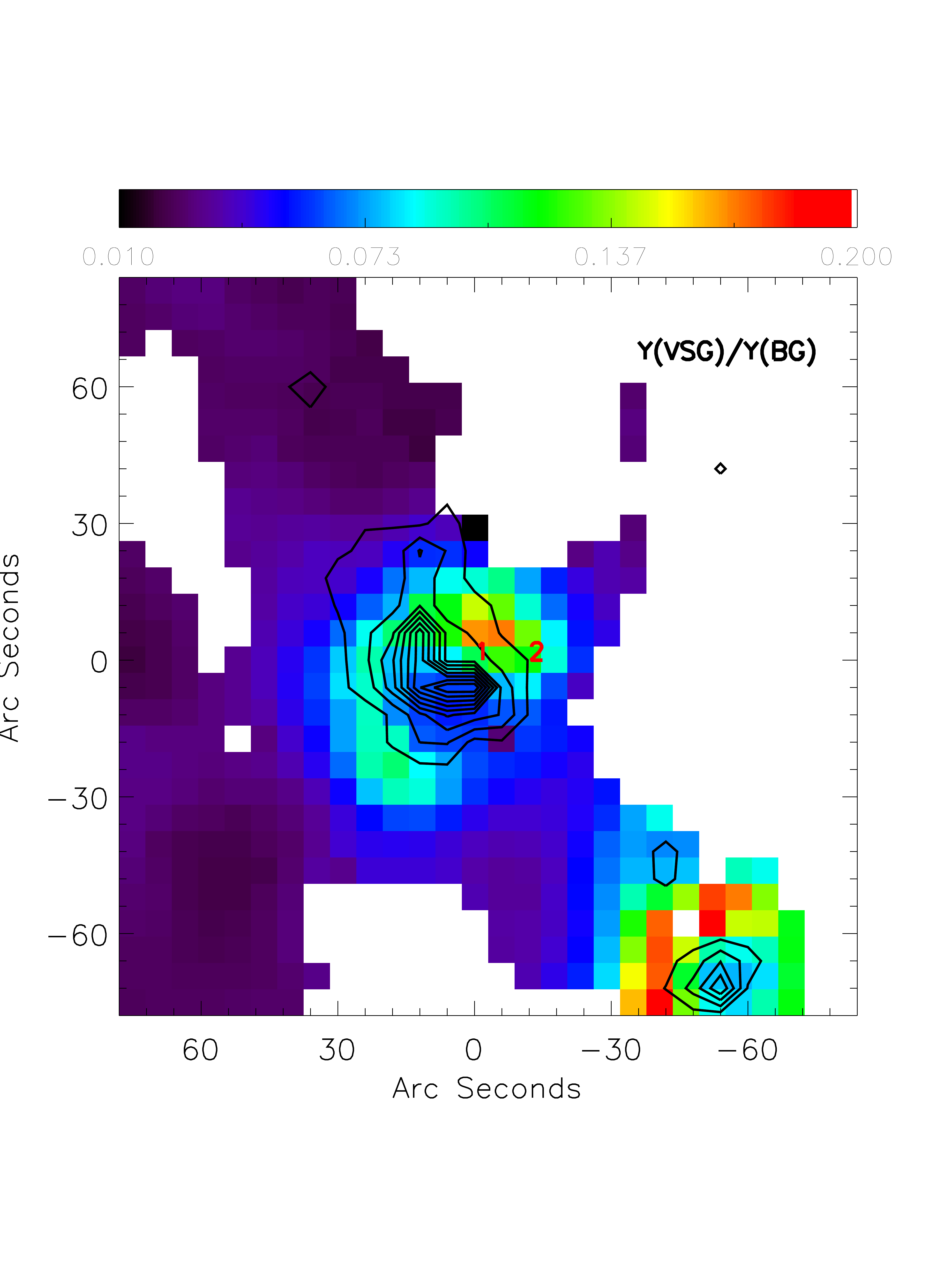}
      \caption{\yvsg/\ybg\ distribution map for NGC~604 (left) and NGC~595 (right) obtained using \DE\ code with a 4\,Myr star cluster ISRF for NGC~604 and for NGC~595 and {\it Desert} dust model. The contours correspond to the continuum-subtracted \ha\ image at high resolution (6\arcsec) and the numbers refer to the location of the expanding ionized gas shells observed by  \citet{1996AJ....112..146Y}  for NGC~604 and \citet{Lagrois:2009p564} for NGC~595. Pixel size is 6\arcsec.}
              \label{Yvsg2bg}%
\end{figure*}

Evolution of the interstellar dust implies a redistribution of the dust mass in the different grain types via coagulation of small grains or fragmentation of large ones. Thus, a map of the relative mass fraction of VSG to BG would show the locations where these processes might take place. In both maps in Fig.~\ref{Yvsg2bg} we can see maxima in \yvsg/\ybg\ that seem to be bordering the \ha\ emission. \citet{1996ApJ...469..740J} predict that the fragmentation of BGs into VSGs occurs  for shocks with velocities of at least of 50\,\kms. Therefore, we expect the presence of expanding structures with similar velocities at the location of \yvsg/\ybg\ enhancements. \citet{1996AJ....112..146Y} studied the kinematics of NGC~604 and found five shell structures within the region with expansion velocities of 40--125\,\kms\ traced by the \ha\ emission line. The approximate locations of the centre of the shells are depicted in Fig.~\ref{Yvsg2bg} with numbers. All of them are within the area where the maxima of \yvsg/\ybg\ are located. The kinematics of the ionised gas in the interior of NGC~595 was studied by \citet{Lagrois:2009p564}. These authors found two expanding shells located close to the position where the maximum of \yvsg/\ybg\ occurs and estimated a mean expansion velocity for both shells of $\sim$20\,\kms. 

\subsection{CO emission}
\begin{figure*}
   \centering
   \includegraphics[width=0.49\textwidth]{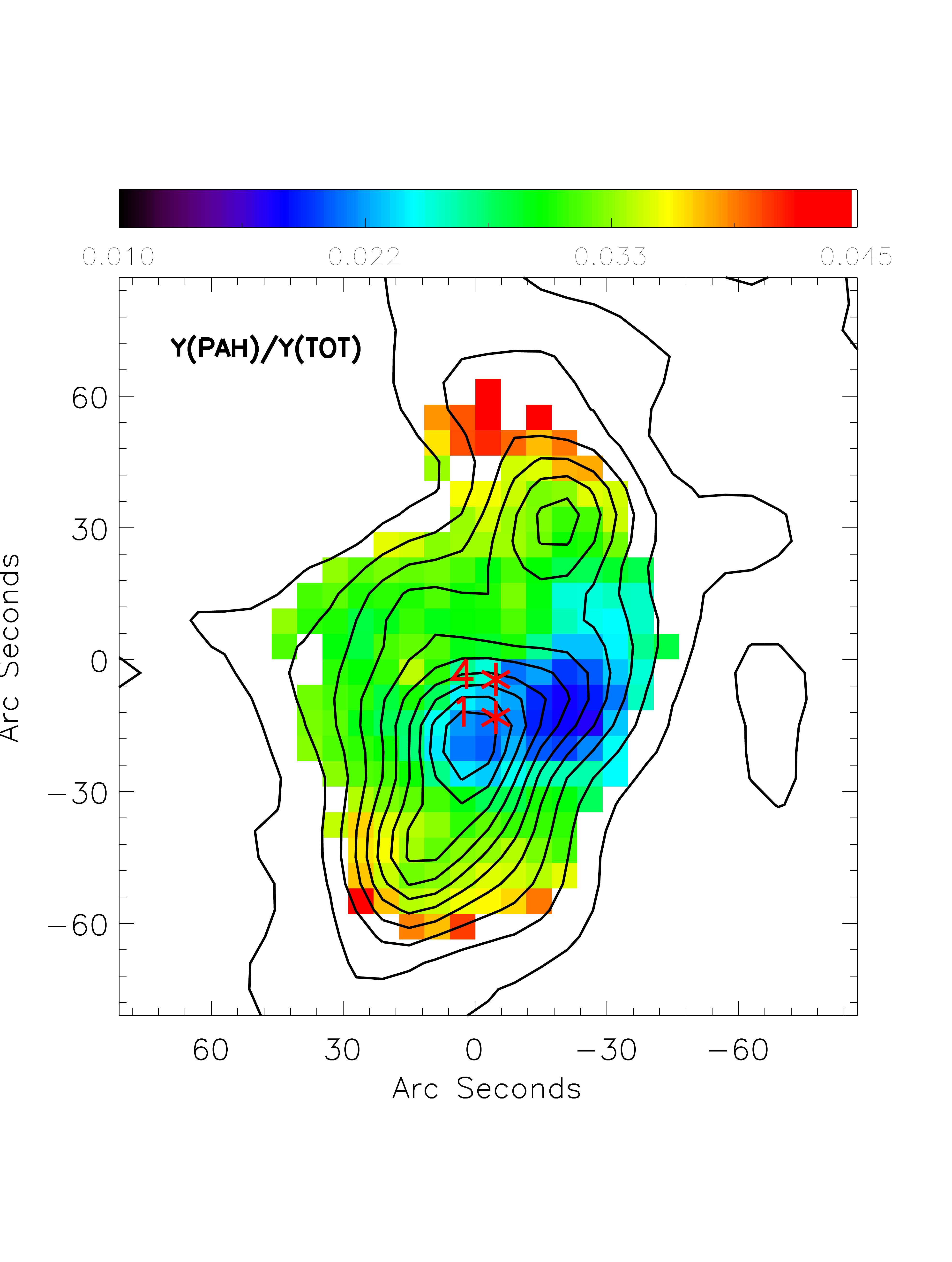} 
   \includegraphics[width=0.49\textwidth]{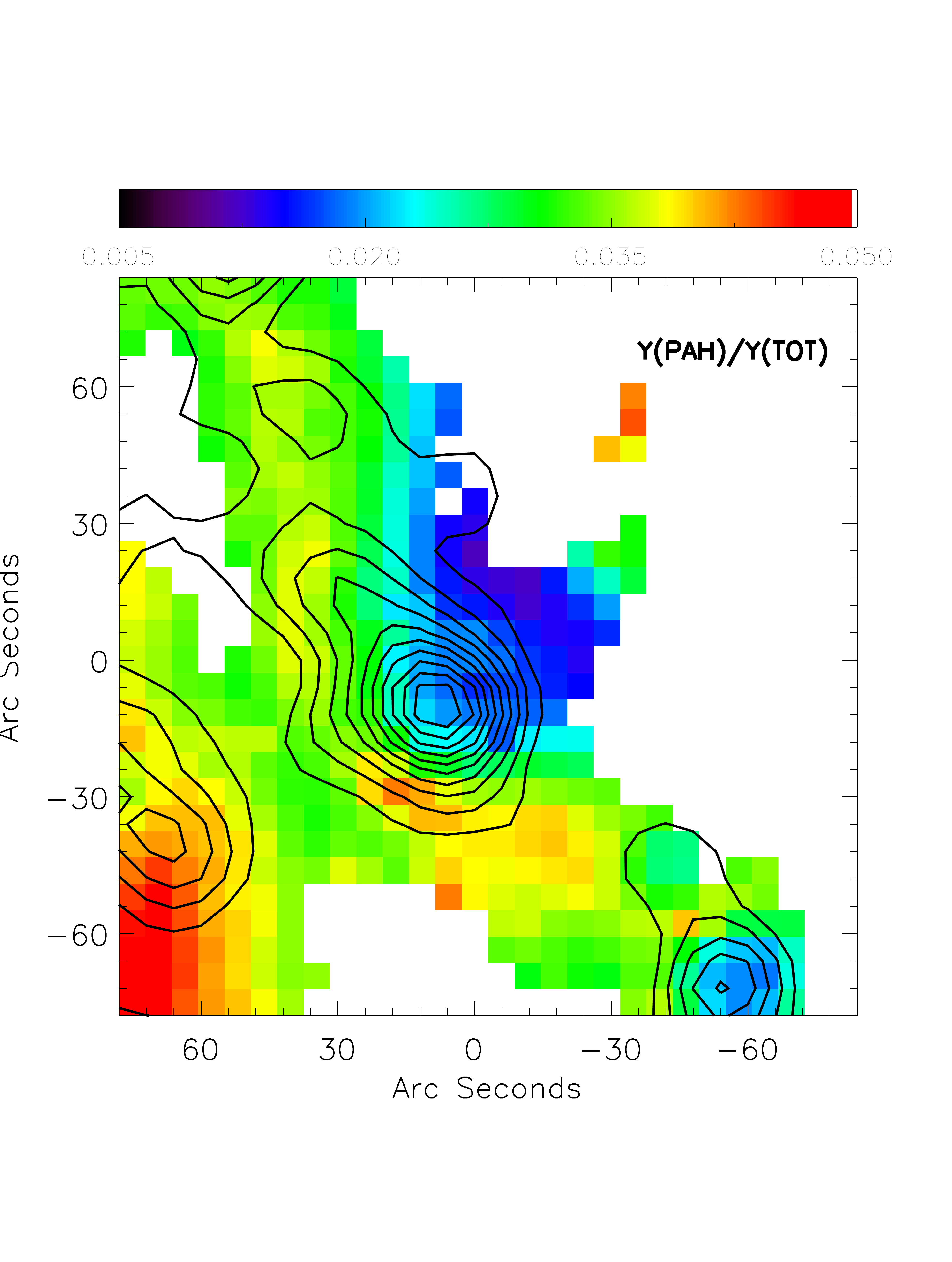}
   \caption{\ypah/\ytot\ distribution map for NGC~604 (left) and NGC~595 (right) as shown in Figs.~\ref{fig:yallrel_4Myr604} and~\ref{fig:yallrel_4Myr595} overlaid with CO(2--1) intensity contours from \citet{2014A&A...567A.118D}. The spatial resolution for both maps are 20\arcsec\ and the pixel size is 6\arcsec.}
              \label{YpahYbg_CO}%
\end{figure*}

In Fig.~\ref{YpahYbg_CO} we compare the \ypah/\ytot\ maps with CO(2--1) intensity distribution from \citet{2014A&A...567A.118D}. We can see that  the \ypah/\ytot\ distribution follows the CO(2--1) emission with the maximum in CO intensity corresponding to a minimum of \ypah/\ytot. The most intense CO(2--1) knot  in NGC~604 is located close to the source 1 of \citet{2012ApJ...761....3M}. These authors observed NGC~604 with the {\it Spitzer Infrared Spectrograph} (IRS) and selected 7 sources of interest based on their IRAC colours. In the left panel of Fig.~\ref{YpahYbg_CO} we show the location of sources 1 and 4, which are the ones exhibiting the highest {\it softness} parameter $\eta\prime$ in the MIR\footnote{$\eta\prime =\rm  \frac{I([Ne\,\textsc {ii}]12.8\,\mu m)/I([Ne\,\textsc {iii}]15.6\,\mu m)}{I([S\,\textsc {iii}]18.7\,\mu m)/I([S\,\textsc {iv}]10.5\,\mu m)}$, as it is defined in \citet{2009MNRAS.400.1721P}}, a tracer of the hardness of the ISRF. Besides, sources 1 and 4 also present the highest (6.2\,\mi+7.7\,\mi+8.6\,\mi)/11.3\,\mi\ ratio. The 11.3\,\mi\ PAH feature is associated to neutral PAHs, while 6.2\,\mi, 7.7\,\mi\ and 8.6\,\mi\ are associated to ionised PAHs. Therefore, at the location of sources 1 and 4, where a minimum in the \ypah/\ytot\ distribution is located, there is the highest ionised/neutral PAH ratio and the hardest spectrum of all the sources analysed by  \citet{2012ApJ...761....3M}. The anti-correlation \ypah/\ytot--hardness of ISRF provides evidence that the PAHs appear to be destroyed in places where the ISRF is strong and hard, as it has also been suggested in previous studies \citep{2006A&A...446..877M,2006ApJ...639..157W,2007ApJ...665..390L}.

The hardness of the ISRF at this particular location within NGC~604 might be produced by newly formed massive stars.  Source 1 in left panel of 
Fig.~\ref{YpahYbg_CO} is related to the maxima in CO(2--1) emission within the region and corresponds to the molecular cloud NMA-8, the most massive one in NGC~604 \citep{2010ApJ...724.1120M}. \citet{2010ApJ...724.1120M} also studied the HCN emission distribution, a tracer of dense regions in molecular clouds where star formation can occur. They found that there is a HCN cloud at the location of NMA-8 whose maximum is shifted from the maximum of NMA-8 towards the centre of the region. They concluded that NMA-8 is a dense molecular cloud that could be in a stage of ongoing massive star formation. At the maxima of NMA-8  \citet{Relano:2009p558} found a difference between the extinction derived from the Balmer emission and the extinction derived from the \ha-24\mi\ ratio. The difference \citep[also reported by][comparing Balmer and radio extinctions]{2004AJ....128.1196M} could be produced if the surface of the molecular cloud has a shell morphology that would produce a lower value of the Balmer extinction. \citet{Relano:2009p558} suggested that there could be embedded star formation at this location using Color Magnitude Diagram analysis of NGC~604.
  
In the bottom-left panels of Figs.~\ref{fig:yallrel_4Myr604} and \ref{fig:yallrel_4Myr595} we compare the CO(2--1) emission distribution with the distribution of \ybg/\ytot \ for NGC~604 and NGC~595, respectively. The maxima of CO intensity is correlated with local maxima of the \ybg/\ytot\ in both regions. This shows that CO is more closely related to the total dust mass, traced by the mass fraction of BGs, than to the PAH abundance in these regions. For NGC~595, the distribution of \ybg\ follows the same radial pattern as the dust-to-gas ratio obtained from the modelling of the region performed by \citet{2011MNRAS.412..675P}. These authors found that, in order to fit the radial trend of the 8\,\mi/24\,\mi\ ratio, the dust-to-gas ratio assumed in the model needed to increase from the centre to the outer parts of the regions. Whether a change in the dust-to-gas ratio is related to a change in the \ybg/\ytot\ due to grain coagulation or reformation requires a detailed study which is outside of the scope of this paper.   

\section{Summary and conclusions}\label{sec:sum}
We analyse the SEDs of a sample of 119 \hii\ regions classified in M\,33 as {\it filled}, {\it mixed} and {\it shell} and {\it clear shell} regions. The performed classification has been based on the \ha\ emission distribution of each object and represents different star-gas-dust spatial configurations within the regions. We update and complete the observed SED of  \citetalias{2013A&A...552A.140R} with new data from \Her\ at 70\,\mi, WISE at 3.4\,\mi, 4.6\,\mi, 12\,\mi, and 22\,\mi, and LABOCA at 870\,\mi. We study the physical properties of the dust under different environments fitting the observed SED of each region with the dust model of  \citet{2011A&A...525A.103C} and analysing the results in terms of the region morphology. We analyse the relative mass for each grain type for the best SED fit of each object in our sample with two different ISRF:  a {\it Mathis} ISRF corresponding to the default solar neighbourhood \citep{Mathis:1983p593} and a more realistic one of a young 4\,Myr cluster with a fixed stellar mass of $10^{4}$\msun. We also use the  approximation of a MBB to describe the emission of the dust and to fit the IR part of the observed SED. Hereafter, we summarise our main results.

\begin{itemize}
\item  All the mass ratios obtained with the modelling were independent of the shape of the ISRF except for PAHs. \ypah/\ytot\ is a factor of 2-10 higher assuming a {\it Mathis} ISRF, which shows the difficulty in constraining the relative abundance of PAHs with the models used here. BGs represent the highest fraction of dust mass in the regions ($\sim$90\%), while the relative mass of the PAHs is in general low (1-10\%).  

\item \yvsg/\ytot\ is higher for {\it filled} and {\it mixed} regions by a factor of $\sim$1.7 than for regions classified as {\it shells} and {\it clear shells}. The difference is the same when we apply the classic {\it Desert} dust model. We interpret this result within the dust evolution framework proposed by  \citet{1996ApJ...469..740J} where large grains can be disrupted by strong shocks. The regions classified as  {\it filled} and {\it mixed} have high-velocity $\sim$\,50-90\,\kms\ expansive structures that can produce an enhancement of the relative fraction of VSGs, while the {\it shells} and {\it clear shells} present a more quiescent environment where the BG can survive or reform. 

\item We derived an estimation of the dust temperature using the equilibrium temperature of the aSil grains provided by \DE. We found no difference in the temperature for both  {\it  filled} and {\it mixed}  regions and {\it shell} objects.  However, a MBB analysis gives temperatures slightly higher than 3$\sigma$ for the  {\it  filled} and {\it mixed}  regions (T=$17.9\pm0.6$\,K,)  than for the {\it shell} objects (T= $15.5\pm0.6$\,K ). This could be due to differences in the total dust opacity of the 
different types of \hii\ regions and/or due to a different spatial distribution of dust and stars within the type of the regions. 

\item We derive the gas-to-dust ratio for the \hii\ regions in our sample from the slope of the dust-gas surface density relation. We find two different slopes in the relation: regions with  $\rm \Sigma_{dust}<0.06M_{\sun}/pc^{2}$, corresponding to the \hi\ diffuse regime, present a gas-to-dust ratio of $390\pm68$ compatible with the expected value if we assume that the gas-to-dust ratio scales linearly with metallicity. Regions with $\rm \Sigma_{dust}\geq0.06M_{\sun}/pc^{2}$, corresponding to a $\rm H_{2}$ molecular phase, present a flatter  dust-gas surface density distribution, with a corresponding gas-to-dust ratio of $67\pm11$. These variations might imply that grain coagulation and/or gas-phase metals incorporation to the dust mass is occurring in the interior of the \hii\ regions in M\,33. 

\item In a spatially resolved analysis, we perform a pixel-by-pixel SED fitting of the whole surface of the two most luminous \hii\ regions in M33: NGC~604 and NGC~595. We derive maps of the relative mass fraction of VSG to BG (\yvsg/\ybg) for NGC~604 and NGC~595. We find local maxima of \yvsg/\ybg\ correlated to expansive ionised gas structures with velocities $\gtrsim $ 50\,\kms\ within the regions. This agrees with the results of  \citet{1996ApJ...469..740J}, who predict the fragmentation of BGs into VSGs due to shocks with velocities of at least 50\,\kms. 

\item \ypah/\ytot\ distribution for NGC~604 and NGC~595 follows the CO(2--1) emission, with the maximum in CO intensity corresponding to minimum of \ypah/\ytot. A deeper analysis using IRS data for NGC~604 shows that the minimum of \ypah/\ytot\ in this region corresponds to a high ionised/neutral PAH ratio and the location of a hard ISRF. This shows evidence that the PAHs are destroyed in places where the ISRF is strong and hard. 

\end{itemize} 

The dust emission in the IR has been shown to be a tracer of the star formation in a galaxy and correlations, albeit with dispersions, have been presented in the literature between the IR bands and other, more direct, star formation rate tracers, as the \ha\ emission coming from the ionised gas heated by recently formed stars. The emission in the 24\,\mi\ band has been particularly successful in tracing the star formation in \hii\ regions, as it corresponds to the warm dust mixed with the hot ionised gas at these locations \citep[see e.g.][]{Calzetti:2005p518}.  However, we show here that the relative fraction of VSGs, which are the main contributors to the 24\,\mi\ band emission, is not constant but changes (a factor of 1.7) depending on the morphology (and the evolutionary state) of the region. This effect introduces an uncertainty in the correlation between the  24\,\mi\  and \ha\ emission, which is expected to contribute to the dispersion in the observed empirical correlation. Understanding the physical properties of the dust, in particular the variation of the VSGs in different \hii\ regions, will help us to better understand the uncertainties in the empirical correlations between SFR tracers.

\begin{acknowledgements}
We would like to thank the referee for her/his useful comments which help to improve the first version of this paper. Part of this research has been supported by the PERG08-GA-2010-276813 from the EC.  This work was partially supported by the Junta de Andaluc\'ia Grant FQM108 and Spanish MEC Grants, AYA-2011-24728 and AYA-2014-53506-P. EPM thanks Spanish MINECO grant AYA-2013-47742-C4-1-P of the Spanish Plan for Astronomy and Astrophysics. This research made use of Montage, funded by the National Aeronautics and Space Administration's Earth Science Technology Office, Computational Technnologies Project, under Cooperative Agreement Number NCC5-626 between NASA and the California Institute of Technology. The code is maintained by the NASA/IPAC Infrared Science Archive. MR would like to thank L. Vestreate and D. Paradis for their help with \DE\ code. MA acknowledges funding by the German Research Foundation (DFG) in the 
framework of the priority programme 1573, "The Physics of the Interstellar 
Medium", through grant number AL 1467/2-1. This research made use of APLpy, an open-source plotting package for Python hosted at http://aplpy.github.com of TOPCAT \& STIL: Starlink Table/VOTable Processing Software \citep{2005ASPC..347...29T} of Matplotlib, a suite of open-source python modules that provide a framework for creating scientific plots.

 \end{acknowledgements}

\bibliographystyle{aa} 
\bibliography{mnras}  

\begin{thebibliography}{78}
\expandafter\ifx\csname natexlab\endcsname\relax\def\natexlab#1{#1}\fi

\bibitem[{{Aniano} {et~al.}(2012){Aniano}, {Draine}, {Calzetti}, {Dale},
  {Engelbracht}, {Gordon}, {Hunt}, {Kennicutt}, {Krause}, {Leroy}, {Rix},
  {Roussel}, {Sandstrom}, {Sauvage}, {Walter}, {Armus}, {Bolatto}, {Crocker},
  {Donovan Meyer}, {Galametz}, {Helou}, {Hinz}, {Johnson}, {Koda}, {Montiel},
  {Murphy}, {Skibba}, {Smith}, \& {Wolfire}}]{2012ApJ...756..138A}
{Aniano}, G., {Draine}, B.~T., {Calzetti}, D., {et~al.} 2012, \apj, 756, 138

\bibitem[{Bernard {et~al.}(2008)Bernard, Reach, Paradis,
  {et~al.}}]{Bernard:2008p587}
Bernard, J.-P., Reach, W.~T., Paradis, D., {et~al.} 2008, AJ, 136, 919

\bibitem[{{Bocchio} {et~al.}(2014){Bocchio}, {Jones}, \&
  {Slavin}}]{2014A&A...570A..32B}
{Bocchio}, M., {Jones}, A.~P., \& {Slavin}, J.~D. 2014, \aap, 570, A32

\bibitem[{{Boquien} {et~al.}(2015){Boquien}, {Calzetti}, {Aalto}, {Boselli},
  {Braine}, {Buat}, {Combes}, {Israel}, {Kramer}, {Lord}, {Rela{\~n}o},
  {Rosolowsky}, {Stacey}, {Tabatabaei}, {van der Tak}, {van der Werf},
  {Verley}, \& {Xilouris}}]{2015A&A...578A...8B}
{Boquien}, M., {Calzetti}, D., {Aalto}, S., {et~al.} 2015, \aap, 578, A8

\bibitem[{Boquien {et~al.}(2011)Boquien, Calzetti, Combes,
  {et~al.}}]{Boquien:2011p764}
Boquien, M., Calzetti, D., Combes, F., {et~al.} 2011, AJ, 142, 111

\bibitem[{{Bresolin}(2011)}]{2011ApJ...730..129B}
{Bresolin}, F. 2011, \apj, 730, 129

\bibitem[{Calzetti {et~al.}(2005)Calzetti, Kennicutt, Bianchi,
  {et~al.}}]{Calzetti:2005p518}
Calzetti, D., Kennicutt, R.~C., Bianchi, L., {et~al.} 2005, ApJ, 633, 871

\bibitem[{{Calzetti} {et~al.}(2007){Calzetti}, {Kennicutt}, {Engelbracht},
  {Leitherer}, {Draine}, {Kewley}, {Moustakas}, {Sosey}, {Dale}, {Gordon},
  {Helou}, {Hollenbach}, {Armus}, {Bendo}, {Bot}, {Buckalew}, {Jarrett}, {Li},
  {Meyer}, {Murphy}, {Prescott}, {Regan}, {Rieke}, {Roussel}, {Sheth}, {Smith},
  {Thornley}, \& {Walter}}]{2007ApJ...666..870C}
{Calzetti}, D., {Kennicutt}, R.~C., {Engelbracht}, C.~W., {et~al.} 2007, \apj,
  666, 870

\bibitem[{Cardelli {et~al.}(1989)Cardelli, Clayton, \&
  Mathis}]{Cardelli:1989p595}
Cardelli, J.~A., Clayton, G.~C., \& Mathis, J.~S. 1989, ApJ, 345, 245

\bibitem[{{Churchwell} {et~al.}(2009){Churchwell}, {Babler}, {Meade},
  {Whitney}, {Benjamin}, {Indebetouw}, {Cyganowski}, {Robitaille}, {Povich},
  {Watson}, \& {Bracker}}]{2009PASP..121..213C}
{Churchwell}, E., {Babler}, B.~L., {Meade}, M.~R., {et~al.} 2009, \pasp, 121,
  213

\bibitem[{{Compi{\`e}gne} {et~al.}(2008){Compi{\`e}gne}, {Abergel},
  {Verstraete}, \& {Habart}}]{2008A&A...491..797C}
{Compi{\`e}gne}, M., {Abergel}, A., {Verstraete}, L., \& {Habart}, E. 2008,
  \aap, 491, 797

\bibitem[{{Compi{\`e}gne} {et~al.}(2011){Compi{\`e}gne}, {Verstraete}, {Jones},
  {Bernard}, {Boulanger}, {Flagey}, {Le Bourlot}, {Paradis}, \&
  {Ysard}}]{2011A&A...525A.103C}
{Compi{\`e}gne}, M., {Verstraete}, L., {Jones}, A., {et~al.} 2011, \aap, 525,
  A103

\bibitem[{{Desert} {et~al.}(1990){Desert}, {Boulanger}, \&
  {Puget}}]{1990A&A...237..215D}
{Desert}, F.-X., {Boulanger}, F., \& {Puget}, J.~L. 1990, \aap, 237, 215

\bibitem[{Draine \& Li(2007)}]{Draine:2007p588}
Draine, B.~T. \& Li, A. 2007, ApJ, 657, 810

\bibitem[{Drissen {et~al.}(2008)Drissen, Crowther, {\'U}beda,
  {et~al.}}]{Drissen:2008p481}
Drissen, L., Crowther, P.~A., {\'U}beda, L., {et~al.} 2008, MNRAS, 389, 1033

\bibitem[{{Druard} {et~al.}(2014){Druard}, {Braine}, {Schuster}, {Schneider},
  {Gratier}, {Bontemps}, {Boquien}, {Combes}, {Corbelli}, {Henkel}, {Herpin},
  {Kramer}, {van der Tak}, \& {van der Werf}}]{2014A&A...567A.118D}
{Druard}, C., {Braine}, J., {Schuster}, K.~F., {et~al.} 2014, \aap, 567, A118

\bibitem[{Eldridge \& Rela{\~n}o(2011)}]{Eldridge:2011p756}
Eldridge, J.~J. \& Rela{\~n}o, M. 2011, MNRAS, 411, 235

\bibitem[{{Ferland} {et~al.}(1998){Ferland}, {Korista}, {Verner}, {Ferguson},
  {Kingdon}, \& {Verner}}]{1998PASP..110..761F}
{Ferland}, G.~J., {Korista}, K.~T., {Verner}, D.~A., {et~al.} 1998, \pasp, 110,
  761

\bibitem[{{Flagey} {et~al.}(2011){Flagey}, {Boulanger}, {Noriega-Crespo},
  {Paladini}, {Montmerle}, {Carey}, {Gagn{\'e}}, \&
  {Shenoy}}]{2011A&A...531A..51F}
{Flagey}, N., {Boulanger}, F., {Noriega-Crespo}, A., {et~al.} 2011, \aap, 531,
  A51

\bibitem[{{Flagey} {et~al.}(2006){Flagey}, {Boulanger}, {Verstraete}, {Miville
  Desch{\^e}nes}, {Noriega Crespo}, \& {Reach}}]{2006A&A...453..969F}
{Flagey}, N., {Boulanger}, F., {Verstraete}, L., {et~al.} 2006, \aap, 453, 969

\bibitem[{{Freedman} {et~al.}(1991){Freedman}, {Wilson}, \&
  {Madore}}]{1991ApJ...372..455F}
{Freedman}, W.~L., {Wilson}, C.~D., \& {Madore}, B.~F. 1991, \apj, 372, 455

\bibitem[{{Galametz} {et~al.}(2015){Galametz}, {Hony}, {Albrecht}, {Galliano},
  {Cormier}, {Lebouteiller}, {Lee}, {Madden}, {Bolatto}, {Bot}, {Hughes},
  {Israel}, {Meixner}, {Oliviera}, {Paradis}, {Pellegrini}, {Roman-Duval},
  {Rubio}, {Sewi{\l}o}, {Fukui}, {Kawamura}, \& {Onishi}}]{2015arXiv151107457G}
{Galametz}, M., {Hony}, S., {Albrecht}, M., {et~al.} 2015, ArXiv e-prints

\bibitem[{{Galliano} {et~al.}(2008){Galliano}, {Dwek}, \&
  {Chanial}}]{2008ApJ...672..214G}
{Galliano}, F., {Dwek}, E., \& {Chanial}, P. 2008, \apj, 672, 214

\bibitem[{{Gratier} {et~al.}(2010){Gratier}, {Braine}, {Rodriguez-Fernandez},
  {Schuster}, {Kramer}, {Xilouris}, {Tabatabaei}, {Henkel}, {Corbelli},
  {Israel}, {van der Werf}, {Calzetti}, {Garcia-Burillo}, {Sievers}, {Combes},
  {Wiklind}, {Brouillet}, {Herpin}, {Bontemps}, {Aalto}, {Koribalski}, {van der
  Tak}, {Wiedner}, {R{\"o}llig}, \& {Mookerjea}}]{2010A&A...522A...3G}
{Gratier}, P., {Braine}, J., {Rodriguez-Fernandez}, N.~J., {et~al.} 2010, \aap,
  522, A3

\bibitem[{{Hermelo} {et~al.}(2016){Hermelo}, {Rela{\~n}o}, {Lisenfeld},
  {Verley}, {Kramer}, {Ruiz-Lara}, {Boquien}, {Xilouris}, \&
  {Albrecht}}]{2016arXiv160302125H}
{Hermelo}, I., {Rela{\~n}o}, M., {Lisenfeld}, U., {et~al.} 2016, ArXiv e-prints

\bibitem[{Hunter {et~al.}(1996)Hunter, Baum, O'Neil,
  {et~al.}}]{Hunter:1996p605}
Hunter, D.~A., Baum, W.~A., O'Neil, E.~J., {et~al.} 1996, Astrophysical Journal
  v.456, 456, 174

\bibitem[{{Jarrett} {et~al.}(2011){Jarrett}, {Cohen}, {Masci}, {Wright},
  {Stern}, {Benford}, {Blain}, {Carey}, {Cutri}, {Eisenhardt}, {Lonsdale},
  {Mainzer}, {Marsh}, {Padgett}, {Petty}, {Ressler}, {Skrutskie}, {Stanford},
  {Surace}, {Tsai}, {Wheelock}, \& {Yan}}]{2011ApJ...735..112J}
{Jarrett}, T.~H., {Cohen}, M., {Masci}, F., {et~al.} 2011, \apj, 735, 112

\bibitem[{{Jarrett} {et~al.}(2013){Jarrett}, {Masci}, {Tsai}, {Petty},
  {Cluver}, {Assef}, {Benford}, {Blain}, {Bridge}, {Donoso}, {Eisenhardt},
  {Koribalski}, {Lake}, {Neill}, {Seibert}, {Sheth}, {Stanford}, \&
  {Wright}}]{2013AJ....145....6J}
{Jarrett}, T.~H., {Masci}, F., {Tsai}, C.~W., {et~al.} 2013, \aj, 145, 6

\bibitem[{{Jones}(2004)}]{2004ASPC..309..347J}
{Jones}, A.~P. 2004, in Astronomical Society of the Pacific Conference Series,
  Vol. 309, Astrophysics of Dust, ed. A.~N. {Witt}, G.~C. {Clayton}, \& B.~T.
  {Draine}, 347

\bibitem[{{Jones} {et~al.}(2013){Jones}, {Fanciullo}, {K{\"o}hler},
  {Verstraete}, {Guillet}, {Bocchio}, \& {Ysard}}]{2013A&A...558A..62J}
{Jones}, A.~P., {Fanciullo}, L., {K{\"o}hler}, M., {et~al.} 2013, \aap, 558,
  A62

\bibitem[{{Jones} {et~al.}(1996){Jones}, {Tielens}, \&
  {Hollenbach}}]{1996ApJ...469..740J}
{Jones}, A.~P., {Tielens}, A.~G.~G.~M., \& {Hollenbach}, D.~J. 1996, \apj, 469,
  740

\bibitem[{{K{\"o}hler} {et~al.}(2012){K{\"o}hler}, {Stepnik}, {Jones},
  {Guillet}, {Abergel}, {Ristorcelli}, \& {Bernard}}]{2012A&A...548A..61K}
{K{\"o}hler}, M., {Stepnik}, B., {Jones}, A.~P., {et~al.} 2012, \aap, 548, A61

\bibitem[{{K{\"o}hler} {et~al.}(2015){K{\"o}hler}, {Ysard}, \&
  {Jones}}]{2015A&A...579A..15K}
{K{\"o}hler}, M., {Ysard}, N., \& {Jones}, A.~P. 2015, \aap, 579, A15

\bibitem[{Kramer {et~al.}(2010)Kramer, Buchbender, Xilouris,
  {et~al.}}]{Kramer:2010p688}
Kramer, C., Buchbender, C., Xilouris, E.~M., {et~al.} 2010, A\&A, 518, L67

\bibitem[{{Kroupa}(2001)}]{2001MNRAS.322..231K}
{Kroupa}, P. 2001, \mnras, 322, 231

\bibitem[{{Krumholz} {et~al.}(2009){Krumholz}, {McKee}, \&
  {Tumlinson}}]{2009ApJ...693..216K}
{Krumholz}, M.~R., {McKee}, C.~F., \& {Tumlinson}, J. 2009, \apj, 693, 216

\bibitem[{Lagrois \& Joncas(2009)}]{Lagrois:2009p564}
Lagrois, D. \& Joncas, G. 2009, ApJ, 700, 1847

\bibitem[{{Lebouteiller} {et~al.}(2007){Lebouteiller}, {Brandl},
  {Bernard-Salas}, {Devost}, \& {Houck}}]{2007ApJ...665..390L}
{Lebouteiller}, V., {Brandl}, B., {Bernard-Salas}, J., {Devost}, D., \&
  {Houck}, J.~R. 2007, \apj, 665, 390

\bibitem[{Leitherer {et~al.}(1999)Leitherer, Schaerer, Goldader,
  {et~al.}}]{Leitherer:1999p491}
Leitherer, C., Schaerer, D., Goldader, J.~D., {et~al.} 1999, ApJS, 123, 3

\bibitem[{{Lu} {et~al.}(2003){Lu}, {Helou}, {Werner}, {Dinerstein}, {Dale},
  {Silbermann}, {Malhotra}, {Beichman}, \& {Jarrett}}]{2003ApJ...588..199L}
{Lu}, N., {Helou}, G., {Werner}, M.~W., {et~al.} 2003, \apj, 588, 199

\bibitem[{{Madden} {et~al.}(2006){Madden}, {Galliano}, {Jones}, \&
  {Sauvage}}]{2006A&A...446..877M}
{Madden}, S.~C., {Galliano}, F., {Jones}, A.~P., \& {Sauvage}, M. 2006, \aap,
  446, 877

\bibitem[{{Ma{\'{\i}}z-Apell{\'a}niz}
  {et~al.}(2004){Ma{\'{\i}}z-Apell{\'a}niz}, {P{\'e}rez}, \&
  {Mas-Hesse}}]{2004AJ....128.1196M}
{Ma{\'{\i}}z-Apell{\'a}niz}, J., {P{\'e}rez}, E., \& {Mas-Hesse}, J.~M. 2004,
  \aj, 128, 1196

\bibitem[{Malumuth {et~al.}(1996)Malumuth, Waller, \&
  Parker}]{Malumuth:1996p494}
Malumuth, E.~M., Waller, W.~H., \& Parker, J.~W. 1996, AJ, 111, 1128

\bibitem[{{Markwardt}(2009)}]{2009ASPC..411..251M}
{Markwardt}, C.~B. 2009, in Astronomical Society of the Pacific Conference
  Series, Vol. 411, Astronomical Data Analysis Software and Systems XVIII, ed.
  D.~A. {Bohlender}, D.~{Durand}, \& P.~{Dowler}, 251

\bibitem[{{Mart{\'{\i}}nez-Galarza} {et~al.}(2012){Mart{\'{\i}}nez-Galarza},
  {Hunter}, {Groves}, \& {Brandl}}]{2012ApJ...761....3M}
{Mart{\'{\i}}nez-Galarza}, J.~R., {Hunter}, D., {Groves}, B., \& {Brandl}, B.
  2012, \apj, 761, 3

\bibitem[{Mathis {et~al.}(1983)Mathis, Mezger, \& Panagia}]{Mathis:1983p593}
Mathis, J.~S., Mezger, P.~G., \& Panagia, N. 1983, Astronomy and Astrophysics
  (ISSN 0004-6361), 128, 212

\bibitem[{{Miura} {et~al.}(2010){Miura}, {Okumura}, {Tosaki}, {Tamura},
  {Kurono}, {Kuno}, {Nakanishi}, {Sakamoto}, {Hasegawa}, \&
  {Kawabe}}]{2010ApJ...724.1120M}
{Miura}, R., {Okumura}, S.~K., {Tosaki}, T., {et~al.} 2010, \apj, 724, 1120

\bibitem[{{Ott}(2010)}]{2010ASPC..434..139O}
{Ott}, S. 2010, in Astronomical Society of the Pacific Conference Series, Vol.
  434, Astronomical Data Analysis Software and Systems XIX, ed. Y.~{Mizumoto},
  K.-I. {Morita}, \& M.~{Ohishi}, 139

\bibitem[{{Ott}(2011)}]{2011ASPC..442..347O}
{Ott}, S. 2011, in Astronomical Society of the Pacific Conference Series, Vol.
  442, Astronomical Data Analysis Software and Systems XX, ed. I.~N. {Evans},
  A.~{Accomazzi}, D.~J. {Mink}, \& A.~H. {Rots}, 347

\bibitem[{{Paradis} {et~al.}(2009){Paradis}, {Bernard}, \&
  {M{\'e}ny}}]{2009A&A...506..745P}
{Paradis}, D., {Bernard}, J.-P., \& {M{\'e}ny}, C. 2009, \aap, 506, 745

\bibitem[{{Paradis} {et~al.}(2011){Paradis}, {Paladini}, {Noriega-Crespo},
  {Lagache}, {Kawamura}, {Onishi}, \& {Fukui}}]{2011ApJ...735....6P}
{Paradis}, D., {Paladini}, R., {Noriega-Crespo}, A., {et~al.} 2011, \apj, 735,
  6

\bibitem[{Pellerin(2006)}]{Pellerin:2006p499}
Pellerin, A. 2006, AJ, 131, 849

\bibitem[{{P{\'e}rez-Montero} {et~al.}(2011){P{\'e}rez-Montero}, {Rela{\~n}o},
  {V{\'{\i}}lchez}, \& {Monreal-Ibero}}]{2011MNRAS.412..675P}
{P{\'e}rez-Montero}, E., {Rela{\~n}o}, M., {V{\'{\i}}lchez}, J.~M., \&
  {Monreal-Ibero}, A. 2011, \mnras, 412, 675

\bibitem[{{P{\'e}rez-Montero} \& {V{\'{\i}}lchez}(2009)}]{2009MNRAS.400.1721P}
{P{\'e}rez-Montero}, E. \& {V{\'{\i}}lchez}, J.~M. 2009, \mnras, 400, 1721

\bibitem[{{Rela{\~n}o} {et~al.}(2010){Rela{\~n}o}, {Monreal-Ibero},
  {V{\'{\i}}lchez}, \& {Kennicutt}}]{2010MNRAS.402.1635R}
{Rela{\~n}o}, M., {Monreal-Ibero}, A., {V{\'{\i}}lchez}, J.~M., \& {Kennicutt},
  R.~C. 2010, \mnras, 402, 1635

\bibitem[{{Rela{\~n}o} {et~al.}(2013){Rela{\~n}o}, {Verley}, {P{\'e}rez},
  {Kramer}, {Calzetti}, {Xilouris}, {Boquien}, {Abreu-Vicente}, {Combes},
  {Israel}, {Tabatabaei}, {Braine}, {Buchbender}, {Gonz{\'a}lez}, {Gratier},
  {Lord}, {Mookerjea}, {Quintana-Lacaci}, \& {van der
  Werf}}]{2013A&A...552A.140R}
{Rela{\~n}o}, M., {Verley}, S., {P{\'e}rez}, I., {et~al.} 2013, \aap, 552, A140

\bibitem[{Rela{\~n}o \& Beckman(2005)}]{Relano:2005p644}
Rela{\~n}o, M. \& Beckman, J.~E. 2005, A{\&}A, 430, 911

\bibitem[{Rela{\~n}o \& Kennicutt(2009)}]{Relano:2009p558}
Rela{\~n}o, M. \& Kennicutt, R.~C. 2009, ApJ, 699, 1125

\bibitem[{{R{\'e}my-Ruyer} {et~al.}(2014){R{\'e}my-Ruyer}, {Madden},
  {Galliano}, {Galametz}, {Takeuchi}, {Asano}, {Zhukovska}, {Lebouteiller},
  {Cormier}, {Jones}, {Bocchio}, {Baes}, {Bendo}, {Boquien}, {Boselli},
  {DeLooze}, {Doublier-Pritchard}, {Hughes}, {Karczewski}, \&
  {Spinoglio}}]{2014A&A...563A..31R}
{R{\'e}my-Ruyer}, A., {Madden}, S.~C., {Galliano}, F., {et~al.} 2014, \aap,
  563, A31

\bibitem[{{Roman-Duval} {et~al.}(2014){Roman-Duval}, {Gordon}, {Meixner},
  {Bot}, {Bolatto}, {Hughes}, {Wong}, {Babler}, {Bernard}, {Clayton}, {Fukui},
  {Galametz}, {Galliano}, {Glover}, {Hony}, {Israel}, {Jameson},
  {Lebouteiller}, {Lee}, {Li}, {Madden}, {Misselt}, {Montiel}, {Okumura},
  {Onishi}, {Panuzzo}, {Reach}, {Remy-Ruyer}, {Robitaille}, {Rubio}, {Sauvage},
  {Seale}, {Sewilo}, {Staveley-Smith}, \& {Zhukovska}}]{2014ApJ...797...86R}
{Roman-Duval}, J., {Gordon}, K.~D., {Meixner}, M., {et~al.} 2014, \apj, 797, 86

\bibitem[{{Roussel}(2013)}]{2013PASP..125.1126R}
{Roussel}, H. 2013, \pasp, 125, 1126

\bibitem[{{Sellgren} {et~al.}(1983){Sellgren}, {Werner}, \&
  {Dinerstein}}]{1983ApJ...271L..13S}
{Sellgren}, K., {Werner}, M.~W., \& {Dinerstein}, H.~L. 1983, \apjl, 271, L13

\bibitem[{{Shetty} {et~al.}(2011){Shetty}, {Glover}, {Dullemond}, \&
  {Klessen}}]{2011MNRAS.412.1686S}
{Shetty}, R., {Glover}, S.~C., {Dullemond}, C.~P., \& {Klessen}, R.~S. 2011,
  \mnras, 412, 1686

\bibitem[{{Siringo} {et~al.}(2009){Siringo}, {Kreysa}, {Kov{\'a}cs},
  {Schuller}, {Wei{\ss}}, {Esch}, {Gem{\"u}nd}, {Jethava}, {Lundershausen},
  {Colin}, {G{\"u}sten}, {Menten}, {Beelen}, {Bertoldi}, {Beeman}, \&
  {Haller}}]{2009A&A...497..945S}
{Siringo}, G., {Kreysa}, E., {Kov{\'a}cs}, A., {et~al.} 2009, \aap, 497, 945

\bibitem[{{Staguhn} {et~al.}(2006){Staguhn}, {Benford}, {Allen}, {Moseley},
  {Sharp}, {Ames}, {Brunswig}, {Chuss}, {Dwek}, {Maher}, {Marx}, {Miller},
  {Navarro}, \& {Wollack}}]{2006SPIE.6275E..1DS}
{Staguhn}, J.~G., {Benford}, D.~J., {Allen}, C.~A., {et~al.} 2006, in Society
  of Photo-Optical Instrumentation Engineers (SPIE) Conference Series, Vol.
  6275, Society of Photo-Optical Instrumentation Engineers (SPIE) Conference
  Series, 1

\bibitem[{{Stephens} {et~al.}(2014){Stephens}, {Evans}, {Xue}, {Chu},
  {Gruendl}, \& {Segura-Cox}}]{2014ApJ...784..147S}
{Stephens}, I.~W., {Evans}, J.~M., {Xue}, R., {et~al.} 2014, \apj, 784, 147

\bibitem[{{Stepnik} {et~al.}(2003){Stepnik}, {Abergel}, {Bernard}, {Boulanger},
  {Cambr{\'e}sy}, {Giard}, {Jones}, {Lagache}, {Lamarre}, {Meny}, {Pajot}, {Le
  Peintre}, {Ristorcelli}, {Serra}, \& {Torre}}]{2003A&A...398..551S}
{Stepnik}, B., {Abergel}, A., {Bernard}, J.-P., {et~al.} 2003, \aap, 398, 551

\bibitem[{Tabatabaei {et~al.}(2007)Tabatabaei, Beck, Kr{\"u}gel,
  {et~al.}}]{Tabatabaei:2007p664}
Tabatabaei, F.~S., Beck, R., Kr{\"u}gel, E., {et~al.} 2007, A{\&}A, 475, 133

\bibitem[{{Taylor}(2005)}]{2005ASPC..347...29T}
{Taylor}, M.~B. 2005, in Astronomical Society of the Pacific Conference Series,
  Vol. 347, Astronomical Data Analysis Software and Systems XIV, ed.
  P.~{Shopbell}, M.~{Britton}, \& R.~{Ebert}, 29

\bibitem[{van~den Bergh(2000)}]{vandenBergh:2000p502}
van~den Bergh, S. 2000, The galaxies of the Local Group

\bibitem[{Verley {et~al.}(2010)Verley, Rela{\~n}o, Kramer,
  {et~al.}}]{Verley:2010p687}
Verley, S., Rela{\~n}o, M., Kramer, C., {et~al.} 2010, A\&A, 518, L68

\bibitem[{{Wei{\ss}} {et~al.}(2009){Wei{\ss}}, {Kov{\'a}cs}, {Coppin}, {Greve},
  {Walter}, {Smail}, {Dunlop}, {Knudsen}, {Alexander}, {Bertoldi}, {Brandt},
  {Chapman}, {Cox}, {Dannerbauer}, {De Breuck}, {Gawiser}, {Ivison}, {Lutz},
  {Menten}, {Koekemoer}, {Kreysa}, {Kurczynski}, {Rix}, {Schinnerer}, \& {van
  der Werf}}]{2009ApJ...707.1201W}
{Wei{\ss}}, A., {Kov{\'a}cs}, A., {Coppin}, K., {et~al.} 2009, \apj, 707, 1201

\bibitem[{{Whitmore} {et~al.}(2011){Whitmore}, {Chandar}, {Kim}, {Kaleida},
  {Mutchler}, {Stankiewicz}, {Calzetti}, {Saha}, {O'Connell}, {Balick}, {Bond},
  {Carollo}, {Disney}, {Dopita}, {Frogel}, {Hall}, {Holtzman}, {Kimble},
  {McCarthy}, {Paresce}, {Silk}, {Trauger}, {Walker}, {Windhorst}, \&
  {Young}}]{2011ApJ...729...78W}
{Whitmore}, B.~C., {Chandar}, R., {Kim}, H., {et~al.} 2011, \apj, 729, 78

\bibitem[{{Wright} {et~al.}(2010){Wright}, {Eisenhardt}, {Mainzer}, {Ressler},
  {Cutri}, {Jarrett}, {Kirkpatrick}, {Padgett}, {McMillan}, {Skrutskie},
  {Stanford}, {Cohen}, {Walker}, {Mather}, {Leisawitz}, {Gautier}, {McLean},
  {Benford}, {Lonsdale}, {Blain}, {Mendez}, {Irace}, {Duval}, {Liu}, {Royer},
  {Heinrichsen}, {Howard}, {Shannon}, {Kendall}, {Walsh}, {Larsen}, {Cardon},
  {Schick}, {Schwalm}, {Abid}, {Fabinsky}, {Naes}, \&
  {Tsai}}]{2010AJ....140.1868W}
{Wright}, E.~L., {Eisenhardt}, P.~R.~M., {Mainzer}, A.~K., {et~al.} 2010, \aj,
  140, 1868

\bibitem[{{Wu} {et~al.}(2006){Wu}, {Charmandaris}, {Hao}, {Brandl},
  {Bernard-Salas}, {Spoon}, \& {Houck}}]{2006ApJ...639..157W}
{Wu}, Y., {Charmandaris}, V., {Hao}, L., {et~al.} 2006, \apj, 639, 157

\bibitem[{{Xilouris} {et~al.}(2012){Xilouris}, {Tabatabaei}, {Boquien},
  {Kramer}, {Buchbender}, {Bertoldi}, {Anderl}, {Braine}, {Verley},
  {Rela{\~n}o}, {Quintana-Lacaci}, {Akras}, {Beck}, {Calzetti}, {Combes},
  {Gonzalez}, {Gratier}, {Henkel}, {Israel}, {Koribalski}, {Lord}, {Mookerjea},
  {Rosolowsky}, {Stacey}, {Tilanus}, {van der Tak}, \& {van der
  Werf}}]{2012A&A...543A..74X}
{Xilouris}, E.~M., {Tabatabaei}, F.~S., {Boquien}, M., {et~al.} 2012, \aap,
  543, A74

\bibitem[{{Yang} {et~al.}(1996){Yang}, {Chu}, {Skillman}, \&
  {Terlevich}}]{1996AJ....112..146Y}
{Yang}, H., {Chu}, Y.-H., {Skillman}, E.~D., \& {Terlevich}, R. 1996, \aj, 112,
  146

\bibitem[{{Ysard} {et~al.}(2013){Ysard}, {Abergel}, {Ristorcelli}, {Juvela},
  {Pagani}, {K{\"o}nyves}, {Spencer}, {White}, \&
  {Zavagno}}]{2013A&A...559A.133Y}
{Ysard}, N., {Abergel}, A., {Ristorcelli}, I., {et~al.} 2013, \aap, 559, A133

\end{thebibliography}
\clearpage
\begin{appendix}
\section{Fluxes}
\longtabL{1}{
\begin{landscape}
\begin{longtable}{cccccccccc}
\caption{\label{tab:phot1} Logarithmic of the fluxes from FUV to \Spi\ 8.0\,\mi\ band of our set of \hii\ regions in mJy\,Hz obtained with centres and apertures given in Table~B.1 in  \citetalias{2013A&A...552A.140R}. FUV, NUV and \ha\ fluxes have been corrected for Galactic extinction assuming a colour excess of $E(B-V)=0.07$\,mag \citep{vandenBergh:2000p502} and \citet{Cardelli:1989p595} extinction law with $R_V=3.1$. All the fluxes have been background subtracted. The errors include the uncertainties in the calibration and background subtraction. We did not take into account negative fluxes higher than the corresponding errors, in these cases the fluxes are represented by ****. }\\
\\
\hline\hline
ID  & F(FUV) & F(NUV) & F(\ha) & F(3.4\mi) & F(3.6\mi) & F(4.5\mi) &  F(4.6\mi) & F(5.8\mi)&F(8.0\mi) \\ 
     & $\log$(mJy\,Hz) &$\log$(mJy\,Hz) &$\log$(mJy\,Hz) &$\log$(mJy\,Hz) &$\log$(mJy\,Hz) &$\log$(mJy\,Hz) &$\log$(mJy\,Hz) &$\log$(mJy\,Hz) &$\log$(mJy\,Hz) \\
   \hline
\endfirsthead
\caption{continued.}\\
\\
\hline\hline
ID  & F(FUV) & F(NUV) & F(\ha) & F(3.4\mi) & F(3.6\mi) & F(4.5\mi) &  F(4.6\mi) & F(5.8\mi)&F(8.0\mi) \\ 
     & $\log$(mJy\,Hz) &$\log$(mJy\,Hz) &$\log$(mJy\,Hz) &$\log$(mJy\,Hz) &$\log$(mJy\,Hz) &$\log$(mJy\,Hz) &$\log$(mJy\,Hz) &$\log$(mJy\,Hz) &$\log$(mJy\,Hz) \\
   \hline
\endhead
\hline
\endfoot
\small
  1 & 16.18$\pm$0.00 & 16.15$\pm$0.00 & 14.22$\pm$0.07 & 14.77$\pm$0.02 & 14.79$\pm$0.03 & 14.54$\pm$0.05 & 14.50$\pm$0.02 & 15.01$\pm$0.05 & 15.24$\pm$0.07 \\
  2 & 15.66$\pm$0.01 & 15.58$\pm$0.01 & 13.67$\pm$0.07 & 14.58$\pm$0.02 & 14.62$\pm$0.03 & 14.52$\pm$0.05 & 14.39$\pm$0.02 & 14.80$\pm$0.05 & 14.92$\pm$0.07 \\
  3 & 16.13$\pm$0.00 & 16.01$\pm$0.00 & 14.11$\pm$0.07 & 14.66$\pm$0.02 & 14.70$\pm$0.03 & 14.47$\pm$0.05 & 14.32$\pm$0.02 & 14.68$\pm$0.05 & 14.87$\pm$0.07 \\
  4 & 15.92$\pm$0.01 & 15.83$\pm$0.01 & 14.11$\pm$0.07 & 14.57$\pm$0.02 & 14.48$\pm$0.04 & 14.25$\pm$0.06 & 14.23$\pm$0.03 & 14.57$\pm$0.06 & 14.73$\pm$0.07 \\
  5 & 15.92$\pm$0.00 & 15.82$\pm$0.00 & 13.70$\pm$0.07 & 14.35$\pm$0.02 & 14.35$\pm$0.03 & 14.15$\pm$0.05 & 14.03$\pm$0.02 & 14.73$\pm$0.05 & 14.96$\pm$0.07 \\
  6 & 16.22$\pm$0.00 & 16.09$\pm$0.00 & 14.23$\pm$0.07 & 14.38$\pm$0.05 & 14.46$\pm$0.05 & 14.44$\pm$0.05 & 14.23$\pm$0.04 & 14.63$\pm$0.07 & 15.10$\pm$0.07 \\
  7 & 16.19$\pm$0.00 & 16.14$\pm$0.00 & 14.65$\pm$0.07 & 14.44$\pm$0.03 & 14.53$\pm$0.03 & 14.48$\pm$0.05 & 14.37$\pm$0.02 & 14.82$\pm$0.05 & 15.20$\pm$0.07 \\
  8 & 15.04$\pm$0.02 & 14.92$\pm$0.02 & 13.47$\pm$0.07 & 13.87$\pm$0.05 & 14.01$\pm$0.04 & 14.06$\pm$0.05 & 13.91$\pm$0.02 & 14.24$\pm$0.05 & 14.45$\pm$0.07 \\
  9 & 15.47$\pm$0.02 & 15.45$\pm$0.02 & 13.65$\pm$0.07 & 14.30$\pm$0.04 & 14.11$\pm$0.08 & 13.68$\pm$0.14 & 13.56$\pm$0.15 & 14.38$\pm$0.07 & 14.35$\pm$0.09 \\
 10 & 16.06$\pm$0.01 & 15.95$\pm$0.01 & 13.77$\pm$0.07 & 14.47$\pm$0.03 & 14.54$\pm$0.03 & 14.27$\pm$0.05 & 14.15$\pm$0.03 & 14.55$\pm$0.05 & 14.65$\pm$0.07 \\
 11 & 15.31$\pm$0.01 & 15.24$\pm$0.02 & 12.85$\pm$0.07 & 13.49$\pm$0.11 & 13.52$\pm$0.11 & 13.55$\pm$0.07 & 13.52$\pm$0.05 & 13.79$\pm$0.08 & 13.98$\pm$0.07 \\
 12 & 15.55$\pm$0.00 & 15.43$\pm$0.01 & 13.66$\pm$0.07 & 14.22$\pm$0.03 & 14.16$\pm$0.05 & 14.00$\pm$0.06 & 13.94$\pm$0.04 & 14.43$\pm$0.05 & 14.59$\pm$0.07 \\
 13 & 15.15$\pm$0.03 & 15.01$\pm$0.03 & 13.33$\pm$0.07 &   ****$\pm$**** &   ****$\pm$**** &   ****$\pm$**** &   ****$\pm$**** & 13.91$\pm$0.15 & 14.13$\pm$0.13 \\
 14 & 15.69$\pm$0.01 & 15.66$\pm$0.01 & 13.53$\pm$0.07 & 13.95$\pm$0.10 & 14.15$\pm$0.06 & 13.85$\pm$0.08 & 13.55$\pm$0.11 & 14.65$\pm$0.05 & 14.90$\pm$0.07 \\
 15 & 15.92$\pm$0.01 & 15.84$\pm$0.00 & 13.74$\pm$0.07 & 14.28$\pm$0.03 & 14.35$\pm$0.03 & 14.04$\pm$0.05 & 13.89$\pm$0.04 & 14.50$\pm$0.05 & 14.74$\pm$0.07 \\
 16 & 15.47$\pm$0.00 & 15.37$\pm$0.01 & 13.63$\pm$0.07 & 15.59$\pm$0.01 & 14.12$\pm$0.05 & 15.24$\pm$0.04 & 15.18$\pm$0.01 & 15.03$\pm$0.04 & 14.83$\pm$0.07 \\
 17 & 16.32$\pm$0.01 & 16.26$\pm$0.00 & 14.40$\pm$0.07 & 14.78$\pm$0.03 & 14.80$\pm$0.04 & 14.60$\pm$0.05 & 14.48$\pm$0.03 & 14.99$\pm$0.06 & 15.28$\pm$0.07 \\
 18 & 16.10$\pm$0.00 & 15.98$\pm$0.00 & 14.11$\pm$0.07 & 14.09$\pm$0.04 & 14.27$\pm$0.03 & 14.13$\pm$0.05 & 14.06$\pm$0.03 & 13.54$\pm$0.26 & 14.55$\pm$0.07 \\
 19 & 15.97$\pm$0.00 & 15.85$\pm$0.00 & 13.77$\pm$0.07 & 14.46$\pm$0.03 & 14.49$\pm$0.04 & 14.25$\pm$0.05 & 14.14$\pm$0.03 & 14.64$\pm$0.05 & 14.89$\pm$0.07 \\
 20 & 16.14$\pm$0.00 & 16.05$\pm$0.00 & 14.21$\pm$0.07 & 15.19$\pm$0.01 & 15.19$\pm$0.02 & 14.95$\pm$0.04 & 14.89$\pm$0.01 & 15.31$\pm$0.04 & 15.54$\pm$0.07 \\
 21 & 16.08$\pm$0.00 & 15.99$\pm$0.00 & 14.29$\pm$0.07 & 15.06$\pm$0.01 & 15.12$\pm$0.02 & 14.93$\pm$0.04 & 14.87$\pm$0.01 & 15.52$\pm$0.04 & 15.74$\pm$0.07 \\
 22 & 16.30$\pm$0.00 & 16.21$\pm$0.00 & 14.36$\pm$0.07 & 14.81$\pm$0.02 & 14.86$\pm$0.02 & 14.64$\pm$0.05 & 14.54$\pm$0.02 & 15.13$\pm$0.04 & 15.38$\pm$0.07 \\
 23 & 15.95$\pm$0.00 & 15.87$\pm$0.00 & 14.38$\pm$0.07 & 14.52$\pm$0.03 & 14.66$\pm$0.03 & 14.55$\pm$0.05 & 14.47$\pm$0.02 & 15.23$\pm$0.04 & 15.52$\pm$0.07 \\
 24 & 15.92$\pm$0.00 & 15.84$\pm$0.01 & 14.06$\pm$0.07 & 14.64$\pm$0.03 & 14.67$\pm$0.03 & 14.41$\pm$0.05 & 14.34$\pm$0.03 & 15.00$\pm$0.05 & 15.27$\pm$0.07 \\
 25 & 16.32$\pm$0.00 & 16.28$\pm$0.00 & 14.45$\pm$0.07 & 14.95$\pm$0.02 & 14.99$\pm$0.03 & 14.79$\pm$0.05 & 14.71$\pm$0.02 & 15.24$\pm$0.04 & 15.48$\pm$0.07 \\
 26 & 15.75$\pm$0.01 & 15.68$\pm$0.01 & 13.74$\pm$0.07 &   ****$\pm$**** & 14.05$\pm$0.15 & 13.80$\pm$0.15 & 13.51$\pm$0.22 & 14.84$\pm$0.05 & 15.15$\pm$0.07 \\
 27 & 15.82$\pm$0.02 & 15.70$\pm$0.02 & 13.63$\pm$0.07 & 13.95$\pm$0.14 & 13.76$\pm$0.25 & 13.62$\pm$0.18 & 13.66$\pm$0.14 & 14.82$\pm$0.06 & 15.09$\pm$0.07 \\
 28 & 15.86$\pm$0.01 & 15.76$\pm$0.01 & 14.62$\pm$0.07 & 15.28$\pm$0.01 & 15.20$\pm$0.02 & 14.98$\pm$0.04 & 14.91$\pm$0.02 & 15.11$\pm$0.05 & 15.27$\pm$0.07 \\
 29 & 16.52$\pm$0.00 & 16.40$\pm$0.00 & 14.57$\pm$0.07 & 15.11$\pm$0.02 & 15.16$\pm$0.03 & 15.15$\pm$0.04 & 15.10$\pm$0.01 & 15.59$\pm$0.04 & 15.87$\pm$0.07 \\
 30 & 16.00$\pm$0.00 & 15.88$\pm$0.00 & 14.44$\pm$0.07 & 14.81$\pm$0.02 & 14.83$\pm$0.03 & 14.60$\pm$0.05 & 14.58$\pm$0.02 & 15.07$\pm$0.05 & 15.36$\pm$0.07 \\
 31 & 15.36$\pm$0.02 & 15.31$\pm$0.02 & 13.53$\pm$0.07 & 14.28$\pm$0.06 & 14.34$\pm$0.06 & 14.07$\pm$0.07 & 13.75$\pm$0.11 & 14.88$\pm$0.05 & 15.06$\pm$0.07 \\
 32 & 15.60$\pm$0.01 & 15.49$\pm$0.02 & 13.63$\pm$0.07 & 14.85$\pm$0.02 & 14.89$\pm$0.03 & 14.71$\pm$0.05 & 14.63$\pm$0.02 & 15.22$\pm$0.04 & 15.55$\pm$0.07 \\
 33 & 15.65$\pm$0.01 & 15.57$\pm$0.01 & 13.21$\pm$0.07 & 14.28$\pm$0.07 & 14.08$\pm$0.12 & 13.95$\pm$0.12 & 13.69$\pm$0.14 & 14.22$\pm$0.15 & 14.27$\pm$0.15 \\
 34 & 15.28$\pm$0.02 & 15.20$\pm$0.02 & 13.29$\pm$0.07 & 14.49$\pm$0.04 & 14.48$\pm$0.04 & 14.10$\pm$0.07 & 13.97$\pm$0.07 & 14.70$\pm$0.05 & 14.94$\pm$0.07 \\
 35 & 15.47$\pm$0.01 & 15.32$\pm$0.01 & 13.85$\pm$0.07 & 14.88$\pm$0.02 & 15.00$\pm$0.02 & 14.95$\pm$0.04 & 14.84$\pm$0.01 & 15.24$\pm$0.04 & 15.43$\pm$0.07 \\
 36 & 15.35$\pm$0.01 & 15.25$\pm$0.01 & 13.14$\pm$0.07 & 14.11$\pm$0.03 & 14.16$\pm$0.03 & 13.76$\pm$0.06 & 13.74$\pm$0.04 & 13.86$\pm$0.09 & 14.21$\pm$0.08 \\
 37 & 15.74$\pm$0.03 & 15.61$\pm$0.04 & 14.02$\pm$0.07 & 15.27$\pm$0.03 & 15.29$\pm$0.03 & 14.97$\pm$0.05 & 14.90$\pm$0.04 & 15.58$\pm$0.05 & 15.87$\pm$0.07 \\
 38 & 15.71$\pm$0.00 & 15.60$\pm$0.01 & 13.48$\pm$0.07 & 14.28$\pm$0.02 & 14.25$\pm$0.04 & 14.02$\pm$0.06 & 13.81$\pm$0.05 & 13.80$\pm$0.18 & 14.43$\pm$0.07 \\
 39 & 14.60$\pm$0.06 & 14.76$\pm$0.05 & 12.90$\pm$0.07 & 14.16$\pm$0.15 & 14.17$\pm$0.15 & 13.89$\pm$0.15 & 13.79$\pm$0.15 & 13.85$\pm$0.22 & 14.52$\pm$0.08 \\
 40 &   ****$\pm$**** &   ****$\pm$**** & 13.23$\pm$0.07 &   ****$\pm$**** &   ****$\pm$**** &   ****$\pm$**** &   ****$\pm$**** & 14.17$\pm$0.15 & 14.21$\pm$0.10 \\
 41 & 15.83$\pm$0.02 & 15.79$\pm$0.02 & 14.37$\pm$0.07 & 15.11$\pm$0.02 & 15.17$\pm$0.03 & 14.99$\pm$0.05 & 14.95$\pm$0.02 & 15.73$\pm$0.04 & 16.03$\pm$0.07 \\
 42 & 15.96$\pm$0.01 & 15.83$\pm$0.01 & 14.40$\pm$0.07 & 14.73$\pm$0.07 & 14.85$\pm$0.05 & 14.77$\pm$0.05 & 14.74$\pm$0.03 & 15.43$\pm$0.05 & 15.76$\pm$0.07 \\
 43 & 15.24$\pm$0.01 & 15.09$\pm$0.01 & 13.11$\pm$0.07 & 14.01$\pm$0.03 & 14.12$\pm$0.03 & 13.92$\pm$0.06 & 13.83$\pm$0.04 & 13.74$\pm$0.15 & 14.12$\pm$0.09 \\
 44 & 16.42$\pm$0.00 & 16.33$\pm$0.00 & 15.09$\pm$0.07 & 15.55$\pm$0.01 & 15.60$\pm$0.02 & 15.45$\pm$0.04 & 15.39$\pm$0.01 & 15.92$\pm$0.04 & 16.21$\pm$0.07 \\
 45 &   ****$\pm$**** &   ****$\pm$**** & 13.60$\pm$0.07 & 14.62$\pm$0.05 & 14.69$\pm$0.03 & 14.39$\pm$0.05 & 14.32$\pm$0.03 & 14.85$\pm$0.05 & 15.12$\pm$0.07 \\
 46 & 15.47$\pm$0.02 & 15.32$\pm$0.02 & 13.32$\pm$0.07 & 14.34$\pm$0.09 & 14.47$\pm$0.07 & 14.37$\pm$0.06 & 14.24$\pm$0.05 & 14.91$\pm$0.05 & 15.20$\pm$0.07 \\
 47 & 14.63$\pm$0.14 & 14.53$\pm$0.14 & 13.80$\pm$0.07 & 14.76$\pm$0.03 & 14.73$\pm$0.04 & 14.63$\pm$0.05 & 14.60$\pm$0.02 & 15.13$\pm$0.05 & 15.41$\pm$0.07 \\
 48 & 15.83$\pm$0.01 & 15.74$\pm$0.01 & 13.99$\pm$0.07 &   ****$\pm$**** & 13.62$\pm$0.19 & 13.60$\pm$0.18 &   ****$\pm$**** & 14.01$\pm$0.20 & 14.73$\pm$0.07 \\
 49 & 14.86$\pm$0.17 & 14.53$\pm$0.32 & 13.70$\pm$0.07 & 15.18$\pm$0.02 & 15.20$\pm$0.03 & 14.81$\pm$0.05 & 14.77$\pm$0.03 & 15.37$\pm$0.05 & 15.56$\pm$0.07 \\
 50 & 15.82$\pm$0.00 & 15.76$\pm$0.01 & 14.14$\pm$0.07 & 14.50$\pm$0.05 & 14.51$\pm$0.05 & 14.28$\pm$0.06 & 14.16$\pm$0.04 & 14.67$\pm$0.05 & 15.00$\pm$0.07 \\
 51 & 16.16$\pm$0.00 & 16.05$\pm$0.00 & 14.15$\pm$0.07 & 14.64$\pm$0.02 & 14.71$\pm$0.03 & 14.53$\pm$0.05 & 14.40$\pm$0.02 & 15.00$\pm$0.05 & 15.26$\pm$0.07 \\
 52 & 15.56$\pm$0.05 & 15.41$\pm$0.07 & 13.68$\pm$0.07 & 14.55$\pm$0.04 & 14.66$\pm$0.04 & 14.43$\pm$0.05 & 14.35$\pm$0.03 & 15.04$\pm$0.05 & 15.33$\pm$0.07 \\
 53 & 16.17$\pm$0.01 & 16.04$\pm$0.01 & 14.06$\pm$0.07 & 14.93$\pm$0.05 & 14.99$\pm$0.05 & 14.70$\pm$0.06 & 14.58$\pm$0.05 & 15.18$\pm$0.06 & 15.46$\pm$0.08 \\
 54 & 16.22$\pm$0.00 & 16.22$\pm$0.00 & 13.88$\pm$0.07 & 14.68$\pm$0.03 & 14.63$\pm$0.03 & 14.05$\pm$0.08 & 14.23$\pm$0.05 & 14.03$\pm$0.15 & 14.80$\pm$0.08 \\
 55 & 16.15$\pm$0.01 & 16.09$\pm$0.01 & 14.46$\pm$0.07 & 15.28$\pm$0.02 & 15.28$\pm$0.03 & 14.97$\pm$0.05 & 14.90$\pm$0.03 & 15.46$\pm$0.05 & 15.77$\pm$0.07 \\
 56 & 15.87$\pm$0.00 & 15.77$\pm$0.00 & 14.03$\pm$0.07 & 14.06$\pm$0.06 & 14.04$\pm$0.08 & 13.92$\pm$0.10 & 13.67$\pm$0.12 & 14.14$\pm$0.15 & 14.50$\pm$0.07 \\
 57 & 15.90$\pm$0.00 & 15.80$\pm$0.01 & 13.77$\pm$0.07 & 14.87$\pm$0.02 & 14.85$\pm$0.03 & 14.61$\pm$0.04 & 14.54$\pm$0.02 & 14.80$\pm$0.04 & 14.98$\pm$0.07 \\
 58 & 16.35$\pm$0.01 & 16.22$\pm$0.02 & 14.41$\pm$0.07 & 15.43$\pm$0.02 & 15.36$\pm$0.03 & 15.07$\pm$0.05 & 15.10$\pm$0.02 & 15.83$\pm$0.04 & 16.13$\pm$0.07 \\
 59 & 15.22$\pm$0.02 & 15.10$\pm$0.03 & 13.44$\pm$0.07 & 14.19$\pm$0.08 & 14.37$\pm$0.06 & 14.16$\pm$0.06 & 14.24$\pm$0.03 & 14.50$\pm$0.06 & 14.84$\pm$0.07 \\
 60 & 15.52$\pm$0.00 & 15.39$\pm$0.01 & 13.64$\pm$0.07 & 13.99$\pm$0.07 & 14.14$\pm$0.06 & 13.96$\pm$0.07 & 13.71$\pm$0.07 & 14.30$\pm$0.06 & 14.70$\pm$0.07 \\
 61 & 16.04$\pm$0.00 & 15.95$\pm$0.00 & 13.73$\pm$0.07 & 13.78$\pm$0.18 & 14.09$\pm$0.10 & 14.12$\pm$0.06 & 13.86$\pm$0.06 & 14.35$\pm$0.06 & 14.71$\pm$0.07 \\
 62 & 15.45$\pm$0.01 & 15.31$\pm$0.01 & 13.69$\pm$0.07 & 14.36$\pm$0.02 & 14.26$\pm$0.04 & 13.99$\pm$0.06 & 14.08$\pm$0.02 & 14.39$\pm$0.06 & 14.67$\pm$0.07 \\
 63 & 16.65$\pm$0.01 & 16.60$\pm$0.01 & 14.51$\pm$0.07 & 16.06$\pm$0.01 & 16.05$\pm$0.02 & 15.77$\pm$0.04 & 15.71$\pm$0.01 & 15.96$\pm$0.05 & 16.16$\pm$0.07 \\
 64 & 15.87$\pm$0.02 & 15.89$\pm$0.02 & 14.21$\pm$0.07 & 14.40$\pm$0.18 & 14.55$\pm$0.13 & 14.30$\pm$0.12 & 14.31$\pm$0.11 & 15.32$\pm$0.05 & 15.62$\pm$0.07 \\
 65 & 16.10$\pm$0.01 & 16.05$\pm$0.00 & 13.88$\pm$0.07 & 13.93$\pm$0.28 & 14.33$\pm$0.11 & 13.97$\pm$0.15 & 13.84$\pm$0.16 & 14.42$\pm$0.15 & 14.54$\pm$0.24 \\
 66 & 15.14$\pm$0.07 & 14.84$\pm$0.15 & 13.63$\pm$0.07 & 14.55$\pm$0.09 & 14.48$\pm$0.10 & 14.08$\pm$0.14 & 14.08$\pm$0.12 & 14.96$\pm$0.06 & 15.31$\pm$0.08 \\
 67 & 15.96$\pm$0.01 & 15.91$\pm$0.01 & 14.48$\pm$0.07 & 14.90$\pm$0.03 & 14.94$\pm$0.04 & 14.76$\pm$0.05 & 14.71$\pm$0.02 & 15.33$\pm$0.05 & 15.62$\pm$0.07 \\
 68 & 15.43$\pm$0.00 & 15.32$\pm$0.01 & 13.37$\pm$0.07 & 13.63$\pm$0.06 & 13.68$\pm$0.06 & 13.52$\pm$0.07 & 13.36$\pm$0.05 & 14.04$\pm$0.06 & 14.41$\pm$0.07 \\
 69 & 15.77$\pm$0.01 & 15.65$\pm$0.01 & 13.70$\pm$0.07 & 15.03$\pm$0.03 & 14.96$\pm$0.04 & 14.58$\pm$0.06 & 14.68$\pm$0.03 & 14.94$\pm$0.05 & 15.06$\pm$0.07 \\
 70 & 16.10$\pm$0.02 & 16.06$\pm$0.02 & 13.84$\pm$0.07 & 15.10$\pm$0.02 & 15.10$\pm$0.03 & 14.83$\pm$0.05 & 14.80$\pm$0.02 & 15.19$\pm$0.05 & 15.50$\pm$0.07 \\
 71 & 14.65$\pm$0.27 & 14.89$\pm$0.14 & 14.04$\pm$0.07 & 14.95$\pm$0.04 & 14.95$\pm$0.05 & 14.73$\pm$0.06 & 14.74$\pm$0.03 & 15.37$\pm$0.05 & 15.69$\pm$0.07 \\
 72 & 15.60$\pm$0.02 & 15.53$\pm$0.02 & 13.16$\pm$0.07 & 14.43$\pm$0.04 & 14.52$\pm$0.04 & 14.27$\pm$0.05 & 14.25$\pm$0.03 & 15.01$\pm$0.05 & 15.32$\pm$0.07 \\
 73 & 16.12$\pm$0.01 & 16.07$\pm$0.01 & 14.27$\pm$0.07 & 14.81$\pm$0.08 & 14.89$\pm$0.07 & 14.66$\pm$0.07 & 14.58$\pm$0.06 & 15.44$\pm$0.05 & 15.75$\pm$0.07 \\
 74 & 15.89$\pm$0.04 & 15.57$\pm$0.09 & 14.37$\pm$0.07 & 15.23$\pm$0.04 & 15.34$\pm$0.04 & 15.09$\pm$0.05 & 15.03$\pm$0.03 & 15.72$\pm$0.05 & 16.03$\pm$0.07 \\
 75 & 15.44$\pm$0.02 & 15.27$\pm$0.02 & 13.80$\pm$0.07 & 13.76$\pm$0.11 & 13.93$\pm$0.09 & 13.77$\pm$0.08 & 13.61$\pm$0.08 & 14.69$\pm$0.05 & 14.98$\pm$0.07 \\
 76 & 15.24$\pm$0.02 & 14.96$\pm$0.04 & 13.59$\pm$0.07 & 14.12$\pm$0.09 & 14.13$\pm$0.09 & 13.84$\pm$0.12 & 13.92$\pm$0.07 & 14.89$\pm$0.05 & 15.20$\pm$0.07 \\
 77 & 15.11$\pm$0.05 & 14.92$\pm$0.07 & 13.63$\pm$0.07 & 14.03$\pm$0.14 & 14.05$\pm$0.15 & 13.79$\pm$0.15 & 13.45$\pm$0.30 & 14.01$\pm$0.33 & 14.59$\pm$0.18 \\
 78 & 16.04$\pm$0.01 & 15.95$\pm$0.01 & 14.29$\pm$0.07 & 15.19$\pm$0.03 & 15.19$\pm$0.03 & 14.81$\pm$0.05 & 14.73$\pm$0.03 & 15.43$\pm$0.05 & 15.80$\pm$0.07 \\
 79 & 16.14$\pm$0.01 & 16.07$\pm$0.01 & 14.47$\pm$0.07 & 15.02$\pm$0.02 & 15.08$\pm$0.03 & 14.90$\pm$0.05 & 14.88$\pm$0.02 & 15.51$\pm$0.04 & 15.79$\pm$0.07 \\
 80 & 15.13$\pm$0.07 & 14.84$\pm$0.13 & 13.39$\pm$0.07 & 14.01$\pm$0.14 & 13.60$\pm$0.36 & 13.39$\pm$0.32 & 13.70$\pm$0.15 & 14.68$\pm$0.06 & 15.10$\pm$0.07 \\
 81 & 15.54$\pm$0.01 & 15.40$\pm$0.02 & 13.58$\pm$0.07 & 13.99$\pm$0.03 & 14.10$\pm$0.04 & 14.01$\pm$0.06 & 13.94$\pm$0.03 & 14.34$\pm$0.07 & 14.51$\pm$0.07 \\
 82 & 14.88$\pm$0.04 & 14.43$\pm$0.14 & 13.27$\pm$0.07 & 14.02$\pm$0.14 & 13.97$\pm$0.16 & 14.06$\pm$0.07 & 13.88$\pm$0.09 & 14.64$\pm$0.05 & 14.99$\pm$0.07 \\
 83 & 15.56$\pm$0.01 & 15.41$\pm$0.01 & 13.44$\pm$0.07 & 13.50$\pm$0.28 & 13.92$\pm$0.12 & 13.58$\pm$0.20 & 13.66$\pm$0.12 & 14.29$\pm$0.12 & 14.42$\pm$0.08 \\
 84 & 15.37$\pm$0.01 & 15.20$\pm$0.02 & 13.29$\pm$0.07 & 13.65$\pm$0.07 & 13.74$\pm$0.08 & 13.58$\pm$0.09 & 13.31$\pm$0.11 & 14.34$\pm$0.06 & 14.59$\pm$0.07 \\
 85 & 15.35$\pm$0.07 & 14.72$\pm$0.30 & 14.01$\pm$0.07 &   ****$\pm$**** & 14.45$\pm$0.15 & 14.06$\pm$0.18 & 13.82$\pm$0.29 & 15.24$\pm$0.05 & 15.66$\pm$0.07 \\
 86 & 16.03$\pm$0.00 & 15.88$\pm$0.01 & 13.78$\pm$0.07 & 14.36$\pm$0.03 & 14.41$\pm$0.04 & 14.25$\pm$0.05 & 14.13$\pm$0.03 & 14.61$\pm$0.06 & 14.90$\pm$0.07 \\
 87 & 16.09$\pm$0.00 & 15.94$\pm$0.00 & 13.75$\pm$0.07 & 14.38$\pm$0.02 & 14.55$\pm$0.03 & 14.20$\pm$0.06 & 14.09$\pm$0.04 &   ****$\pm$**** & 14.08$\pm$0.09 \\
 88 & 15.43$\pm$0.00 & 15.31$\pm$0.01 & 12.99$\pm$0.07 & 13.53$\pm$0.10 & 13.50$\pm$0.12 & 13.29$\pm$0.15 & 13.38$\pm$0.08 & 13.69$\pm$0.15 & 14.26$\pm$0.07 \\
 89 & 15.33$\pm$0.01 & 15.19$\pm$0.01 & 13.11$\pm$0.07 & 13.54$\pm$0.06 & 13.27$\pm$0.20 & 13.30$\pm$0.16 & 13.33$\pm$0.09 & 14.28$\pm$0.07 & 14.10$\pm$0.07 \\
 90 & 16.56$\pm$0.00 & 16.46$\pm$0.00 & 14.63$\pm$0.07 & 14.98$\pm$0.07 & 15.00$\pm$0.07 & 14.86$\pm$0.06 & 14.82$\pm$0.05 & 15.38$\pm$0.06 & 15.63$\pm$0.07 \\
 91 & 16.01$\pm$0.00 & 15.85$\pm$0.00 & 14.41$\pm$0.07 & 14.91$\pm$0.02 & 15.00$\pm$0.02 & 14.82$\pm$0.04 & 14.75$\pm$0.01 & 15.37$\pm$0.04 & 15.63$\pm$0.07 \\
 92 & 15.50$\pm$0.03 & 15.45$\pm$0.03 & 13.45$\pm$0.07 & 14.09$\pm$0.15 & 14.08$\pm$0.15 &   ****$\pm$**** & 13.39$\pm$0.31 & 14.50$\pm$0.07 & 14.93$\pm$0.07 \\
 93 & 15.90$\pm$0.01 & 15.78$\pm$0.02 & 13.25$\pm$0.07 & 14.40$\pm$0.10 & 14.09$\pm$0.21 & 13.99$\pm$0.15 & 13.73$\pm$0.21 & 14.43$\pm$0.09 & 14.64$\pm$0.09 \\
 94 & 14.30$\pm$0.04 & 14.52$\pm$0.02 & 13.01$\pm$0.07 & 14.42$\pm$0.03 & 14.33$\pm$0.04 & 14.18$\pm$0.05 & 14.14$\pm$0.02 & 14.43$\pm$0.05 & 14.60$\pm$0.07 \\
 95 & 15.63$\pm$0.01 & 15.49$\pm$0.01 & 13.44$\pm$0.07 & 13.84$\pm$0.09 & 13.98$\pm$0.07 & 13.91$\pm$0.07 & 13.67$\pm$0.07 & 14.18$\pm$0.07 & 14.48$\pm$0.07 \\
 96 & 15.59$\pm$0.01 & 15.45$\pm$0.01 & 13.54$\pm$0.07 & 13.81$\pm$0.15 & 13.70$\pm$0.22 & 13.58$\pm$0.16 & 13.64$\pm$0.10 & 14.03$\pm$0.14 & 14.22$\pm$0.11 \\
 97 & 15.91$\pm$0.00 & 15.81$\pm$0.00 & 14.23$\pm$0.07 & 14.95$\pm$0.02 & 14.81$\pm$0.03 & 14.69$\pm$0.05 & 14.67$\pm$0.02 & 15.22$\pm$0.04 & 15.46$\pm$0.07 \\
 98 & 17.22$\pm$0.00 & 17.13$\pm$0.00 & 15.59$\pm$0.07 & 15.89$\pm$0.01 & 15.97$\pm$0.02 & 15.82$\pm$0.04 & 15.75$\pm$0.01 & 16.46$\pm$0.04 & 16.77$\pm$0.07 \\
 99 & 15.97$\pm$0.00 & 15.83$\pm$0.00 & 14.33$\pm$0.07 & 14.37$\pm$0.03 & 14.51$\pm$0.04 & 14.40$\pm$0.06 & 14.17$\pm$0.04 & 14.53$\pm$0.08 & 14.84$\pm$0.07 \\
100 & 15.35$\pm$0.01 & 15.20$\pm$0.01 & 13.29$\pm$0.07 & 13.50$\pm$0.18 & 13.84$\pm$0.09 & 13.58$\pm$0.09 & 13.63$\pm$0.07 & 14.79$\pm$0.04 & 15.10$\pm$0.07 \\
101 & 15.98$\pm$0.02 & 15.83$\pm$0.02 & 13.57$\pm$0.07 & 13.71$\pm$0.42 & 13.72$\pm$0.42 &   ****$\pm$**** & 13.82$\pm$0.15 & 14.75$\pm$0.06 & 14.88$\pm$0.10 \\
102 & 15.02$\pm$0.02 & 14.88$\pm$0.02 & 13.41$\pm$0.07 & 14.39$\pm$0.02 & 14.43$\pm$0.03 & 14.22$\pm$0.05 & 14.13$\pm$0.02 & 14.64$\pm$0.05 & 14.99$\pm$0.07 \\
103 & 15.64$\pm$0.01 & 15.59$\pm$0.01 & 13.70$\pm$0.07 & 14.58$\pm$0.06 & 14.35$\pm$0.11 & 14.27$\pm$0.09 & 14.38$\pm$0.04 & 14.74$\pm$0.06 & 15.10$\pm$0.07 \\
104 & 15.90$\pm$0.01 & 15.77$\pm$0.01 & 14.01$\pm$0.07 & 14.35$\pm$0.04 & 14.44$\pm$0.04 & 14.21$\pm$0.05 & 14.12$\pm$0.03 & 14.95$\pm$0.05 & 15.30$\pm$0.07 \\
105 & 15.29$\pm$0.01 & 15.18$\pm$0.01 & 13.33$\pm$0.07 & 13.41$\pm$0.15 & 13.45$\pm$0.16 & 13.43$\pm$0.15 & 13.32$\pm$0.15 & 13.82$\pm$0.15 & 14.28$\pm$0.07 \\
106 & 15.89$\pm$0.01 & 15.78$\pm$0.01 & 14.14$\pm$0.07 & 14.63$\pm$0.03 & 14.70$\pm$0.03 & 14.47$\pm$0.05 & 14.41$\pm$0.02 & 15.21$\pm$0.04 & 15.51$\pm$0.07 \\
107 & 15.62$\pm$0.01 & 15.49$\pm$0.01 & 13.38$\pm$0.07 & 14.35$\pm$0.03 & 14.42$\pm$0.04 & 14.23$\pm$0.06 & 14.01$\pm$0.05 & 14.15$\pm$0.13 & 14.08$\pm$0.09 \\
108 & 15.44$\pm$0.01 & 15.34$\pm$0.01 & 13.16$\pm$0.07 & 14.09$\pm$0.04 & 14.05$\pm$0.07 & 13.71$\pm$0.09 & 13.61$\pm$0.08 & 13.97$\pm$0.15 & 13.86$\pm$0.13 \\
109 & 14.97$\pm$0.01 & 14.71$\pm$0.02 & 13.53$\pm$0.07 & 14.02$\pm$0.03 & 14.15$\pm$0.03 & 13.97$\pm$0.05 & 13.92$\pm$0.02 & 14.62$\pm$0.05 & 14.96$\pm$0.07 \\
110 & 15.81$\pm$0.01 & 15.62$\pm$0.01 & 13.54$\pm$0.07 & 14.23$\pm$0.03 & 14.23$\pm$0.05 & 14.04$\pm$0.09 & 13.95$\pm$0.07 & 14.25$\pm$0.15 & 14.05$\pm$0.11 \\
111 & 15.71$\pm$0.00 & 15.59$\pm$0.00 & 13.72$\pm$0.07 & 13.97$\pm$0.05 & 14.04$\pm$0.05 & 14.00$\pm$0.05 & 14.03$\pm$0.02 & 14.68$\pm$0.05 & 14.91$\pm$0.07 \\
112 & 15.67$\pm$0.01 & 15.53$\pm$0.01 & 12.95$\pm$0.07 & 14.23$\pm$0.02 & 14.21$\pm$0.03 & 14.07$\pm$0.05 & 14.01$\pm$0.03 & 13.82$\pm$0.15 & 13.83$\pm$0.09 \\
113 & 15.43$\pm$0.00 & 15.31$\pm$0.01 & 13.48$\pm$0.07 & 14.19$\pm$0.03 & 14.03$\pm$0.05 & 13.92$\pm$0.06 & 13.99$\pm$0.02 & 14.23$\pm$0.06 & 14.54$\pm$0.07 \\
114 & 14.52$\pm$0.02 & 14.39$\pm$0.02 & 12.19$\pm$0.07 &   ****$\pm$**** &   ****$\pm$**** & 13.14$\pm$0.15 &   ****$\pm$**** &   ****$\pm$**** & 13.28$\pm$0.16 \\
115 & 14.88$\pm$0.01 & 14.68$\pm$0.03 & 13.16$\pm$0.07 &   ****$\pm$**** &   ****$\pm$**** & 13.38$\pm$0.15 & 13.23$\pm$0.15 & 14.06$\pm$0.09 & 14.01$\pm$0.08 \\
116 & 15.24$\pm$0.01 & 15.13$\pm$0.01 & 13.19$\pm$0.07 & 14.02$\pm$0.03 & 14.04$\pm$0.05 & 13.76$\pm$0.07 & 13.78$\pm$0.04 & 13.74$\pm$0.15 & 13.88$\pm$0.08 \\
117 & 16.09$\pm$0.00 & 15.92$\pm$0.01 & 14.07$\pm$0.07 & 14.25$\pm$0.04 & 14.28$\pm$0.06 & 14.37$\pm$0.05 & 14.19$\pm$0.03 & 14.76$\pm$0.05 & 14.85$\pm$0.07 \\
118 & 15.53$\pm$0.01 & 15.43$\pm$0.01 & 13.62$\pm$0.07 & 14.31$\pm$0.02 & 14.19$\pm$0.04 & 14.02$\pm$0.05 & 13.83$\pm$0.03 & 14.15$\pm$0.06 & 14.28$\pm$0.07 \\
119 & 14.73$\pm$0.02 & 14.56$\pm$0.04 & 13.07$\pm$0.07 & 14.48$\pm$0.02 & 13.20$\pm$0.33 & 13.77$\pm$0.07 & 14.13$\pm$0.03 &   ****$\pm$**** &   ****$\pm$**** \\

\hline
\end{longtable}
\end{landscape}
}

\clearpage
\longtabL{2}{
\begin{landscape}
\begin{longtable}{cccccccccc}
\caption{\label{tab:phot2} Logarithmic of the fluxes from {\it W4} 12\,\mi\ to LABOCA 870\,\mi\ bands of our set of \hii\ regions in mJy\,Hz obtained with centres and apertures given in Table~B.1 in  \citetalias{2013A&A...552A.140R}. All the fluxes have been background subtracted. The errors include the uncertainties in the calibration and background subtraction. We did not take into account negative fluxes higher than the corresponding errors, in these cases the fluxes are represented by ****. }\\
\\
\hline\hline
ID  & F(12\,\mi) & F(22\,\mi) & F(24\,\mi)&F(70\mi)& F(70\mi\ (pacs)) & F(100\mi)&F(160\mi)&F(250\mi) & F(870\mi) \\
   & $\log$(mJy\,Hz) &$\log$(mJy\,Hz) &$\log$(mJy\,Hz) &$\log$(mJy\,Hz) &$\log$(mJy\,Hz) &$\log$(mJy\,Hz) &$\log$(mJy\,Hz) &$\log$(mJy\,Hz) &$\log$(mJy) \\
   \hline
\endfirsthead
\caption{continued.}\\
\\
\hline\hline
ID  & F(12\,\mi) & F(22\,\mi) & F(24\,\mi)&F(70\mi)& F(70\mi\ (pacs)) & F(100\mi)&F(160\mi)&F(250\mi) & F(870\mi) \\
   & $\log$(mJy\,Hz) &$\log$(mJy\,Hz) &$\log$(mJy\,Hz) &$\log$(mJy\,Hz) &$\log$(mJy\,Hz) &$\log$(mJy\,Hz) &$\log$(mJy\,Hz) &$\log$(mJy\,Hz) &$\log$(mJy\,Hz) \\
\\
   \hline
\endhead
\hline
\endfoot
\small
  1 & 14.77$\pm$0.02 & 14.83$\pm$0.03 & 14.88$\pm$0.02 & 15.67$\pm$0.02 & 15.74$\pm$0.02 & 15.88$\pm$0.02 & 15.74$\pm$0.02 & 15.35$\pm$0.07 &   ****$\pm$**** \\
  2 & 14.64$\pm$0.02 & 14.40$\pm$0.03 & 14.45$\pm$0.03 & 15.11$\pm$0.02 & 14.83$\pm$0.04 & 15.18$\pm$0.04 & 15.27$\pm$0.03 & 15.01$\pm$0.07 & 13.25$\pm$0.09 \\
  3 & 14.47$\pm$0.03 & 14.56$\pm$0.03 & 14.55$\pm$0.03 & 15.26$\pm$0.02 & 15.32$\pm$0.02 & 15.64$\pm$0.02 & 15.41$\pm$0.02 & 15.04$\pm$0.07 & 13.14$\pm$0.11 \\
  4 & 14.38$\pm$0.03 & 14.03$\pm$0.05 & 13.80$\pm$0.15 & 15.04$\pm$0.03 & 15.29$\pm$0.03 & 15.33$\pm$0.05 & 15.30$\pm$0.03 & 14.89$\pm$0.07 &   ****$\pm$**** \\
  5 & 14.56$\pm$0.02 & 14.54$\pm$0.03 & 14.57$\pm$0.02 & 15.18$\pm$0.03 & 15.31$\pm$0.02 & 15.42$\pm$0.03 & 15.36$\pm$0.03 & 15.01$\pm$0.07 & 12.91$\pm$0.12 \\
  6 & 14.73$\pm$0.03 & 14.49$\pm$0.03 & 14.41$\pm$0.05 & 15.34$\pm$0.03 & 15.42$\pm$0.03 & 15.46$\pm$0.04 & 15.37$\pm$0.03 & 15.16$\pm$0.07 & 13.26$\pm$0.11 \\
  7 & 14.87$\pm$0.02 & 15.07$\pm$0.02 & 15.08$\pm$0.02 & 15.77$\pm$0.02 & 15.89$\pm$0.02 & 15.87$\pm$0.02 & 15.68$\pm$0.02 & 15.27$\pm$0.07 & 13.62$\pm$0.08 \\
  8 & 14.04$\pm$0.03 & 13.74$\pm$0.05 & 13.99$\pm$0.04 & 14.61$\pm$0.03 & 14.47$\pm$0.08 & 14.31$\pm$0.17 & 14.89$\pm$0.03 & 14.58$\pm$0.07 & 12.72$\pm$0.18 \\
  9 & 14.01$\pm$0.07 & 13.92$\pm$0.05 & 13.94$\pm$0.09 & 14.54$\pm$0.07 &   ****$\pm$**** & 14.66$\pm$0.13 & 15.01$\pm$0.04 & 14.71$\pm$0.07 & 12.92$\pm$0.15 \\
 10 & 14.29$\pm$0.04 & 14.08$\pm$0.04 & 14.26$\pm$0.03 & 15.02$\pm$0.03 & 14.96$\pm$0.04 & 15.38$\pm$0.03 & 15.03$\pm$0.04 & 14.64$\pm$0.07 & 12.82$\pm$0.14 \\
 11 & 13.20$\pm$0.16 &   ****$\pm$**** & 13.54$\pm$0.05 & 13.80$\pm$0.16 & 13.75$\pm$0.21 & 14.17$\pm$0.15 & 14.01$\pm$0.15 & 14.04$\pm$0.10 &   ****$\pm$**** \\
 12 & 14.19$\pm$0.04 & 14.13$\pm$0.04 & 14.16$\pm$0.04 & 14.88$\pm$0.03 & 15.14$\pm$0.02 & 15.01$\pm$0.06 & 15.09$\pm$0.04 & 14.66$\pm$0.08 & 12.97$\pm$0.11 \\
 13 & 14.10$\pm$0.06 & 14.12$\pm$0.05 & 13.80$\pm$0.10 & 14.69$\pm$0.05 & 15.06$\pm$0.04 & 15.00$\pm$0.06 & 14.80$\pm$0.07 & 14.69$\pm$0.08 &   ****$\pm$**** \\
 14 & 14.43$\pm$0.03 & 14.13$\pm$0.03 & 14.21$\pm$0.03 & 14.93$\pm$0.03 & 14.87$\pm$0.04 & 15.17$\pm$0.03 & 15.22$\pm$0.03 & 14.88$\pm$0.07 & 13.43$\pm$0.06 \\
 15 & 14.32$\pm$0.03 & 14.14$\pm$0.03 & 14.22$\pm$0.03 & 14.95$\pm$0.03 & 14.87$\pm$0.04 & 14.96$\pm$0.05 & 15.15$\pm$0.03 & 14.70$\pm$0.07 & 13.18$\pm$0.07 \\
 16 & 14.35$\pm$0.02 & 13.93$\pm$0.04 & 14.10$\pm$0.04 & 14.69$\pm$0.03 & 14.83$\pm$0.04 & 14.98$\pm$0.07 & 14.78$\pm$0.04 & 14.63$\pm$0.07 & 12.80$\pm$0.13 \\
 17 & 14.87$\pm$0.04 & 14.93$\pm$0.03 & 14.94$\pm$0.02 & 15.64$\pm$0.03 & 15.77$\pm$0.03 & 15.93$\pm$0.03 & 15.78$\pm$0.03 & 15.36$\pm$0.07 & 13.62$\pm$0.06 \\
 18 & 14.36$\pm$0.02 & 14.34$\pm$0.03 & 14.44$\pm$0.03 & 15.16$\pm$0.02 &   ****$\pm$**** & 15.38$\pm$0.03 &   ****$\pm$**** & 14.83$\pm$0.07 &   ****$\pm$**** \\
 19 & 14.40$\pm$0.03 & 14.34$\pm$0.03 & 14.41$\pm$0.02 & 15.02$\pm$0.03 & 15.06$\pm$0.03 & 15.23$\pm$0.03 & 15.21$\pm$0.03 & 14.76$\pm$0.07 & 12.77$\pm$0.14 \\
 20 & 15.14$\pm$0.02 & 15.01$\pm$0.03 & 14.99$\pm$0.02 & 15.71$\pm$0.02 & 15.86$\pm$0.02 & 16.05$\pm$0.02 & 15.86$\pm$0.02 & 15.41$\pm$0.07 & 13.62$\pm$0.06 \\
 21 & 15.39$\pm$0.02 & 15.51$\pm$0.02 & 15.50$\pm$0.02 & 15.97$\pm$0.02 & 16.16$\pm$0.02 & 16.19$\pm$0.02 & 16.00$\pm$0.02 & 15.59$\pm$0.07 & 13.71$\pm$0.06 \\
 22 & 14.98$\pm$0.02 & 14.98$\pm$0.03 & 14.97$\pm$0.02 & 15.70$\pm$0.02 & 15.84$\pm$0.02 & 15.92$\pm$0.02 & 15.77$\pm$0.02 & 15.36$\pm$0.07 & 13.40$\pm$0.06 \\
 23 & 15.15$\pm$0.02 & 15.38$\pm$0.02 & 15.40$\pm$0.02 & 15.86$\pm$0.02 & 16.00$\pm$0.02 & 16.09$\pm$0.02 & 15.86$\pm$0.02 & 15.39$\pm$0.07 & 13.59$\pm$0.05 \\
 24 & 14.87$\pm$0.02 & 14.81$\pm$0.03 & 14.80$\pm$0.02 & 15.54$\pm$0.02 & 15.64$\pm$0.02 & 15.80$\pm$0.02 & 15.68$\pm$0.02 & 15.36$\pm$0.07 & 13.76$\pm$0.05 \\
 25 & 15.08$\pm$0.02 & 15.26$\pm$0.02 & 15.28$\pm$0.02 & 15.79$\pm$0.02 & 15.91$\pm$0.02 & 15.98$\pm$0.02 & 15.78$\pm$0.02 & 15.33$\pm$0.07 & 13.20$\pm$0.08 \\
 26 & 14.70$\pm$0.03 & 14.37$\pm$0.04 & 14.54$\pm$0.03 & 15.12$\pm$0.03 & 15.07$\pm$0.04 & 15.36$\pm$0.05 & 15.49$\pm$0.02 & 15.11$\pm$0.07 & 13.37$\pm$0.07 \\
 27 & 14.66$\pm$0.02 & 14.33$\pm$0.06 & 14.40$\pm$0.04 & 15.10$\pm$0.04 & 15.22$\pm$0.04 & 15.42$\pm$0.04 & 15.15$\pm$0.06 & 14.74$\pm$0.10 & 13.16$\pm$0.07 \\
 28 & 14.88$\pm$0.02 & 14.96$\pm$0.03 & 14.95$\pm$0.02 & 15.73$\pm$0.02 & 15.86$\pm$0.02 & 15.82$\pm$0.03 & 15.67$\pm$0.02 & 15.32$\pm$0.07 &   ****$\pm$**** \\
 29 & 15.88$\pm$0.02 & 16.43$\pm$0.02 & 16.34$\pm$0.02 & 16.47$\pm$0.02 & 16.66$\pm$0.02 & 16.58$\pm$0.02 & 16.21$\pm$0.02 & 15.73$\pm$0.07 & 14.05$\pm$0.06 \\
 30 & 14.96$\pm$0.02 & 14.93$\pm$0.03 & 14.95$\pm$0.02 & 15.68$\pm$0.02 & 15.76$\pm$0.02 & 15.95$\pm$0.02 & 15.79$\pm$0.02 & 15.42$\pm$0.07 & 13.72$\pm$0.05 \\
 31 & 14.64$\pm$0.03 & 14.41$\pm$0.03 & 14.25$\pm$0.04 & 15.22$\pm$0.03 & 15.31$\pm$0.04 & 15.47$\pm$0.03 & 15.48$\pm$0.03 & 15.10$\pm$0.07 & 13.53$\pm$0.06 \\
 32 & 15.12$\pm$0.02 & 14.91$\pm$0.03 & 14.88$\pm$0.02 & 15.57$\pm$0.02 & 15.65$\pm$0.02 & 15.80$\pm$0.02 & 15.83$\pm$0.02 & 15.53$\pm$0.07 & 13.87$\pm$0.05 \\
 33 & 13.90$\pm$0.15 & 14.03$\pm$0.07 & 13.97$\pm$0.15 &   ****$\pm$**** & 14.75$\pm$0.06 &   ****$\pm$**** &   ****$\pm$**** & 14.47$\pm$0.10 &   ****$\pm$**** \\
 34 & 14.55$\pm$0.04 & 14.42$\pm$0.04 & 14.35$\pm$0.04 & 15.17$\pm$0.03 & 15.16$\pm$0.02 & 15.34$\pm$0.05 & 15.21$\pm$0.04 & 14.77$\pm$0.08 & 13.01$\pm$0.07 \\
 35 & 15.01$\pm$0.02 & 14.96$\pm$0.02 & 15.02$\pm$0.02 & 15.44$\pm$0.02 & 15.56$\pm$0.02 & 15.72$\pm$0.02 & 15.62$\pm$0.02 & 15.22$\pm$0.07 & 13.56$\pm$0.06 \\
 36 & 13.79$\pm$0.06 & 13.62$\pm$0.05 & 13.55$\pm$0.06 & 14.40$\pm$0.04 & 14.32$\pm$0.07 & 14.58$\pm$0.09 & 14.70$\pm$0.04 & 14.24$\pm$0.08 & 12.55$\pm$0.10 \\
 37 & 15.43$\pm$0.02 & 15.19$\pm$0.03 & 15.11$\pm$0.02 & 15.83$\pm$0.03 & 15.96$\pm$0.02 & 16.14$\pm$0.02 & 16.09$\pm$0.02 & 15.63$\pm$0.07 & 13.66$\pm$0.06 \\
 38 & 14.01$\pm$0.05 & 13.08$\pm$0.26 & 13.97$\pm$0.07 & 14.74$\pm$0.04 & 14.26$\pm$0.20 & 15.01$\pm$0.06 & 14.94$\pm$0.03 & 14.53$\pm$0.07 & 13.20$\pm$0.15 \\
 39 & 13.85$\pm$0.10 & 13.59$\pm$0.10 & 14.00$\pm$0.05 & 14.53$\pm$0.06 &   ****$\pm$**** & 14.61$\pm$0.15 & 14.86$\pm$0.04 & 14.56$\pm$0.07 & 12.82$\pm$0.10 \\
 40 & 13.80$\pm$0.14 & 13.94$\pm$0.08 & 13.96$\pm$0.15 & 13.77$\pm$0.40 &   ****$\pm$**** &   ****$\pm$**** &   ****$\pm$**** & 14.65$\pm$0.07 & 12.59$\pm$0.43 \\
 41 & 15.60$\pm$0.02 & 15.59$\pm$0.03 & 15.59$\pm$0.02 & 16.15$\pm$0.02 & 16.33$\pm$0.02 & 16.38$\pm$0.02 & 16.24$\pm$0.02 & 15.76$\pm$0.07 & 13.75$\pm$0.06 \\
 42 & 15.54$\pm$0.02 & 15.94$\pm$0.02 & 15.94$\pm$0.02 & 16.07$\pm$0.02 & 16.31$\pm$0.02 & 16.26$\pm$0.02 & 15.95$\pm$0.03 & 15.35$\pm$0.07 & 13.47$\pm$0.06 \\
 43 & 13.48$\pm$0.14 &   ****$\pm$**** & 13.57$\pm$0.15 & 13.83$\pm$0.22 & 14.52$\pm$0.08 & 14.57$\pm$0.15 & 14.66$\pm$0.05 &   ****$\pm$**** & 12.76$\pm$0.24 \\
 44 & 15.91$\pm$0.02 & 16.25$\pm$0.02 & 16.26$\pm$0.02 & 16.55$\pm$0.02 & 16.79$\pm$0.02 & 16.79$\pm$0.02 & 16.48$\pm$0.02 & 15.90$\pm$0.07 & 13.94$\pm$0.05 \\
 45 & 14.75$\pm$0.03 & 14.64$\pm$0.03 & 14.53$\pm$0.03 & 15.16$\pm$0.04 & 15.27$\pm$0.04 & 15.37$\pm$0.05 & 15.26$\pm$0.04 & 14.81$\pm$0.08 & 12.89$\pm$0.08 \\
 46 & 14.73$\pm$0.03 & 14.45$\pm$0.03 & 14.55$\pm$0.03 & 15.16$\pm$0.03 & 15.31$\pm$0.03 & 15.42$\pm$0.04 & 15.44$\pm$0.03 & 15.03$\pm$0.07 & 13.38$\pm$0.06 \\
 47 & 15.03$\pm$0.03 & 15.10$\pm$0.03 & 15.17$\pm$0.02 & 15.53$\pm$0.03 & 15.77$\pm$0.03 & 15.82$\pm$0.03 & 15.59$\pm$0.03 & 15.08$\pm$0.07 & 13.11$\pm$0.06 \\
 48 & 14.23$\pm$0.05 & 14.20$\pm$0.04 & 14.33$\pm$0.06 & 15.21$\pm$0.03 & 15.23$\pm$0.03 & 15.43$\pm$0.04 & 15.32$\pm$0.03 & 15.05$\pm$0.07 & 13.24$\pm$0.15 \\
 49 & 15.09$\pm$0.05 & 15.33$\pm$0.03 & 15.38$\pm$0.02 & 15.58$\pm$0.05 & 15.96$\pm$0.03 & 15.96$\pm$0.04 & 15.79$\pm$0.04 & 15.26$\pm$0.07 & 13.44$\pm$0.06 \\
 50 & 14.57$\pm$0.03 & 14.71$\pm$0.03 & 14.73$\pm$0.02 & 15.36$\pm$0.02 & 15.48$\pm$0.02 & 15.61$\pm$0.03 & 15.40$\pm$0.03 & 14.96$\pm$0.07 & 12.91$\pm$0.11 \\
 51 & 14.87$\pm$0.02 & 15.06$\pm$0.03 & 14.47$\pm$0.03 & 15.68$\pm$0.02 & 15.79$\pm$0.02 & 15.84$\pm$0.02 & 15.60$\pm$0.02 & 15.14$\pm$0.07 & 12.79$\pm$0.15 \\
 52 & 14.91$\pm$0.03 & 14.77$\pm$0.03 & 14.79$\pm$0.02 & 15.44$\pm$0.03 & 15.62$\pm$0.03 & 15.75$\pm$0.03 & 15.54$\pm$0.03 & 15.05$\pm$0.07 & 13.20$\pm$0.06 \\
 53 & 15.00$\pm$0.05 & 15.08$\pm$0.03 & 15.18$\pm$0.02 & 15.46$\pm$0.05 & 15.75$\pm$0.04 & 15.80$\pm$0.05 & 15.59$\pm$0.05 & 14.97$\pm$0.10 & 12.81$\pm$0.17 \\
 54 & 14.47$\pm$0.04 & 14.48$\pm$0.04 & 14.48$\pm$0.04 & 15.29$\pm$0.03 & 14.92$\pm$0.07 & 15.39$\pm$0.05 & 15.15$\pm$0.05 & 14.64$\pm$0.09 & 12.81$\pm$0.16 \\
 55 & 15.34$\pm$0.02 & 15.53$\pm$0.02 & 15.55$\pm$0.02 & 16.04$\pm$0.02 & 16.19$\pm$0.02 & 16.23$\pm$0.02 & 15.99$\pm$0.02 & 15.46$\pm$0.07 & 13.52$\pm$0.06 \\
 56 & 14.12$\pm$0.06 & 14.52$\pm$0.03 & 14.54$\pm$0.04 & 15.08$\pm$0.03 & 15.41$\pm$0.03 & 14.83$\pm$0.12 & 15.11$\pm$0.03 & 14.86$\pm$0.07 & 12.92$\pm$0.29 \\
 57 & 14.57$\pm$0.02 & 14.57$\pm$0.03 & 14.60$\pm$0.02 & 15.29$\pm$0.02 & 15.35$\pm$0.02 & 15.52$\pm$0.02 & 15.26$\pm$0.02 & 14.91$\pm$0.07 & 13.01$\pm$0.08 \\
 58 & 15.68$\pm$0.02 & 15.60$\pm$0.03 & 15.62$\pm$0.02 & 16.09$\pm$0.02 & 16.30$\pm$0.02 & 16.43$\pm$0.02 & 16.19$\pm$0.03 & 15.66$\pm$0.07 & 13.73$\pm$0.06 \\
 59 & 14.35$\pm$0.04 & 13.92$\pm$0.07 & 14.21$\pm$0.04 & 14.72$\pm$0.05 & 14.62$\pm$0.07 & 14.97$\pm$0.06 & 15.04$\pm$0.04 & 14.53$\pm$0.08 & 12.82$\pm$0.11 \\
 60 & 14.22$\pm$0.03 & 14.08$\pm$0.03 & 14.21$\pm$0.03 & 14.92$\pm$0.03 & 14.87$\pm$0.04 & 15.22$\pm$0.03 & 14.99$\pm$0.03 & 14.68$\pm$0.07 & 12.49$\pm$0.15 \\
 61 & 14.19$\pm$0.03 & 14.20$\pm$0.03 & 14.16$\pm$0.04 & 14.83$\pm$0.03 & 14.78$\pm$0.04 & 15.08$\pm$0.04 & 14.97$\pm$0.03 & 14.43$\pm$0.08 & 12.72$\pm$0.13 \\
 62 & 14.32$\pm$0.03 & 13.95$\pm$0.05 & 14.14$\pm$0.04 & 14.85$\pm$0.03 & 14.87$\pm$0.03 & 14.03$\pm$0.41 & 15.13$\pm$0.03 & 14.88$\pm$0.07 & 12.69$\pm$0.15 \\
 63 & 15.69$\pm$0.02 & 15.51$\pm$0.03 & 15.54$\pm$0.02 & 16.17$\pm$0.02 & 16.41$\pm$0.02 & 16.47$\pm$0.03 & 16.21$\pm$0.03 & 15.59$\pm$0.07 & 13.50$\pm$0.06 \\
 64 & 15.20$\pm$0.03 & 15.16$\pm$0.03 & 15.18$\pm$0.02 & 15.77$\pm$0.03 & 15.92$\pm$0.03 & 16.06$\pm$0.03 & 15.82$\pm$0.03 & 15.29$\pm$0.07 & 13.47$\pm$0.07 \\
 65 & 14.31$\pm$0.15 & 14.36$\pm$0.06 & 13.96$\pm$0.16 & 15.15$\pm$0.05 & 15.19$\pm$0.06 & 15.10$\pm$0.13 & 15.13$\pm$0.09 & 14.39$\pm$0.18 & 12.83$\pm$0.16 \\
 66 & 15.02$\pm$0.04 & 14.76$\pm$0.04 & 14.74$\pm$0.03 & 14.60$\pm$0.22 & 15.20$\pm$0.06 & 15.43$\pm$0.07 & 15.47$\pm$0.05 & 15.11$\pm$0.07 & 13.10$\pm$0.08 \\
 67 & 15.19$\pm$0.02 & 15.37$\pm$0.03 & 15.38$\pm$0.02 & 15.85$\pm$0.02 & 16.05$\pm$0.02 & 16.13$\pm$0.02 & 15.87$\pm$0.02 & 15.35$\pm$0.07 & 13.42$\pm$0.06 \\
 68 & 14.01$\pm$0.03 & 14.07$\pm$0.03 & 14.03$\pm$0.03 & 14.63$\pm$0.03 & 14.80$\pm$0.03 & 14.64$\pm$0.05 & 14.78$\pm$0.03 & 14.42$\pm$0.07 & 12.66$\pm$0.15 \\
 69 & 14.68$\pm$0.03 & 14.42$\pm$0.04 & 14.40$\pm$0.04 & 15.21$\pm$0.03 & 15.26$\pm$0.03 & 14.80$\pm$0.15 & 15.31$\pm$0.04 & 15.08$\pm$0.07 & 13.21$\pm$0.10 \\
 70 & 15.07$\pm$0.03 & 14.89$\pm$0.03 & 14.82$\pm$0.02 & 15.57$\pm$0.02 & 15.48$\pm$0.03 & 15.79$\pm$0.03 & 15.73$\pm$0.03 & 15.33$\pm$0.07 & 13.47$\pm$0.07 \\
 71 & 15.25$\pm$0.03 & 15.16$\pm$0.03 & 15.17$\pm$0.02 & 15.63$\pm$0.03 & 15.88$\pm$0.03 & 16.02$\pm$0.03 & 15.82$\pm$0.03 & 15.35$\pm$0.07 & 13.36$\pm$0.06 \\
 72 & 14.85$\pm$0.03 & 14.57$\pm$0.03 & 14.57$\pm$0.03 & 15.29$\pm$0.03 & 15.34$\pm$0.03 & 15.62$\pm$0.03 & 15.45$\pm$0.04 & 15.06$\pm$0.07 & 13.19$\pm$0.06 \\
 73 & 15.21$\pm$0.03 & 15.07$\pm$0.03 & 15.11$\pm$0.02 & 15.81$\pm$0.02 & 15.91$\pm$0.02 & 16.00$\pm$0.03 & 15.94$\pm$0.03 & 15.43$\pm$0.07 & 13.57$\pm$0.06 \\
 74 & 15.55$\pm$0.02 & 15.52$\pm$0.03 & 15.52$\pm$0.02 & 16.09$\pm$0.02 & 16.26$\pm$0.02 & 16.39$\pm$0.02 & 16.23$\pm$0.02 & 15.74$\pm$0.07 & 13.82$\pm$0.05 \\
 75 & 14.51$\pm$0.03 & 14.28$\pm$0.03 & 14.35$\pm$0.03 & 14.98$\pm$0.03 & 15.11$\pm$0.03 & 15.33$\pm$0.03 & 15.25$\pm$0.03 & 14.86$\pm$0.07 & 12.88$\pm$0.08 \\
 76 & 14.72$\pm$0.02 & 14.48$\pm$0.03 & 14.48$\pm$0.02 & 15.12$\pm$0.02 & 15.31$\pm$0.02 & 15.45$\pm$0.02 & 15.43$\pm$0.02 & 15.14$\pm$0.07 & 13.49$\pm$0.06 \\
 77 & 13.93$\pm$0.29 & 14.12$\pm$0.11 & 14.21$\pm$0.09 & 14.59$\pm$0.16 & 14.99$\pm$0.08 & 15.30$\pm$0.06 & 15.03$\pm$0.09 & 14.47$\pm$0.15 & 12.72$\pm$0.15 \\
 78 & 15.34$\pm$0.02 & 15.14$\pm$0.03 & 15.14$\pm$0.02 & 15.76$\pm$0.02 & 15.86$\pm$0.03 & 16.06$\pm$0.03 & 15.98$\pm$0.02 & 15.52$\pm$0.07 & 13.61$\pm$0.06 \\
 79 & 15.34$\pm$0.02 & 15.42$\pm$0.03 & 15.43$\pm$0.02 & 15.99$\pm$0.02 & 16.15$\pm$0.02 & 16.25$\pm$0.02 & 16.00$\pm$0.03 & 15.51$\pm$0.07 & 13.58$\pm$0.06 \\
 80 & 14.68$\pm$0.04 & 14.38$\pm$0.05 & 14.46$\pm$0.04 & 14.90$\pm$0.06 & 15.19$\pm$0.03 & 15.47$\pm$0.04 & 15.38$\pm$0.04 & 15.02$\pm$0.07 & 12.61$\pm$0.18 \\
 81 & 14.12$\pm$0.04 & 13.97$\pm$0.04 & 13.68$\pm$0.14 & 14.35$\pm$0.09 &   ****$\pm$**** & 14.93$\pm$0.06 & 15.04$\pm$0.03 & 14.65$\pm$0.07 & 12.88$\pm$0.14 \\
 82 & 14.60$\pm$0.03 & 14.37$\pm$0.03 & 14.36$\pm$0.03 & 15.05$\pm$0.03 & 15.02$\pm$0.06 & 15.23$\pm$0.04 & 15.27$\pm$0.03 & 14.81$\pm$0.07 & 12.75$\pm$0.13 \\
 83 & 13.98$\pm$0.09 & 13.84$\pm$0.08 & 14.06$\pm$0.10 & 14.86$\pm$0.03 & 14.69$\pm$0.07 &   ****$\pm$**** & 15.23$\pm$0.03 & 14.77$\pm$0.07 & 13.12$\pm$0.18 \\
 84 & 14.13$\pm$0.03 & 14.05$\pm$0.03 & 14.16$\pm$0.04 & 14.84$\pm$0.03 & 14.80$\pm$0.04 & 14.84$\pm$0.07 & 15.07$\pm$0.03 & 14.74$\pm$0.07 & 13.15$\pm$0.07 \\
 85 & 15.11$\pm$0.04 & 15.23$\pm$0.03 & 15.26$\pm$0.02 & 15.68$\pm$0.03 & 15.97$\pm$0.02 & 16.06$\pm$0.03 & 15.89$\pm$0.03 & 15.44$\pm$0.07 & 13.04$\pm$0.10 \\
 86 & 14.43$\pm$0.04 & 14.51$\pm$0.03 & 14.42$\pm$0.03 & 15.00$\pm$0.04 & 15.22$\pm$0.03 & 15.39$\pm$0.03 & 15.20$\pm$0.04 & 14.75$\pm$0.08 & 13.08$\pm$0.07 \\
 87 & 13.69$\pm$0.15 & 13.55$\pm$0.15 & 13.93$\pm$0.14 & 14.83$\pm$0.03 &   ****$\pm$**** & 15.35$\pm$0.04 &   ****$\pm$**** & 14.62$\pm$0.07 &   ****$\pm$**** \\
 88 & 13.84$\pm$0.04 & 13.53$\pm$0.07 & 13.34$\pm$0.19 & 14.17$\pm$0.06 &   ****$\pm$**** & 14.38$\pm$0.15 & 14.81$\pm$0.04 & 14.38$\pm$0.07 & 12.67$\pm$0.15 \\
 89 &   ****$\pm$**** &   ****$\pm$**** & 13.38$\pm$0.21 &   ****$\pm$**** &   ****$\pm$**** & 14.73$\pm$0.09 &   ****$\pm$**** & 14.39$\pm$0.07 &   ****$\pm$**** \\
 90 & 15.24$\pm$0.04 & 15.46$\pm$0.03 & 15.47$\pm$0.02 & 15.96$\pm$0.03 & 16.12$\pm$0.02 & 16.15$\pm$0.03 & 15.95$\pm$0.03 & 15.43$\pm$0.07 & 13.57$\pm$0.06 \\
 91 & 15.29$\pm$0.02 & 15.49$\pm$0.02 & 15.50$\pm$0.02 & 15.78$\pm$0.02 & 15.98$\pm$0.02 & 16.05$\pm$0.02 & 15.88$\pm$0.02 & 15.44$\pm$0.07 & 13.54$\pm$0.05 \\
 92 & 14.53$\pm$0.02 & 13.97$\pm$0.06 & 14.23$\pm$0.03 & 14.82$\pm$0.05 & 14.62$\pm$0.12 & 15.14$\pm$0.05 & 14.90$\pm$0.07 & 14.61$\pm$0.08 & 12.60$\pm$0.15 \\
 93 &   ****$\pm$**** &   ****$\pm$**** &   ****$\pm$**** &   ****$\pm$**** &   ****$\pm$**** & 14.72$\pm$0.15 &   ****$\pm$**** & 14.25$\pm$0.12 & 12.83$\pm$0.15 \\
 94 & 14.11$\pm$0.04 & 13.91$\pm$0.03 & 13.83$\pm$0.04 & 14.42$\pm$0.04 & 14.12$\pm$0.08 & 14.53$\pm$0.08 & 14.52$\pm$0.09 & 14.29$\pm$0.08 &   ****$\pm$**** \\
 95 & 13.96$\pm$0.05 & 13.91$\pm$0.05 & 14.08$\pm$0.03 & 14.76$\pm$0.03 & 14.95$\pm$0.03 & 14.98$\pm$0.05 & 14.81$\pm$0.04 & 14.39$\pm$0.08 & 12.25$\pm$0.24 \\
 96 & 13.68$\pm$0.15 & 13.39$\pm$0.15 & 13.47$\pm$0.21 & 14.44$\pm$0.10 &   ****$\pm$**** & 15.24$\pm$0.05 & 14.62$\pm$0.09 & 14.29$\pm$0.10 & 12.73$\pm$0.15 \\
 97 & 15.03$\pm$0.02 & 15.13$\pm$0.02 & 15.12$\pm$0.02 & 15.71$\pm$0.02 & 15.83$\pm$0.02 & 15.88$\pm$0.02 & 15.82$\pm$0.02 & 15.45$\pm$0.07 & 13.60$\pm$0.06 \\
 98 & 16.43$\pm$0.02 & 16.76$\pm$0.02 & 16.74$\pm$0.02 & 17.09$\pm$0.02 & 17.29$\pm$0.02 & 17.28$\pm$0.02 & 17.01$\pm$0.02 & 16.47$\pm$0.07 & 14.53$\pm$0.05 \\
 99 & 14.25$\pm$0.05 & 14.53$\pm$0.03 & 14.65$\pm$0.03 & 15.31$\pm$0.02 & 15.39$\pm$0.03 & 15.48$\pm$0.03 & 15.39$\pm$0.02 & 14.92$\pm$0.07 & 12.99$\pm$0.15 \\
100 & 14.66$\pm$0.02 & 14.33$\pm$0.03 & 14.42$\pm$0.02 & 14.97$\pm$0.03 & 15.03$\pm$0.03 & 15.28$\pm$0.03 & 15.37$\pm$0.02 & 15.07$\pm$0.07 & 13.27$\pm$0.06 \\
101 & 14.24$\pm$0.12 & 14.07$\pm$0.12 & 14.01$\pm$0.15 & 15.03$\pm$0.07 & 15.07$\pm$0.08 & 15.22$\pm$0.06 & 14.52$\pm$0.34 & 14.43$\pm$0.21 & 13.21$\pm$0.10 \\
102 & 14.65$\pm$0.02 & 14.91$\pm$0.03 & 14.92$\pm$0.02 & 15.32$\pm$0.02 & 15.45$\pm$0.02 & 15.39$\pm$0.03 & 15.20$\pm$0.03 & 14.91$\pm$0.07 & 13.12$\pm$0.08 \\
103 & 14.67$\pm$0.03 & 14.44$\pm$0.03 & 14.47$\pm$0.05 & 15.33$\pm$0.02 & 14.07$\pm$0.33 & 15.39$\pm$0.04 & 15.43$\pm$0.03 & 15.07$\pm$0.07 & 13.10$\pm$0.15 \\
104 & 14.80$\pm$0.02 & 14.71$\pm$0.03 & 14.80$\pm$0.02 & 15.41$\pm$0.02 & 15.57$\pm$0.02 & 15.70$\pm$0.02 & 15.64$\pm$0.02 & 15.20$\pm$0.07 & 13.34$\pm$0.06 \\
105 & 14.05$\pm$0.04 & 13.93$\pm$0.04 & 14.06$\pm$0.05 & 14.55$\pm$0.03 & 14.71$\pm$0.06 & 14.71$\pm$0.08 & 14.46$\pm$0.07 & 14.47$\pm$0.07 & 12.66$\pm$0.15 \\
106 & 15.07$\pm$0.02 & 15.10$\pm$0.03 & 15.14$\pm$0.02 & 15.70$\pm$0.02 & 15.83$\pm$0.02 & 15.97$\pm$0.02 & 15.79$\pm$0.02 & 15.37$\pm$0.07 & 13.50$\pm$0.06 \\
107 & 13.84$\pm$0.09 & 13.94$\pm$0.05 & 13.39$\pm$0.38 & 14.29$\pm$0.09 & 14.22$\pm$0.15 &   ****$\pm$**** & 14.77$\pm$0.05 & 14.29$\pm$0.09 &   ****$\pm$**** \\
108 & 13.41$\pm$0.13 & 13.11$\pm$0.36 & 13.45$\pm$0.16 &   ****$\pm$**** & 14.02$\pm$0.20 & 14.60$\pm$0.12 & 14.48$\pm$0.12 & 13.97$\pm$0.13 &   ****$\pm$**** \\
109 & 14.51$\pm$0.02 & 14.28$\pm$0.03 & 14.33$\pm$0.02 & 14.89$\pm$0.03 & 14.88$\pm$0.03 & 15.29$\pm$0.03 & 15.16$\pm$0.03 & 14.94$\pm$0.07 &   ****$\pm$**** \\
110 & 14.12$\pm$0.08 & 13.68$\pm$0.14 & 14.02$\pm$0.15 & 14.79$\pm$0.04 &   ****$\pm$**** &   ****$\pm$**** &   ****$\pm$**** & 14.61$\pm$0.07 &   ****$\pm$**** \\
111 & 14.46$\pm$0.02 & 14.28$\pm$0.03 & 14.26$\pm$0.03 & 14.96$\pm$0.02 & 15.09$\pm$0.03 & 15.18$\pm$0.03 & 15.23$\pm$0.02 & 14.92$\pm$0.07 & 13.02$\pm$0.10 \\
112 & 13.31$\pm$0.19 & 13.49$\pm$0.09 & 13.54$\pm$0.16 & 14.16$\pm$0.07 &   ****$\pm$**** & 14.71$\pm$0.07 &   ****$\pm$**** & 13.62$\pm$0.18 &   ****$\pm$**** \\
113 & 14.05$\pm$0.04 & 13.91$\pm$0.04 & 13.96$\pm$0.05 & 14.67$\pm$0.03 & 14.84$\pm$0.04 & 14.74$\pm$0.07 & 14.99$\pm$0.03 & 14.53$\pm$0.07 & 12.56$\pm$0.15 \\
114 & 13.01$\pm$0.15 & 12.94$\pm$0.14 & 13.19$\pm$0.15 & 13.26$\pm$0.34 & 14.26$\pm$0.07 & 14.24$\pm$0.11 & 13.81$\pm$0.15 & 13.50$\pm$0.13 & 12.40$\pm$0.15 \\
115 & 13.35$\pm$0.16 & 13.47$\pm$0.10 & 13.26$\pm$0.28 &   ****$\pm$**** & 14.18$\pm$0.11 & 14.71$\pm$0.09 & 14.49$\pm$0.05 & 14.27$\pm$0.07 & 12.69$\pm$0.15 \\
116 & 13.41$\pm$0.12 & 13.20$\pm$0.15 & 13.40$\pm$0.18 & 13.66$\pm$0.18 &   ****$\pm$**** &   ****$\pm$**** & 14.07$\pm$0.15 & 13.94$\pm$0.11 & 12.83$\pm$0.11 \\
117 & 14.44$\pm$0.03 & 14.04$\pm$0.05 & 14.19$\pm$0.05 & 15.10$\pm$0.03 & 15.07$\pm$0.04 & 15.24$\pm$0.04 & 15.21$\pm$0.03 & 14.94$\pm$0.07 & 12.76$\pm$0.17 \\
118 & 13.72$\pm$0.04 & 13.44$\pm$0.10 & 13.77$\pm$0.06 & 14.62$\pm$0.04 & 14.51$\pm$0.07 & 14.93$\pm$0.04 & 14.63$\pm$0.03 & 14.32$\pm$0.07 &   ****$\pm$**** \\
119 & 13.10$\pm$0.27 &   ****$\pm$**** & 13.55$\pm$0.15 & 14.47$\pm$0.04 & 14.64$\pm$0.04 &   ****$\pm$**** & 13.64$\pm$0.27 & 13.85$\pm$0.11 &   ****$\pm$**** \\

\hline
\end{longtable}
\end{landscape}
}

\section{Comparison of fluxes}
\begin{figure*}
\centering
\includegraphics[width=\textwidth]{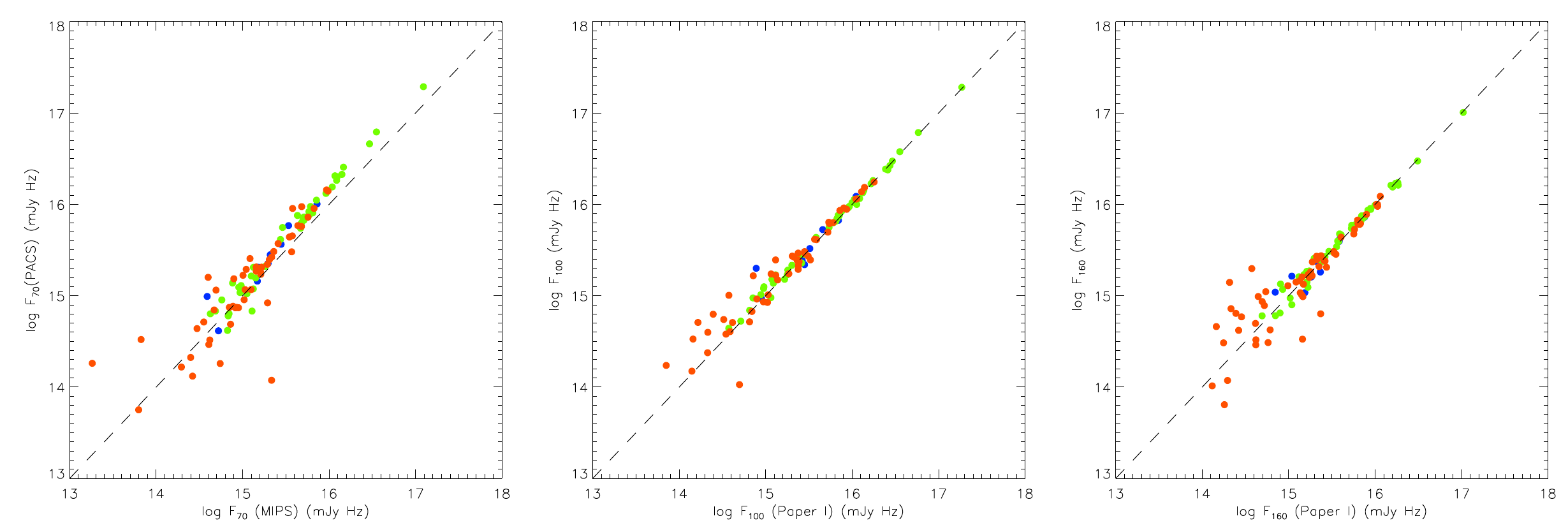}   
\includegraphics[width=\textwidth]{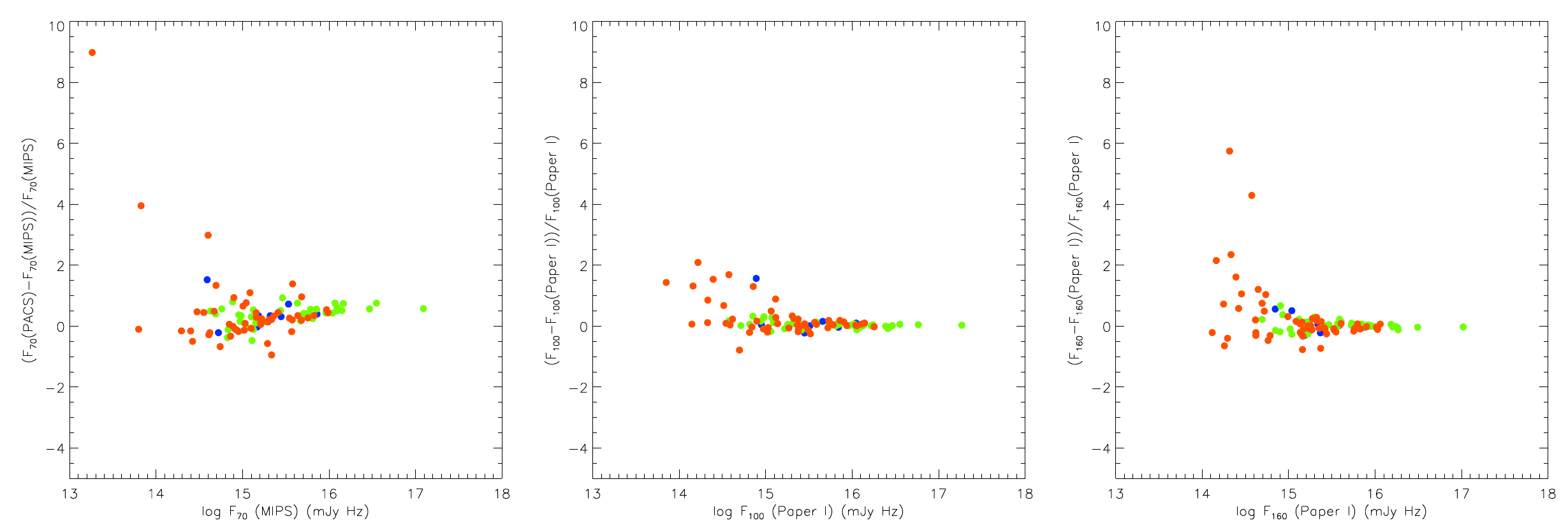}   
\caption{Left column: Comparison of the MIPS and PACS 70\,\mi\ fluxes. Middle and right columns: Comparison of the PACS 100\,\mi\ and 160\,\mi\ fluxes presented in this paper and those published in \citetalias{2013A&A...552A.140R}.}
\label{comp_pacs}
\end{figure*}

In Fig.~\ref{comp_pacs} we show the comparison of the PACS 100\,\mi\ and 160\,\mi\ obtained here with the ones published in \citetalias{2013A&A...552A.140R}. The low luminosity regions tend to show higher fluxes than in previous images, which shows that the new images recover the missed flux in the previous ones. The comparison of PACS and MIPS 70\,\mi\ fluxes is shown in the left panels of Fig.~\ref{comp_pacs}. In general PACS 70\,\mi\  fluxes are higher than MIPS 70\,\mi\ ones for bright regions. This trend is the same as it was found in \citet{2012ApJ...756..138A} for NGC~628 and NGC~6946. 

\section{SED fitting for {\it Desert} dust model}\label{app:compSED}
\begin{figure*}
   \centering
  \includegraphics[width=0.49\textwidth]{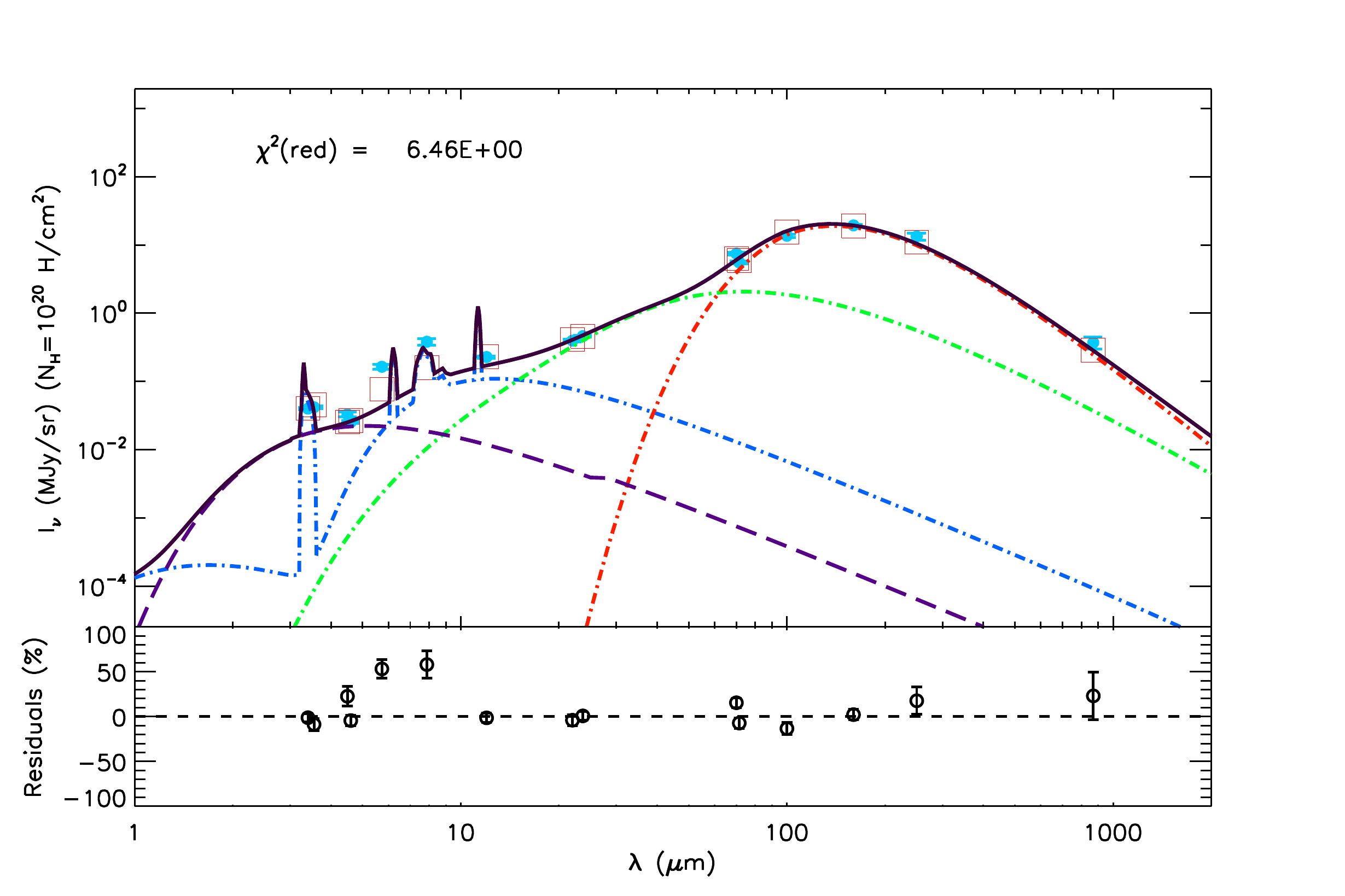} 
  \includegraphics[width=0.49\textwidth]{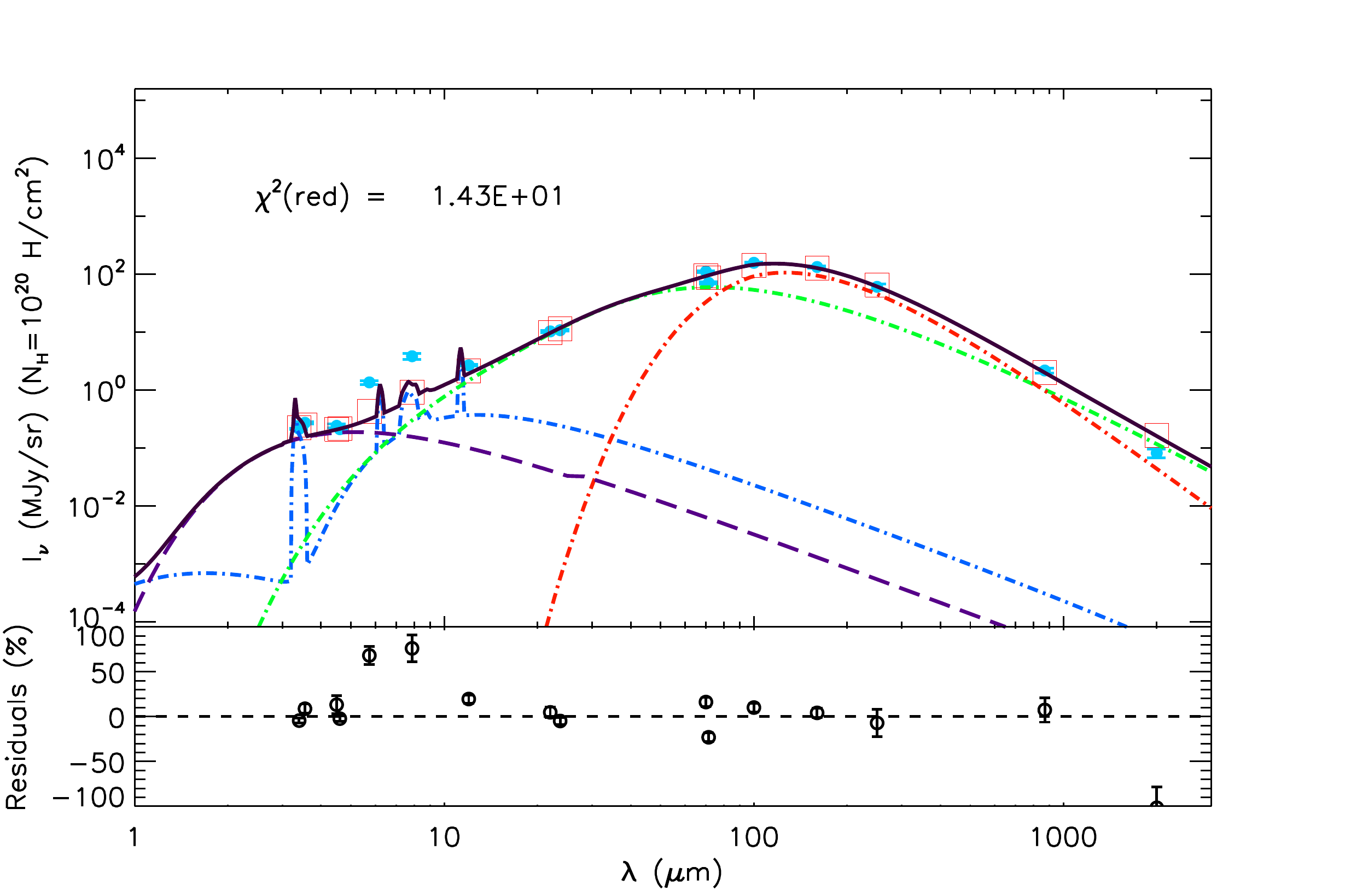} 
  \includegraphics[width=0.49\textwidth]{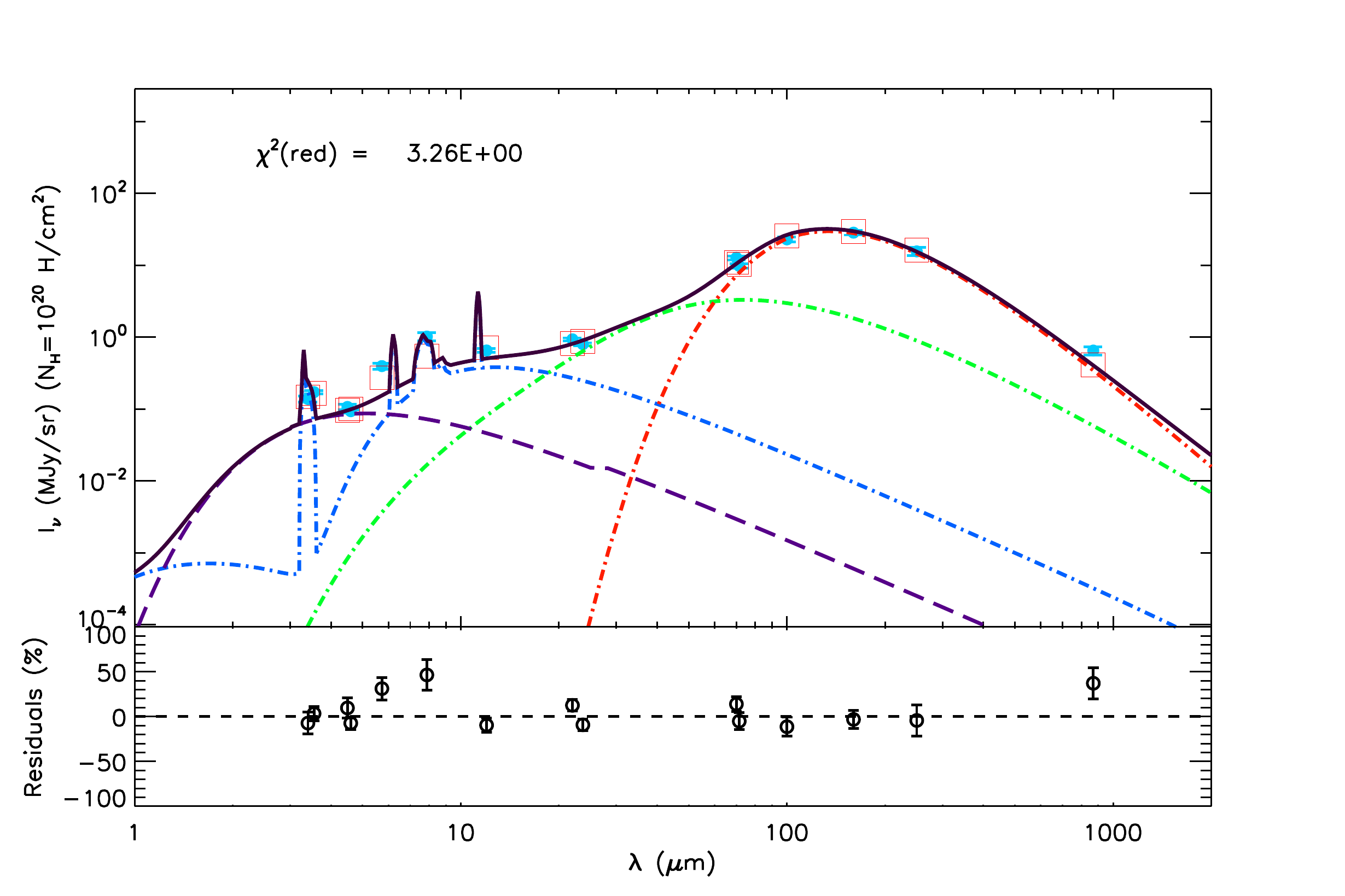} 
  \includegraphics[width=0.49\textwidth]{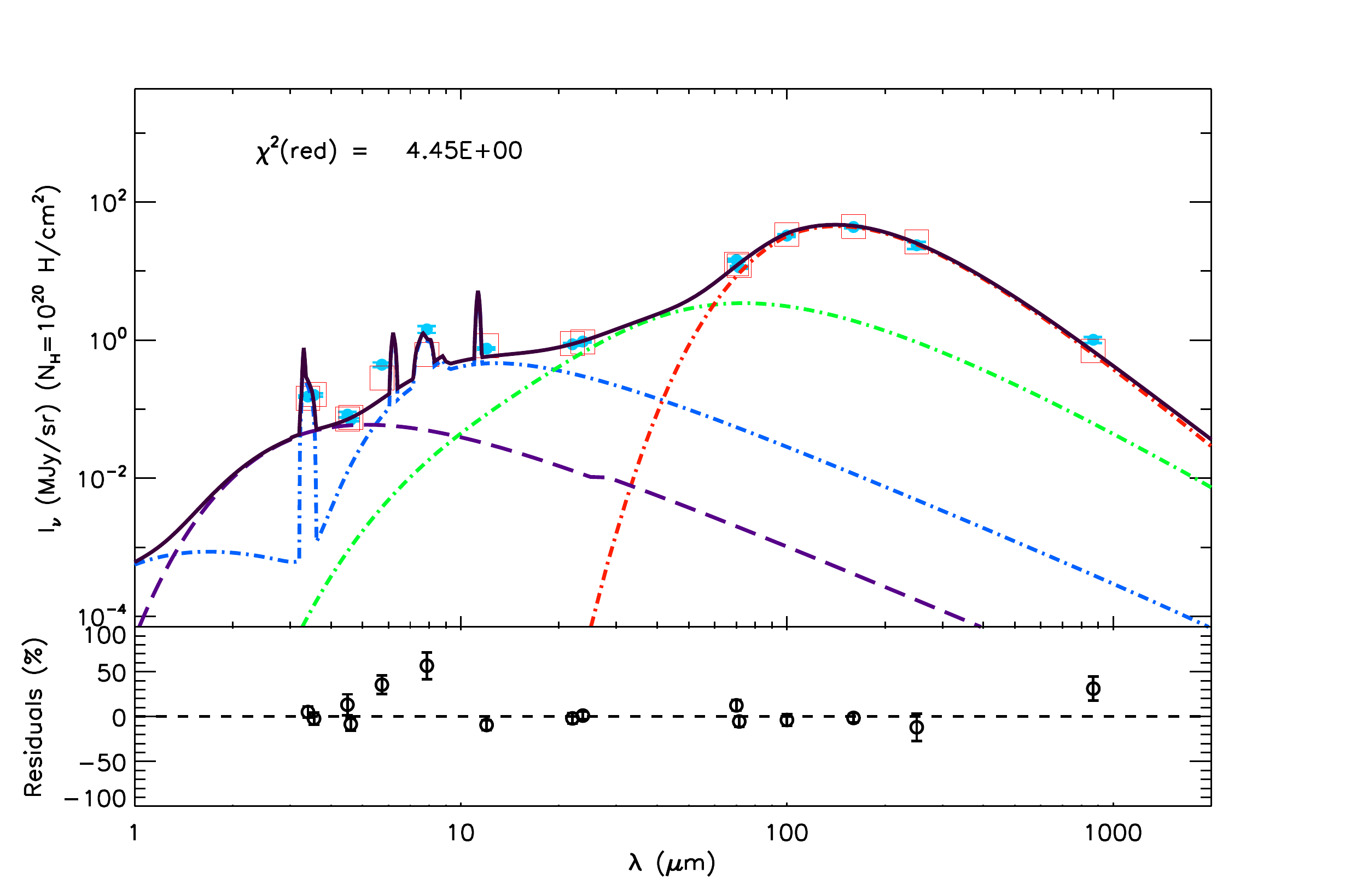}   
   \caption{SED fittings for the 4 regions shown in Fig.~\ref{fig:sed} but fitted with a  4\,Myr star cluster ISRF and {\it Desert} dust model. {\it Light blue points}: observed data with the errors, {\it red squares}: modelled broad-band fluxes, {\it dashed-dot blue line}: PAH emission, {\it dashed-dot green line}: VSG emission, {\it dashed-dot red line}: BG emission and {\it dashed purple line}: NIR continuum. The SED is well fitted at most bands except in the 6-9\,\mi\ wavelength range where the  {\it Desert} dust model under-predicts the PAH emission.}
              \label{fig:sed:desert}%
    \end{figure*}

\begin{figure*}
   \centering
\includegraphics[width=\textwidth]{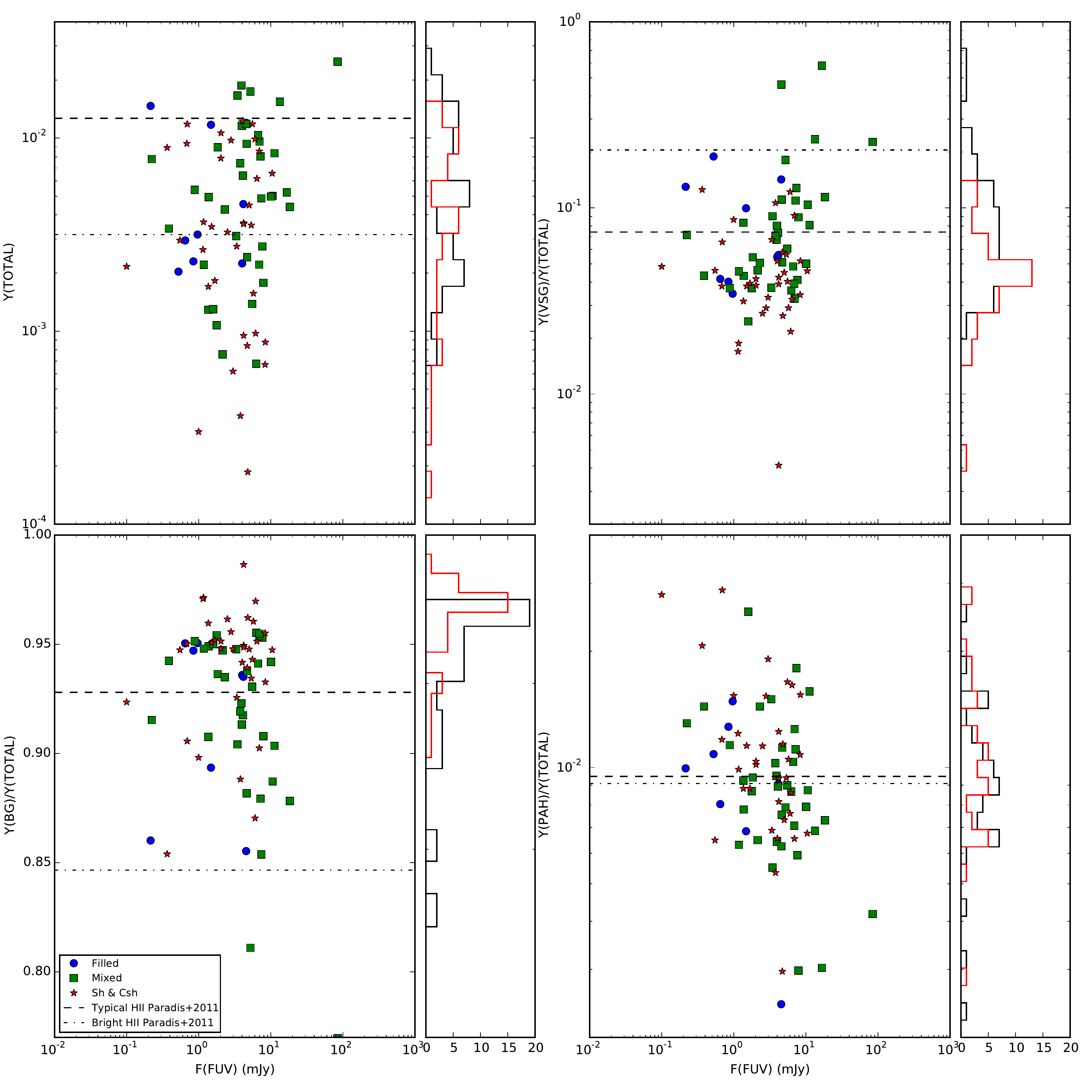}
   \caption{Dust mass abundances obtained fitting the SED of each region using 4\,Myr ISRF and {\it Desert} dust model. Only fits with a $\chi^{2}_{\rm red}<20$ are taken into account. The colour code is the same as in Fig~\ref{fig:Y4myr}. {\it Dot and dashed lines} correspond to the values obtained by \citet{2011ApJ...735....6P} for {\it typical} and {\it bright} \hii\ regions in the LMC modelled assuming a 4\,Myr star cluster ISRF and  {\it Desert} dust model.}
              \label{fig:Y4myr:desert}%
    \end{figure*}

We show here the results of the SED fitting using the classical dust model from \citet{1990A&A...237..215D}. This model does not distinguish between neutral and ionised PAHs and therefore it is not able to fit properly the 6-9\,\mi\ wavelength range corresponding to ionised PAHs. All the SEDs in Fig.~\ref{fig:sed:desert} show high residuals in this wavelength range. However, as we can see from Fig.~\ref{fig:Y4myr:desert}, {\it Desert} dust model also predicts a higher \yvsg\ for {\it filled} and {\it mixed} than for shells and clear shells, showing that the destruction of BGs into VSGs by strong shocks is mainly regulated by the size of the grains.

\section{Robustness of the best fit}
\begin{figure*}
\centering
\includegraphics[width=\textwidth]{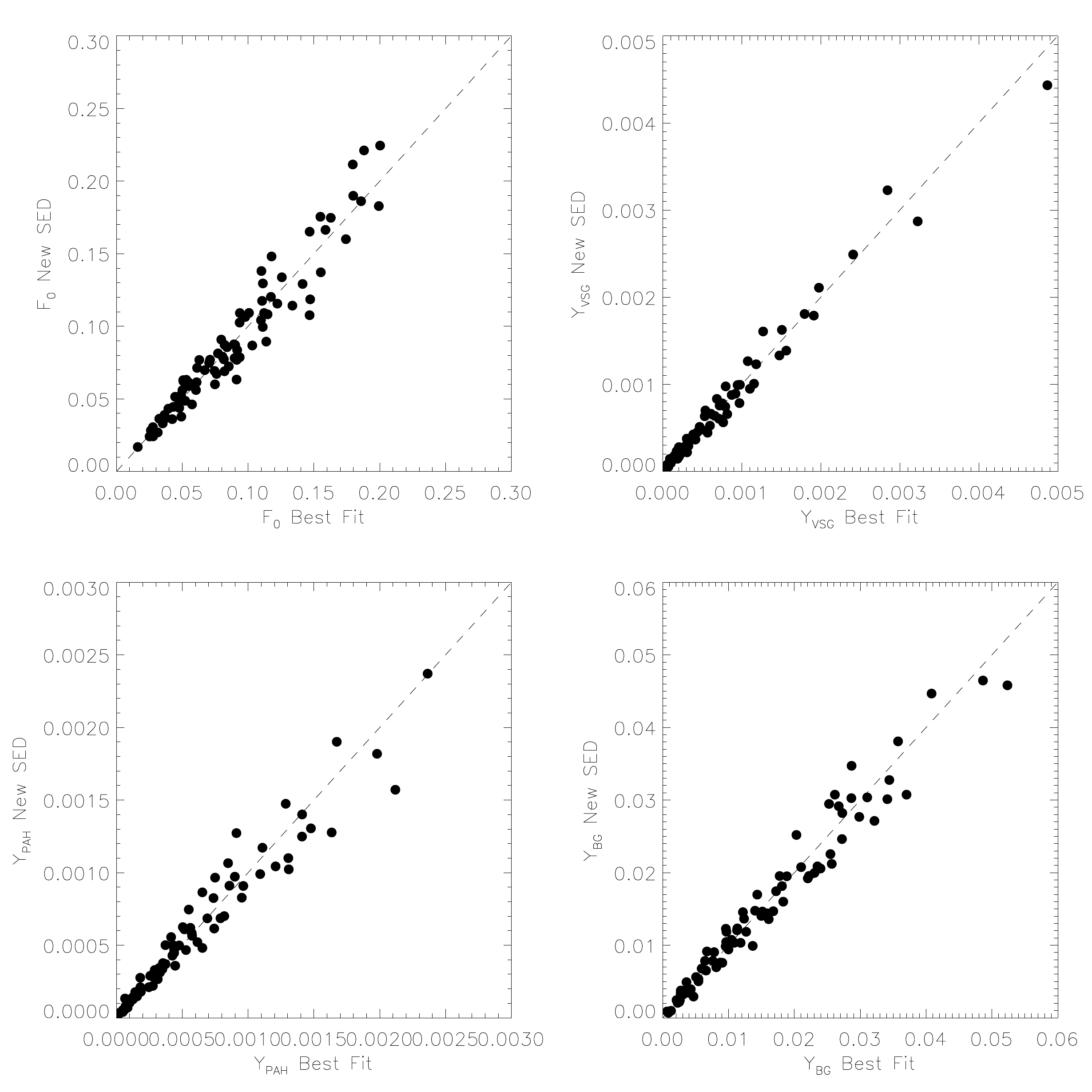}   
\caption{Robustness of the best fit for the $F_{0}$ (top-left), \yvsg\ (top-right), \ypah\ (bottom-left), and \ybg\ (bottom-right). We use the best fit SED given by the minimisation method ($\rm Y_{Best Fit}$) and create a new SED choosing a random flux value in each band from a Gaussian distribution having the best fit flux as a mean value and the observational uncertainty as $\sigma$. The best fit parameters of the new fit ($\rm Y_{New}$) are then compared with those obtained initially from the minimisation method.}
\label{robustfit}
\end{figure*}

The SED fitting is obtained based on the Levenberg-Marquardt minimisation method. We have explored the robustness of the fit using the best fit SED given by the minimisation method and creating a new {\it fake } SED which is then fitted with the same initial parameters. We generate the new fluxes of the  {\it fake} SED in each band choosing a random flux value from a Gaussian distribution having the best fit flux as a mean value and the observational uncertainty as $\sigma$. The best fit parameters of the new fit is then compared with those obtained initially. In Fig.~\ref{robustfit} we show the comparison for the parameters in the fit for each \hii\ region. The relative abundance of the grains derived from the fit are quite robust, however the differences in the ISRF scale seem to increase for higher values of $F_{0}$.

 \end{appendix}

\end{document}